\documentclass[review]{elsarticle}

\usepackage[colorlinks,citecolor=blue,linktoc=all,linkcolor=cyan]{hyperref}
\usepackage{graphicx}

\usepackage[T1]{fontenc}
\usepackage{dsfont}               % use mathds instead of mathbb for outline fonts
\usepackage{mathrsfs}             % provides mathscr without overwriting mathcal
\usepackage{slashed}              % For Dirac slash notation.
\usepackage{amsmath}
\usepackage{amssymb}
\usepackage{amsbsy}
\usepackage{amsfonts}
\usepackage{verbatim}
\usepackage{braket}

\numberwithin{equation}{section}
\numberwithin{table}{section}
\numberwithin{figure}{section}

\journal{Progress in Particle and Nuclear Physics}

\usepackage{titlesec}
\usepackage{sectsty}
\titleformat{\section}{\normalfont\Large\bfseries}{\thesection}{1em}{}
\titleformat{\subsection}{\normalfont\large\bfseries}{\thesubsection}{1em}{}
\titleformat{\subsubsection}{\normalfont\normalsize\bfseries}{\thesubsubsection}{1em}{}
\biboptions{sort&compress}
\begin{document}
	
	\begin{frontmatter}
        \title{Probing Quantum Phenomena through Photoproduction in Relativistic Heavy-Ion Collisions}
			
		    %authors, affiliations, corresponding author mention 
            \author[addr2]{James Daniel Brandenburg}
            \author[addr3]{Spencer R. Klein}
            \author[addr4,addr7]{Zhangbu Xu\corref{mycorrespondingauthor}}
            \cortext[mycorrespondingauthor]{Corresponding author}
		\ead{zxu22@kent.edu}
            \author[addr5]{Shuai Yang}
		\author[addr1]{Wangmei Zha\corref{my2correspondingauthor}}
            \cortext[my2correspondingauthor]{Corresponding author}
		\ead{first@ustc.edu.cn}
            \author[addr6]{Jian Zhou}
		
            \address[addr2]{The Ohio State University, Columbus, Ohio, USA 43210}
            \address[addr3]{Lawrence Berkeley National Laboratory, Berkeley, California, USA  94720}
            \address[addr4]{Physics Department, Kent State University, Kent, Ohio, USA 44242}
            \address[addr7]{Physics Department, Brookhaven National Laboratory, Upton, New York, USA 11973}
            \address[addr5]{Institute of Quantum Matter, South China Normal University, Guangzhou, China 510006}
		\address[addr1]{Department of Modern Physics, University of Science and Technology of China, Hefei P.R. China 230026}
            \address[addr6]{Key Laboratory of Particle Physics and Particle Irradiation (MOE), Institute of Frontier and Interdisciplinary Science, Shandong University, Qingdao, China 266237}
		
		\begin{abstract}
Photoproduction in ultra-peripheral relativistic heavy-ion collisions displays many unique features, often involving quantum mechanical coherence and two-source interference between photon emission from the two ions.  We review the recent experimental results from RHIC and the LHC and theoretical studies of coherent vector meson photoproduction, emphasizing the quantum mechanical aspects of the interactions and the entanglement between the final state particles. These studies enrich our understanding of non-local realism, underscore the critical role of the polarization of the photon source, quantum interference and nuclear effect on the gluon distribution. It paves a way for quantitatively probing the quantum nature of these high-energy nuclear collisions.
		\end{abstract}

	\end{frontmatter}

	\thispagestyle{empty}
	\tableofcontents
	
        \newpage
	\section{Introduction}\label{intro}
        \subsection{History from classical to quantum}\label{sub:intro1}
         \hspace{2em}The development of quantum mechanics was a seminal point in the annals of physics.  With its wave functions and wave-particle duality principles, it resolved many challenges posed by new data that could not be explained by classical physics.
Quantum theory permeates modern physics, and is making significant inroads into disciplines such as chemistry~\cite{levine2014quantum}, biology~\cite{Engel:2007zzb}, and contemporary technology~\cite{nielsen2010quantum}. Quantum mechanics has demonstrated unparalleled success, becoming one of the most rigorously tested physical theories ever developed~\cite{Chao1930PhysRev.36.1519,goworek201480,feynmanQuantumElectrodynamics1998}.

However, beneath its triumphant success lie some profoundly counter-intuitive concepts and notions. R. P. Feynman famously asserted, ``I think I can safely say that nobody today understands quantum physics''~\cite{feynman2017character}, while Roger Penrose remarked that the theory ``makes absolutely no sense''~\cite{penrose2021road}. As recently as 2012, Steven Weinberg commented that ``There is now in my opinion no entirely satisfactory interpretation of quantum mechanics''~\cite{PhysRevA.85.062116}. The challenge lies in the arcane nature of the wave function, its perplexing collapse, the intricacies of state superposition, and the enigmatic phenomenon of entanglement. These aspects continue to provoke animated debates within the scientific community regarding their mechanisms~\cite{PhysRevA.85.062116,bassi2013models}\ and the philosophical underpinnings of the theory itself~\cite{tegmark1998interpretation,ballentine1970statistical,jammer1974philosophy}.  Various interpretations have been proposed to tackle these foundational questions. The Copenhagen interpretation~\cite{Bohr:1928vqa,Bohr1949-BOHDWE,aHeisenberg:1927zz,Born:1926yhp}, pilot wave theories (e.g., Bohmian mechanics~\cite{dürr2009bohmian}), many worlds theories~\cite{Everett:1957hd}, QBism~\cite{fuchs2014introduction}, and retro-causal interpretations~\cite{Cramer:1986zz} are diverse efforts to grapple with these inherent mysteries.  Each interpretation introduces unique perspectives and frameworks, contributing to the ongoing discourse surrounding these fundamental issues. While some physicists engage in probing the foundational intricacies of quantum mechanics, others adopt a pragmatic `Shut up and calculate' approach~\cite{mermin2004could}, emphasizing the efficacy of the quantum formalism in solving problems, while downplaying the need for (or admitting the futility of)  exhaustive contemplation of the theory's philosophical underpinnings. Nevertheless, recent strides in addressing an array of experimental intricacies, including issues related to freedom-of-choice, fair-sampling, communication (or locality), coincidence, and memory loopholes~\cite{Scheidl:2008cgy, Hensen:2015ccp, Giustina:2013yhz, Rauch:2018rvx, Ansmann:2009ims, Rowe:2001kop}, have sparked a renewed surge of curiosity in unraveling the foundational aspects of quantum mechanics. 
        \subsection{Wave-particle duality and quantum entanglement}\label{sub:intro2}
         \hspace{2em}The double-slit experiment is a quintessential demonstration of quantum mechanics, encapsulating the mystery of wave-particle duality.   As articulated by Feynman~\cite{feynman1963feynman}, this experiment represents the solitary enigma that defines quantum mechanics distinctly apart from classical theories, such as electrodynamics:

\begin{quote}
I will take just this one experiment, which has been designed to contain all of the mystery of quantum mechanics, to put you up against the paradoxes and mysteries and peculiarities of nature one hundred per cent. Any other situation in quantum mechanics, it turns out, can always be explained by saying, 'You remember the case of the experiment with the two holes? It's the same thing'. 
\end{quote}

The quantum double-slit thought experiment dates back to 1909 when G. I. Taylor performed an experiment with low-intensity light~\cite{Taylor1909}, strategically reducing incident light levels to the point where only a single photon was in the interferometer at once.  In 1961 Claus Jönsson expanded the experiment to coherent electron beams and multiple slits~\cite{jonsson1961elektroneninterferenzen}. Another significant milestone occurred in 1974, when Italian physicists Merli, Missiroli, and Pozzi conducted a related experiment with single electrons~\cite{merli1976statistical}. This underscored the statistical nature of the interference pattern as anticipated by quantum theory. 
In 2012, Stefano Frabboni and colleagues showcased the double-slit interference pattern using nanofabricated slits and fast single-electron detection allowing the build-up of the interference pattern from each individual electron to be observed~\cite{FRABBONI201273}. 
These experiments, pivotal in modern electron diffraction and imaging, underscore the enduring significance of the double-slit experiment in unraveling the intricacies of wave-particle duality.  In 2018 Marco Giammarchi's group at the Positron Laboratory in Como, Italy, demonstrated single-particle interference of antimatter~\cite{Ariga:2018rjq}, performing the double-slit experiment with positrons. Studies in recent decades have used a diverse set of particles, including neutrons~\cite{Zeilinger:1988zz}, helium atoms~\cite{Carnal:1991zz}, C$_{60}$ fullerenes~\cite{Arndt:1999kyb}, Bose-Einstein condensates~\cite{andrews1997observation}, and even biological molecules~\cite{hackermuller2003wave}.  These diverse experimental pursuits showcase the adaptability of the double-slit experiment. To that end, a major portion of this review will discuss a variation of the double-slit experiment in a unique femto-scale environment. 

In addition to wave-particle duality, quantum entanglement is another defining feature of quantum mechanics not seen in any classical theory. Famously referred to by Einstein as ``spooky action at a distance''~\cite{heimann1973reviews}, quantum entanglement describes the peculiar correlation between particles where measuring the state of one instantaneously constrains the state of its entangled counterpart, even when separated by vast (spacelike) distances. This phenomenon challenges classical notions of locality and realism, providing profound insights into the nature of quantum correlations. One of the first attempts to explore entanglement was in 1949 by C S. Wu and I. Shaknov, who looked at the angular correlations between two photons from positron annihilation, to study their polarization correlations.  Unfortunately, they were not able to observe non-classical correlations \cite{PhysRev.77.136}, but a later experiment in the 1980s~\cite{Aspect:1982fx} found evidence against local hidden-variable theories and affirmed the non-local character of quantum entanglement. Advances in experimental techniques have since enabled researchers to entangle different types of particles, including photons~\cite{Wang:2016hzz}, electrons~\cite{Hensen:2015ccp}, and even atoms~\cite{Duan:2001hoq}. These experiments typically involve creating entangled pairs and measuring their properties in different locations, reinforcing the non-separability implied by quantum entanglement. The development of quantum technologies, such as quantum cryptography and quantum communication, relies heavily on the principles of entanglement for secure information transmission.

Traditionally in quantum mechanics, interference only occurs for ``indistinguishable'' states. A variation of the double-slit setup, so-called `which-way' experiments, demonstrate the destruction of the observed double-slit interference pattern when information is gained about which slit a particle traversed~\cite{qmww} - effectively making the paths distinguishable. For this reason, all the double-slit and entanglement experiments mentioned above used a source of identical particles, whether they be bosons (e.g. photons) or fermions (e.g. electrons).
However, recent experimental and theory work has explored the potential for recovering interference effects between distinguishable quantum states through phase locking and entanglement~\cite{CotlerPhysRevLett.127.103601,NoelPhysRevLett.75.1252}. 
In particular, a phenomenon~\cite{COTLER2021168346} called ``Entanglement Enabled Intensity Interferometry'' ($E^2I^2$) has demonstrated the ability to recover quantum mechanical wavefunction interference in cases that would traditionally not yield interference - between photons of different wavelengths.  An analog of $E^2I^2$ in relativistic heavy ion collisions will be discussed throughout this review.

        \subsection{Quantum mechanics in nuclear physics}\label{sub:intro3}
         \hspace{2em}Researchers have sought visual representations of their subject in almost every field, from images of tiny proteins~\cite{nogales2015cryo} to massive black holes~\cite{collaboration2019first}. Such images can be powerful tools for understanding and provide insights into complex systems that are otherwise difficult to conceptualize. An essential goal of nuclear physics is to image the internal structure of nucleons and nuclei, to better understand how they form and how their emergent properties arise~\cite{Arneodo:1992wf,thomas2010structure,H1:2015ubc}. The wave-like behavior of light, and quantum phenomena in general, have provided non-classical methods of imaging physical processes in a multitude of fields of science. 
For instance, X-ray crystallography is an example of a classical optical diffraction-based imaging technique that utilizes the wavelike nature of light to illuminate the arrangement of microscopic structures and has led to numerous Nobel prizes~\cite{Nobel1914,Nobel1915,Nobel1962,Nobel1962-1,Nobel1982,Nobel2009}.
Based on similar principles, elastic electron-proton scattering, where a high-energy electrons interact with a proton, has long been utilized to `image' the internal structure and dynamics of the proton. This and similar methods provide valuable information about the distribution of quarks and the electromagnetic properties of the proton~\cite{PhysRev.142.922, Carlson:2015jba}. 
Elastic scattering experiments involving protons, neutrons, and light nuclei with heavier nuclei offer a unique avenue for uncovering the structure and dynamics within larger nuclear systems. These experiments utilize diffraction patterns and angular distributions to explore nuclear shell structure, collective excitations, and halo nuclei~\cite{KEELEY2009396,Tanihata:2013jwa,duan2022elastic}.

Low energy nuclear interactions reveal quantum effects in 
a wide range of phenomena, including nuclear reactions, particle decays, and exotic nuclear states. Quantum tunneling plays a crucial role in nuclear fusion, allowing particles to overcome potential barriers and facilitate reactions~\cite{Hagino:2012cu}. Nuclear fission, where heavy nuclei split into lighter fragments, also relies on quantum tunneling, highlighting the intricate relationship between nuclear structure and dynamics~\cite{Balantekin:1997yh}.

Ultra-relativistic collisions of heavy nuclei enable the study of quantum effects in high energy nuclear interactions. High energy heavy-ion collisions were originally developed as a laboratory for producing the extreme energy density and temperature required to create the quark-gluon plasma (QGP)~\cite{Braun-Munzinger:2007edi}. However, the high nuclear charges and relativistic velocities of heavy-ions also produce highly Lorentz contracted electromagnetic fields - the strongest electric and magnetic fields in nature. These fields are equivalent to a large flux of high-energy photons. This review will explore the utilization of these high energy photons to probe nuclear matter through coherent photoproduction - where a photon interacts with a target nucleus to produce a hadronic final state. 
These processes are most cleanly studied in ``Ultra-Peripheral Collisions''~\cite{Bertulani:2005ru} (UPCs) where ions interact via their long-range electromagnetic fields without direct contact. These processes are sensitive to the distribution of gluons within nuclei and test the strong force under unusual conditions - without color exchange.

\begin{figure}[htbp]
\centering
\includegraphics[width=0.95\textwidth]{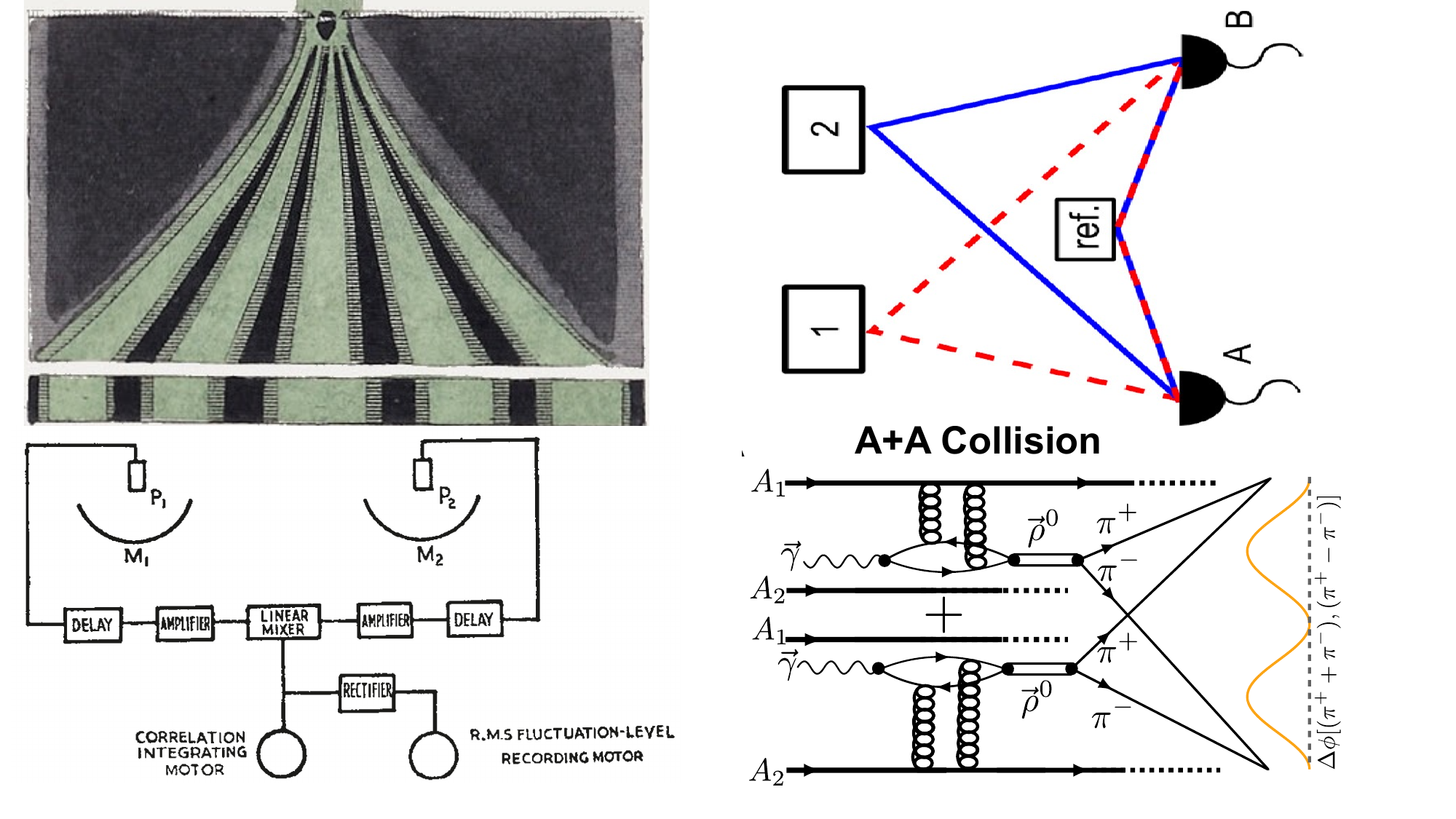}
\caption{Top left: Observing interference pattern in the classical Young's double-slit experiment~\cite{Young1807bhlitem63005}; Bottom left: Detecting two photon coincidence in the HBT Intensity Interference~\cite{HanburyBrown:1956bqd,Baym:1997ce}; Top right: Detecting two photons with difference wavelengths from sources 1 and 2 at detectors A and B in the Entangled Enhanced Intensity Interference~\cite{COTLER2021168346}; Bottom right: Detecting one pair of $\pi^\pm$ in the Entangled Enhanced Nuclear Tomography~\cite{STAR:2022wfe}. Plots are adapted from the references.}
\label{fig:4exp}
\end{figure}

In particular, the quantum phenomena of $E^2I^2$~\cite{COTLER2021168346} is cited in a recent publication by the STAR Collaboration~\cite{STAR:2022wfe}. Same effect was observed by the ALICE Collaboration with an impact parameter dependence in the ultra-peripheral collisions~\cite{ALICE:2024ife}. As will be discussed below, photoproduction in heavy-ion collisions shares many similarities with the double-slit experiment and the recently proposed $E^2I^2$ phenomena. Nevertheless, photoproduction has several distinct difference as depicted in Fig.~\ref{fig:4exp}: traditional double-slit experiments used one photon source passing through two possible slits (top left panel)~\cite{Young1807bhlitem63005}; the Hanbury-Brown and Twiss (HBT) intensity interferometry effect~\cite{HanburyBrown:1956bqd,Baym:1997ce} (bottom left) detects a pair of photon (or identical pion) correlations from a distant source (stars, nuclear collisions); $E^2I^2$ (top right), a variation of HBT, detects the correlation between two photons of distinctly different wavelength after entanglement and manipulation~\cite{COTLER2021168346,CotlerPhysRevLett.127.103601}, while in diffractive photonuclear process in UPCs~\cite{STAR:2022wfe} (bottom right), a pair of \textit{distinguishable} particles (a $\pi^{+}$ and $\pi^-$) from two sources of $\rho^0$ that decay at a distance displays interference.  

The exploration of quantum phenomena in nuclear physics continues to push the boundaries of our understanding, from classical analogies to high-energy probes. The study of quantum effects in heavy-ion collisions, especially through coherent photoproduction, offers a window into the fundamental interactions governing nuclear matter. This review primarily focuses on the quantum phenomena observed in photoproduction processes in heavy-ion collisions. Through detailed analysis and discussion, we aim to provide a comprehensive understanding of how these quantum effects manifest and what they reveal about the fundamental interactions in nuclear matter.

        \subsection{ Objectives and Scope of the Review}\label{sub:intro4}
        
\hspace{2em}Recent experimental studies at both RHIC and LHC have found distinctive spin interference patterns from $\rho^0$ photoproduction in relativistic heavy-ion collisions, 
These findings are the first instances of Fermi-scale single-particle double-slit interference experiments, which pave the way to significantly enhance our understanding of quantum phenomena in high-energy environments. This intriguing observation, exhibiting features of the double-slit experiment and quantum entanglement, serves as the primary motivation for this comprehensive review. The aim of the review is to review the theoretical formulation and experimental development of photoproduction in high-energy nuclear experiments. 

The review is structured to provide a coherent journey through the subject matter. Beginning with Section~\ref{intro}, we embark on an exploration of the historical development of quantum mechanisms. Section~\ref{second} navigates through progressions in the investigation of photoproduction within high-energy collisions. Following this groundwork, Section~\ref{third} immerses readers in the observations of quantum phenomena related to vector meson photoproduction, with a special emphasis on the recently discovered spin-interference pattern.
Section~\ref{fourth} delves into a detailed discussion on coherence requirements, observational effects, and their intriguing connection to the domain of quantum entanglement. As we look towards the future and broader perspectives, Section~\ref{fifth} explores potential developments in the field and related topics that shape the ongoing research landscape.
Our review concludes in Section~\ref{sec:sum} with succinct summaries, encapsulating the key insights and contributions discussed throughout. Through this narrative structure, we aim to provide readers with a holistic understanding of the quantum mechanisms governing photoproduction within the dynamic setting of relativistic heavy-ion collisions.
 
	\section{Photoproduction in high energy collisions}\label{second}
  \hspace{2em}In high-energy nuclear physics, photons can attain energies on the order of gigaelectronvolts (GeV), far surpassing the electronvolt (eV) scale typically associated with ordinary light. Such high-energy photons can result from bremsstrahlung of a high-energy electron beam or from Compton backscattering of light by high-energy electrons~\cite{Bauer:1977iq}. High energy photons are also produced in heavy-ion collisions by the highly Lorentz-contracted electromagnetic fields induced by charged particles moving at velocities close to the speed of light~\cite{vonWeizsacker:1934nji,Williams:1934ad}. Possessing exceedingly short wavelengths, photons with energy at the GeV-scale and above are capable of probing the fundamental properties of matter at the femtoscale. 
At these energy scales, the quantum fluctuations of photons in the vacuum become significant.
In its bare state, the photon can only be involved in electromagnetic interactions, such as Compton scattering. However, the photon can also fluctuate into a virtual $q\bar{q}$ pair, which may then scatter with the target via the strong nuclear force. With quantum numbers $J^{PC} = 1^{--}$, the photon preferentially fluctuates into a vector meson.
The wave function of a photon can be expressed through a Fock decomposition~\cite{Schuler:1993td}:
\begin{equation}
\label{Fock_photon_wave}
|\gamma \rangle = C_{\rm{bare}}|\gamma_{\rm{bare}}\rangle + C_{V}|V\rangle + ... + C_{q}|q\bar{q}\rangle.
\end{equation}
Here, $C_{\rm{bare}} \approx 1$ and $C_{V} \sim \sqrt{\alpha}$ ($V = \rho, \omega, \phi, J/\psi, ...$), and $\alpha\approx 1/137$ is the fine-structure constant. The coefficient $C_V$ is determined by the photon-vector meson coupling, $f_V$, through $C_V = \sqrt{4\pi \alpha}/f_V$, where the coupling $f_V$ can be extracted from the vector meson leptonic decay width $\Gamma(V \rightarrow e^{+}e^{-})$.  The fluctuated product subsequently interacts with the target particle and emerges as real, a phenomenon known as photoproduction. In a related process, Compton scattering, the $q \overline q$ dipole will fluctuate back into a photon after scattering \cite{Belitsky:2001ns}. For all of these processes, at high energies, the lifetime of these fluctuations is longer than the time required for the dipole to pass through the nucleus.   In this section, we will briefly review the experimental measurements of photoproduction in high-energy collisions.

	\subsection{Experimental measurements of photoproduction processes in e/$\gamma$+p/A collisions}\label{sub: chapter3:first}
      \hspace{2em}High energy photon beams can be generated via bremsstrahlung by directing a high-energy electron beam onto a selected radiator material. The choice of radiator material and precise control over the energy and trajectory of the electron beam are crucial for optimizing the characteristics of the resulting photon beam, including its energy spectrum and intensity~\cite{Kellie:2005wy}. To select a narrow-band photon beam, monoenergetic electrons can be directed into a a thin photon-generating target, and the momentum of the outgoing electron can be measured, producing tagged photons of known energy~\cite{Alvensleben:1970it}. Linearly polarized photon beams can be made using diamond radiators through coherent bremsstrahlung~\cite{Criegee:1968fxl}.

High-energy photon beams can also be produced when light interacts with high-energy electrons via Compton back-scattering. The kinematics of this two-body interaction dictates that the photon energy is related only to its emission angle relative to the electron direction and to the electron energy.  By constraining the photon scattering angle, a nearly monochromatic high-energy photon beam can be generated~\cite{Bingham:1970ks}. An intriguing characteristic of the Compton process is its polarization: if the incident light is polarized, the high-energy segment of the resulting Compton spectrum becomes highly polarized in a manner consistent with the incident light~\cite{Ballam:1970qn}. Similarly, a monochromatic photon beam can also be produced by positron annihilation, where the resulting photons maintain a high degree of energy consistency~\cite{Ballam:1968du}.

In the late 1950's, physicists demonstrated the feasibility of probing high-energy photoproduction in hydrogen through the utilization of the track chamber technique, as evidenced by early experiments conducted with a hydrogen-diffusion cloud chamber~\cite{PhysRev.113.1323}. Subsequent breakthroughs ensued with the inception of photoproduction measurements employing a hydrogen bubble chamber by the Cambridge Bubble Chamber Group~\cite{crouch1964gamma}. As noted earlier, the photons can emanate from either bremsstrahlung or Compton back-scattering. Determination of the incident photon beam luminosity can be achieved through the observation of dielectron pair events in conjunction with the known pair production cross-section in the target~\cite{aachen1966photoproduction}, or alternatively, via utilization of a meticulously calibrated quantameter~\cite{Davier:1968xd}. Analysis of the photon energy spectrum can be undertaken by examining either the energy spectrum of dielectron pairs or through a "3C" (3 Constraints) fit of $p\pi^{+}\pi^{-}$ events. The principal advantages of track chamber experiments lie in their ability to provide an over-constrained kinematic fit and to facilitate comprehensive observation of the angular distribution of decay products. This over-constraint permits unambiguous elimination of inelastic events wherein the recoiling target is fragmented. However, these experiments suffer from a limited number of events and consequently, poor statistical precision. To address this limitation, the counter-spectrometer system utilized by the Harvard group at the Cambridge Electron Accelerator~\cite{Alexander:1965zz} offers a solution. This system typically comprises two spectrometers, each equipped with a half-quadrupole magnet and a series of scintillation counters to ascertain the momentum of the tracks originating from the decay of photoproduction products. Particle identification in this system is achieved through the utilization of Cerenkov counters and shower counters. The two spectrometers are independently movable, allowing for the observation of the angular distribution of the decay daughters by adjusting their positions in the horizontal plane. and significant background of secondary particles scattered from the magnets, which can obscure the desired signals and introduce uncertainties into the measurements.

Inelastic lepton scattering experiments bring an additional dimension to photoproduction, by allowing experiments with virtual (space-like) photons.  Early lepton scattering experiments~\cite{Bloom:1969kc,Breidenbach:1969kd,Miller:1971qb} - characterized by a fixed-target setup, encompassing an incident electron/muon beam, a target, a spectrometer positioned at a fixed angle, and a variable energy configuration for the spectrometer - laid the foundation for subsequent advancements in the field. 

The pursuit of higher center-of-mass collision energies led to pioneering collider experiments at the HERA (Hadron Electron Ring Anlage), $ep$ collider~\cite{Voss:1994sr}. HERA's distinctive configuration comprises separate synchrotron rings for electrons and protons, spanning approximately 6.3 km in circumference, %10-30 meters beneath the Earth's surface, 
where 30 GeV electrons collide with 920 GeV protons. 
Housing four experimental areas, HERA serves as a key facility for $ep$ interactions. Two of these areas were dedicated collision points for the ZEUS and H1 collaborations, while the others were used for fixed target experiments. Photoproduction measurements were primarily carried out by ZEUS and H1. These general-purpose detectors boast nearly hermetic calorimetric coverage with a distinct forward-backward asymmetry tailored to accommodate the boosted $ep$ center-of-mass in the direction of the proton beam. Equipped with central tracking detectors operating within a magnetic field, sampling calorimeters responsive to both electromagnetic and hadronic constituents, H1 and ZEUS detectors offered extensive detection capabilities for their era. Additionally, they feature muon detectors and forward/backward electron and proton taggers. The determination of luminosity at HERA hinges upon the bremsstrahlung reaction ($ep \rightarrow ep \gamma$), with both ZEUS and H1 meticulously equipped to measure this process - one targeting electrons and the other photons.

The kinematics of the process $ep \rightarrow epX$ are governed by several key variables: the square of the $ep$ center-of-mass energy $s = (p+k)^{2}$; the negative four-momentum transfer squared at the lepton vertex (carried by the virtual photon) $Q^2 = -q^2= (k-k^{\prime})^2$; the four-momentum transfer squared at the proton vertex $t = (p-p^{\prime})^2$; and the inelasticity $y = (p \cdot q)/(p \cdot k)$. Here, $k$, $k^{\prime}$, $p$, $p^{\prime}$, and $q$ represent the incident and scattered electron, incoming and outgoing proton, and exchanged photon, respectively, with $X$ denoting the reaction product. Neglecting the proton mass (appropriate at high energies), the center-of-mass energy of the photon-proton system is $W_{\gamma p}=(p+q)^{2} = ys -Q^2$. In the photoproduction limit ($Q^2 \rightarrow 0$), both the incident electron and proton scatter at small angles, eluding direct detection. Under such conditions, the square of the $\gamma p$ center-of-mass energy can be reconstructed utilizing the variable $W^2_{\gamma p,rec} = s y^{rec}$, where $y^{rec}$ is the reconstructed inelasticity. Specifically, $y^{rec}= (E_{e} - p_{z, X})/(2E_{e})$, with $E_{X}$ and $p_{z, X}$ denoting the reconstructed energy and momentum along the proton beam direction (z-axis) of the photoproduction product, and $E_e$ representing the electron beam energy. Moreover, the variable $t$ can be approximated from the transverse momentum of $X$ in the laboratory frame, as expressed by the observable $t_{rec} = - p^{2}_{T, X}$. The reconstructed variables $W_{\gamma p,rec}$ and $t_{rec}$ only approximately align with the variables $W_{\gamma p}$ and $t$, owing to their definitions and the smearing effects induced by the detector and beam energty spread. To accurately derive the $\gamma p$ cross section, precise knowledge of the effective photon flux induced by the electron beam is imperative.  This flux is:
\begin{equation}
\label{chapter3-1_eq}
n_{\gamma}^{e}(x) = \frac{\alpha}{2\pi}[\frac{1+(1-x)^2}{x}\log\frac{(1-x)(sx-m_{X}^{2})}{m_{e}^2 x^2}],
\end{equation}
where $x$ is the fraction of electron energy carried by the photon, and $m_{X}$ denotes the mass of the photoproduction product. Deriving this formula involves multiple considerations, including the treatment of photon $Q^2$ and the finite beam size, which imposes an upper limit on impact parameter~\cite{Budnev:1975poe}.

\begin{figure}[!htbp]
    \centering
    \includegraphics[width=0.8\textwidth]{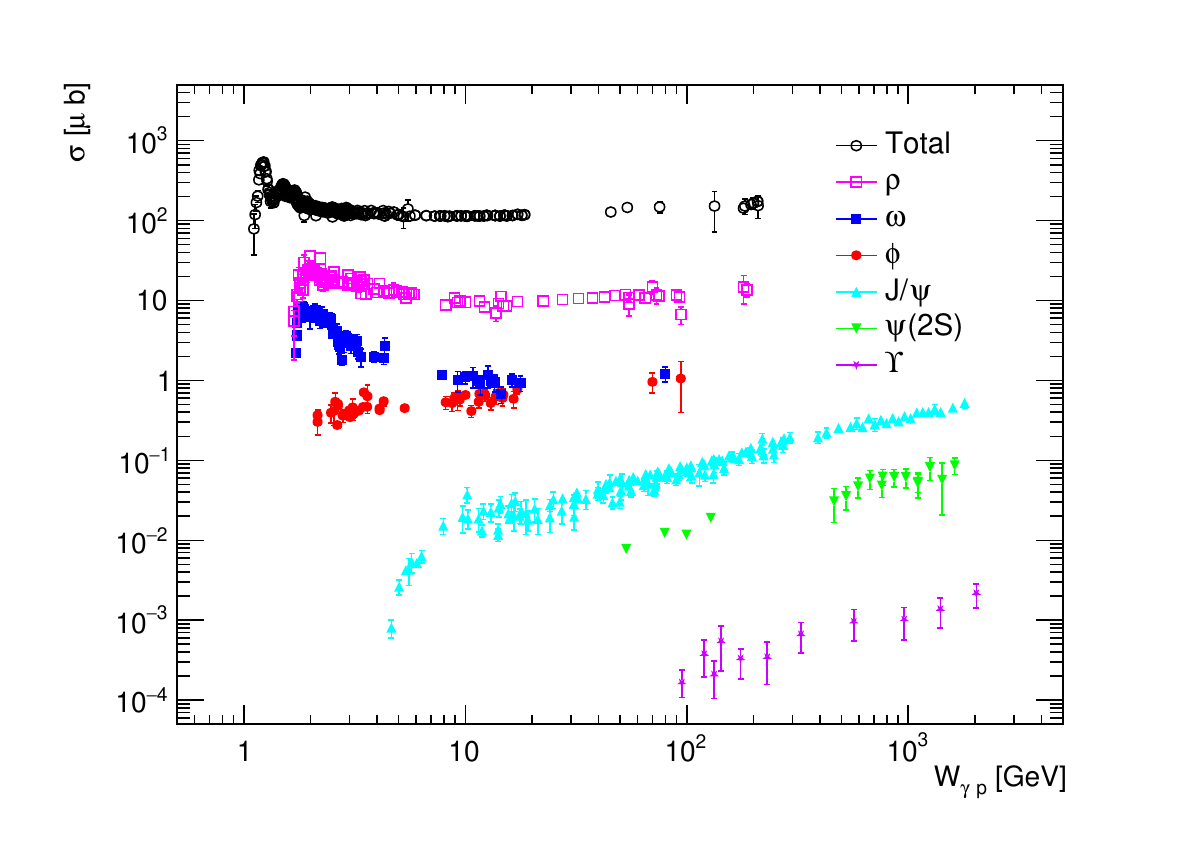} 
    \caption{The world-wide experimental data~\cite{Caldwell:1970fw,Armstrong:1971ns,Alexander:1974zc,Michalowski:1977eg,Armstrong:1972sa,Joos:1976nm,Eisenberg:1975jr,Ballam:1971yd,Caldwell:1978yb,Bloom:1969pn,H1:1995hmw,H1:1992nyi,Bingham:1973fu,Caldwell:1973bu,ZEUS:2010yir,ZEUS:1994mec,ZEUS:1992lda,ZEUS:2001wan,ZEUS:1994mec,Aachen-Hamburg-Heidelberg-Munich:1975jed,Clifft:1976be,Bonn-CERN-EcolePoly-Glasgow-Lancaster-Manchester-Orsay-Paris-Rutherford-Sheffield:1982qiv,PhysRev.146.994,crouch1967photoproduction,Cassel:1981sx,Berger:1972an,Aachen-Berlin-Bonn-Hamburg-Heidelberg-Munich:1968rzt,Francis:1976sq,Egloff:1979mg,Chapin:1985mf,Gladding:1973cac,H1:1999pji,H1:1996prv,H1:2020lzc,HERMES:2000jnb,Bodenkamp:1984dg,Wu:2005wf,Struczinski:1972ond,Anderson:1969kn,Davier:1968dt,Ballam:1972eq,Eisenberg:1971wy,Ballam:1971wq,Ballam:1972eq,ZEUS:1996bgb,ZEUS:1998xpo,ZEUS:1997rof,ZEUS:1999ptu,ZEUS:1995bfs,Clark:1979xs,Gittelman:1975ix,ZEUS:1997kpo,Nash:1976ic,Binkley:1981kv,Denby:1983az,EuropeanMuon:1979nky,H1:2000kis,H1:1996kyo,H1:1999ujo,H1:1996gwv,Camerini:1975cy,ZEUS:1995kab,Strakovsky:2014wja,Breakstone:1981wk,BrownHarvardMITPadovaWeizmannInstituteBubbleChamberGroup:1967zz,CLAS:2002cdi,Egloff:1979xg,Busenitz:1989gq,PhysRevLett.26.155,Eisenberg:1969xc,Eisenberg:1969kk,Eisenberg:1970xe,OmegaPhoton:1983huz,ZEUS:1996zse,Barber:1981fj,Besch:1974rp,CLAS:2000kid,Behrend:1978ik,LAMP2Group:1978mgh,LEPS:2009nuw,OmegaPhoton:1983huz,ZEUS:2005bhf,H1:2002yab,H1:1997fol,ZEUS:1998cdr,ZEUS:2009asc} on the total and vector meson photoproduction as a function of $W_{\gamma p}$. 
    }
\label{fig:cross:all}
\end{figure}

\begin{figure}[!htbp]
    \centering
    \includegraphics[width=0.8\textwidth]{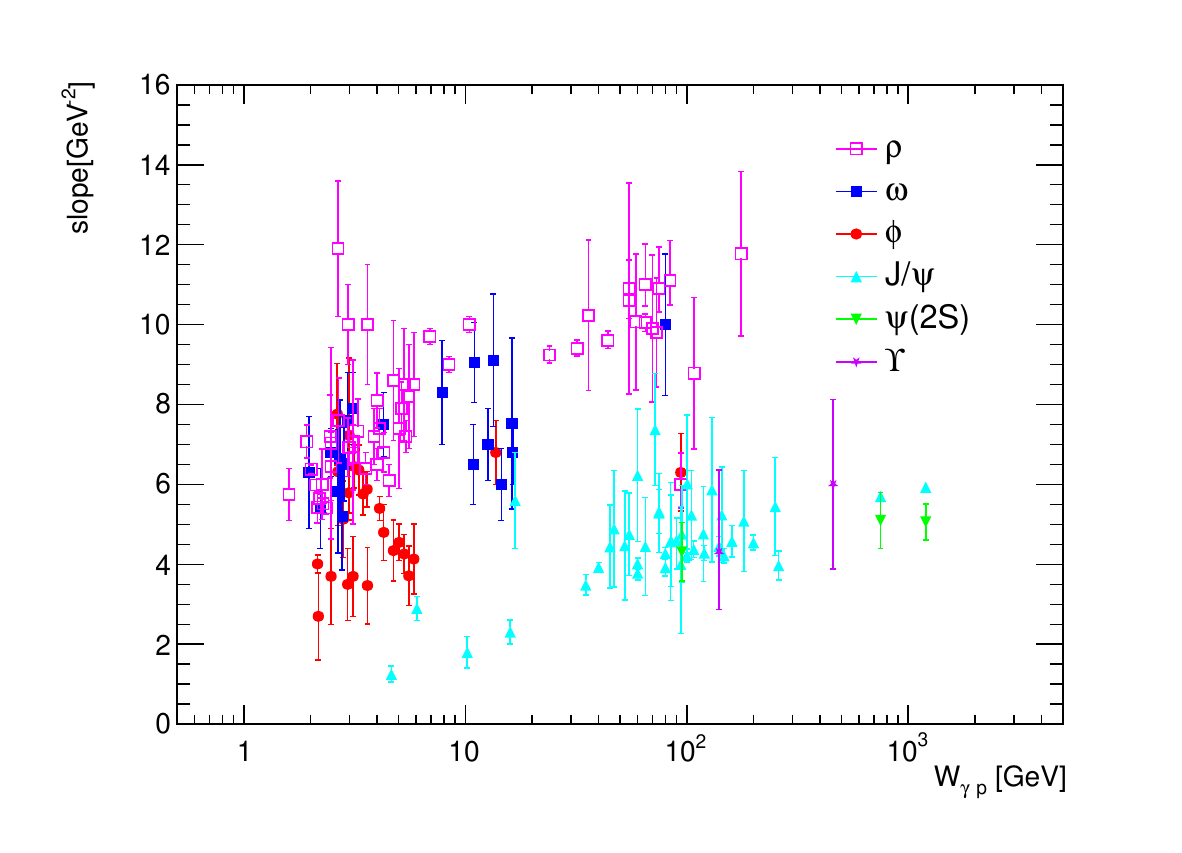} 
    \caption{The world-wide experimental data~\cite{Aachen-Hamburg-Heidelberg-Munich:1975jed,Clifft:1976be,Bonn-CERN-EcolePoly-Glasgow-Lancaster-Manchester-Orsay-Paris-Rutherford-Sheffield:1982qiv,PhysRev.146.994,crouch1967photoproduction,Cassel:1981sx,Berger:1972an,Aachen-Berlin-Bonn-Hamburg-Heidelberg-Munich:1968rzt,Francis:1976sq,Egloff:1979mg,Chapin:1985mf,Gladding:1973cac,H1:1999pji,H1:1996prv,H1:2020lzc,HERMES:2000jnb,Bodenkamp:1984dg,Wu:2005wf,Struczinski:1972ond,Anderson:1969kn,Davier:1968dt,Ballam:1972eq,Eisenberg:1971wy,Ballam:1971wq,Ballam:1972eq,ZEUS:1996bgb,ZEUS:1998xpo,ZEUS:1997rof,ZEUS:1999ptu,ZEUS:1995bfs,Clark:1979xs,Gittelman:1975ix,ZEUS:1997kpo,Nash:1976ic,Binkley:1981kv,Denby:1983az,EuropeanMuon:1979nky,H1:2000kis,H1:1996kyo,H1:1999ujo,H1:1996gwv,Camerini:1975cy,ZEUS:1995kab,Strakovsky:2014wja,Breakstone:1981wk,BrownHarvardMITPadovaWeizmannInstituteBubbleChamberGroup:1967zz,CLAS:2002cdi,Egloff:1979xg,Busenitz:1989gq,PhysRevLett.26.155,Eisenberg:1969xc,Eisenberg:1969kk,Eisenberg:1970xe,OmegaPhoton:1983huz,ZEUS:1996zse,Barber:1981fj,Besch:1974rp,CLAS:2000kid,Behrend:1978ik,LAMP2Group:1978mgh,LEPS:2009nuw,OmegaPhoton:1983huz,ZEUS:2005bhf,H1:2002yab,H1:1997fol,ZEUS:1998cdr,ZEUS:2009asc} on the exponential slopes of  the $-t$ distribution of vector meson photoproduction as a function of $W_{\gamma p}$. 
    }
\label{fig:slope:all}
\end{figure}

The experimental data on the energy dependence of vector meson photoproduction cross sections are summarized in Fig.~\ref{fig:cross:all}. For light mesons, the energy dependence far above threshold is well described by the quark-vector meson dominance (quark-VMD) model~\cite{Bauer:1977iq}, in which the photon couples to a quark-antiquark pair that then forms a vector meson interacting with the hadron.  There, the light vector meson ($\rho$, $\omega$, and $\phi$) cross sections increase slowly with energy roughly as $W_{\gamma p}^{0.22}$.  This may be ascribed to a photon-Pomeron reaction, where the Pomeron carries the colorless strong force, and has the same quantum numbers as the vacuum.  In perturbative Quantum Chromodynamics (pQCD), Pomeron is mostly gluonic, and must be composed of at least two gluons, to preserve color neutrality.  

For the $\rho^0$ and $\omega$ mesons, an additional component appears at lower energies, exhibiting a behavior characterized by approximately $W_{\gamma p}^{-1.2}$ for the $\rho^0$ and $W_{\gamma p}^{-2.0}$ for the $\omega$. This additional component is due to photon-Reggeon interactions, involving the exchange of Reggeonized meson trajectories, so the exchanged particles are mostly quark-antiquark pairs.  These trajectories can operate when the target baryon and the final state mesons share at least one quark, {\it i. e.} for mesons that contain $u$ or $d$ quarks.  These exchanges can accommodate a wider range of quantum numbers than the Pomeron. In contrast, the heavy-meson cross-sections rise faster with energy, and an empirical formula provides a more accurate description. 
For $J/\psi$ photoproduction, the cross section roughly follows $W_{\gamma p}^{(0.7-0.8)}$. This energy dependence highlights the different underlying dynamics for the photoproduction of light and heavy vector mesons. The steeper rise in the cross section for heavy mesons points to a higher sensitivity to the gluon distribution in the proton, consistent with the expectations from pQCD.

The experimental data on the $|t|$ dependence of vector meson photoproduction is summarized in Fig.~\ref{fig:slope:all}. The $t$-dependence of the differential cross section at small-$t$ is usually parameterized in
terms of the slope $b_{V}$ ($\frac{d\sigma}{dt} = A\exp(-b_{V}t)$) of the diffractive peak. The fitted values of $b_V$ depend slightly on the range of $t$, therefore a more refined parameterization $\frac{d\sigma}{dt} = A\exp(-b_{V}|t|-c|t|^2)$ with a curvature parameter $(c)$ allows one to extend the fits of the experimental data up to $|t| \sim 1 {\rm GeV}^2$ . The allowance for the curvature $c$ does not shift the value of fitted $b_V$ significantly. Additionally, alternative parameterizations of the effective slope are available in the literature, such as defining $b_V$ as the derivative of the logarithm of the differential cross section at $t = 0$: $b_V = \frac{1}{\sigma} \frac{d\sigma}{dt}|_{t=0}$ \cite{Ivanov:2004ax}.
The decomposition of the diffraction slope into the target transition, beam transition and the $t$-channel exchange components is exact in the simple Regge model~\cite{Collins:1977jy}. Similar decomposition holds also in the $k_{\perp}$ factorization~\cite{Catani:1990eg} and color dipole approaches~\cite{Mueller:1993rr} despite breaking of the strict Regge factorization. In the photoproduction the key point is the shrinkage of the $q \bar{q}$ state of the photon in going from light to heavy quarks, accompanied by the related decrease of the radius of the ground state vector meson, which entail the hierarchy of diffraction slopes $b_{J/\psi} < b_{\phi} < b_{\rho}$. It would be worth noting that as the photon energy increases, the proton becomes "blacker," causing $b_V$ to vary slowly with $W_{\gamma p}$, thereby reflecting the energy evolution of the interaction profile.

\begin{figure}[!htbp]
    \centering
    \includegraphics[width=0.9\textwidth]{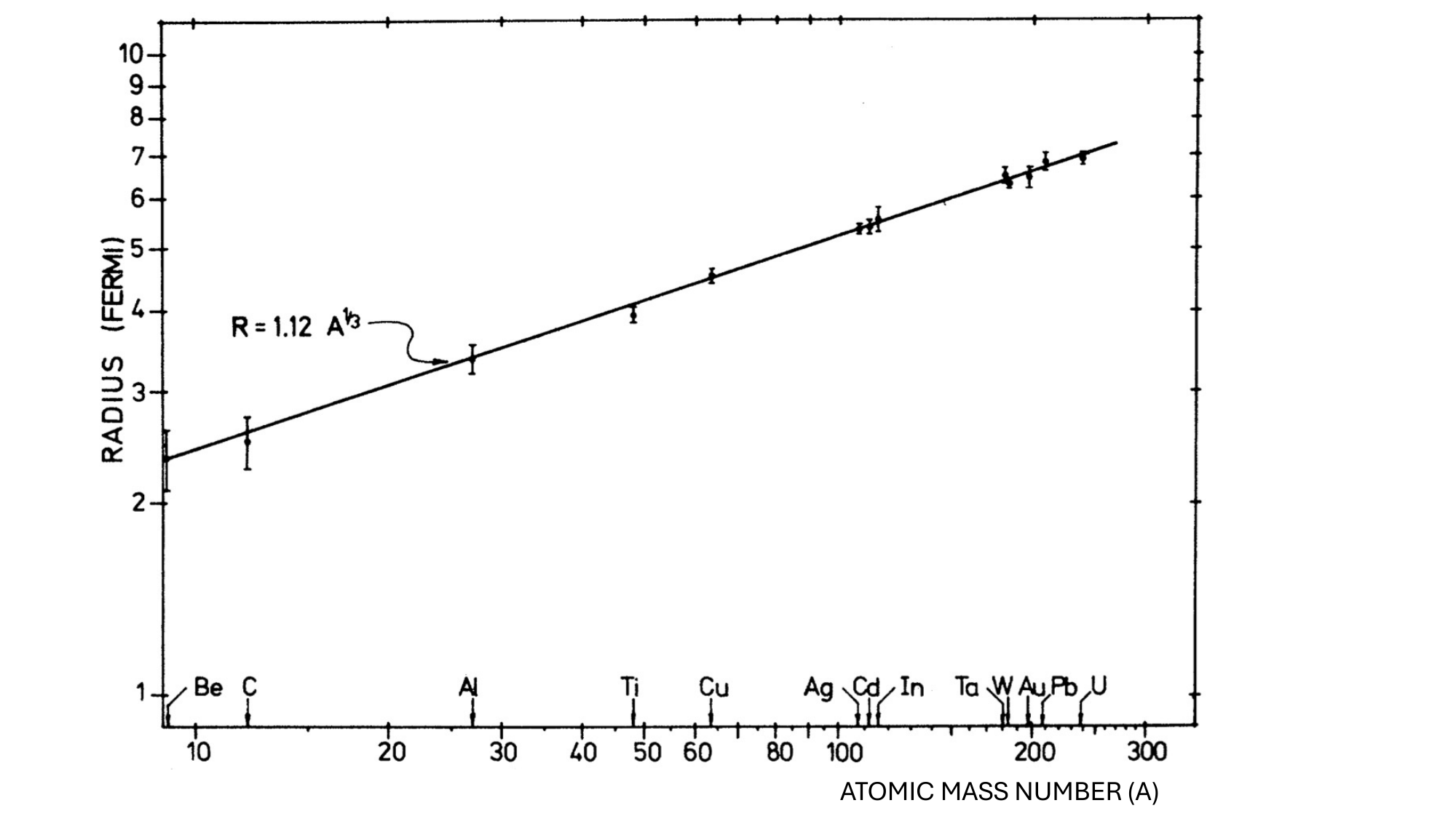} 
    \caption{Extracted radius from a multi-dimensional fits to the diffractive $\rho^0$ photoproduction as a function of atomic mass number. Figure from Ref~\cite{Alvensleben:1970uv}.
    }
\label{fig:DESY1970rho2RvsA}
\end{figure}

Experimental work at the Cambridge Electron Accelerator~\cite{Lanzerotti:1968PhysRev.166.1365} in 1968 showed for the first time diffraction-like coherent photoproduction of $\rho^0$ using 5.5 GeV and 3 GeV photons on proton and carbon targets. Subsequently, exclusive coherent photonuclear production of $\rho^0$ was used at various facilities to determine the nuclear radii in the early 1970s ~\cite{Alvensleben:1970uw,Alvensleben:1970uv,TingPhysRevLett.27.888,CornellPhysRevD.4.2683} with high-energy photon incident on nuclear targets. The $|t|$ distribution from the diffractive process was fitted with the Fourier-Transform of the Woods-Saxon nuclear density distribution~\cite{Alvensleben:1970uv}. A scaling of $R=(1.12\pm0.02)A^{1/3}$ was extracted from the experimental data at DESY~\cite{Alvensleben:1970uw,Alvensleben:1970uv} with $R_{\rm Au}=6.45\pm0.27$ ${\rm fm}$ and $R_{\rm U}=6.90\pm0.14$ ${\rm fm}$ as shown in Fig.~\Ref{fig:DESY1970rho2RvsA} while another experiment at Cornell~\cite{CornellPhysRevD.4.2683} obtained $R_{\rm Au}=6.74\pm0.06$ ${\rm fm}$. These values are used for comparison to recent RHIC results from ultra-peripheral A+A collisions in later sections. The cross section ratios between coherent and incoherent interactions for photonuclear $\rho^0$ production have also been used as an effective tool to study the color transparency at HERA and JLab in later experiments~\cite{JAINColor199667,CLASELFASSI2012326}. 

	\subsection{Photoproduction in heavy-ion collisions}\label{sub:chater3:second}
        \subsubsection{The Weizsäcker-Williams photons in relativistic heavy-ion collisions}\label{sec3-2-1}
        
\begin{figure}[!htbp]
    \centering
    \includegraphics[width=0.4\textwidth]{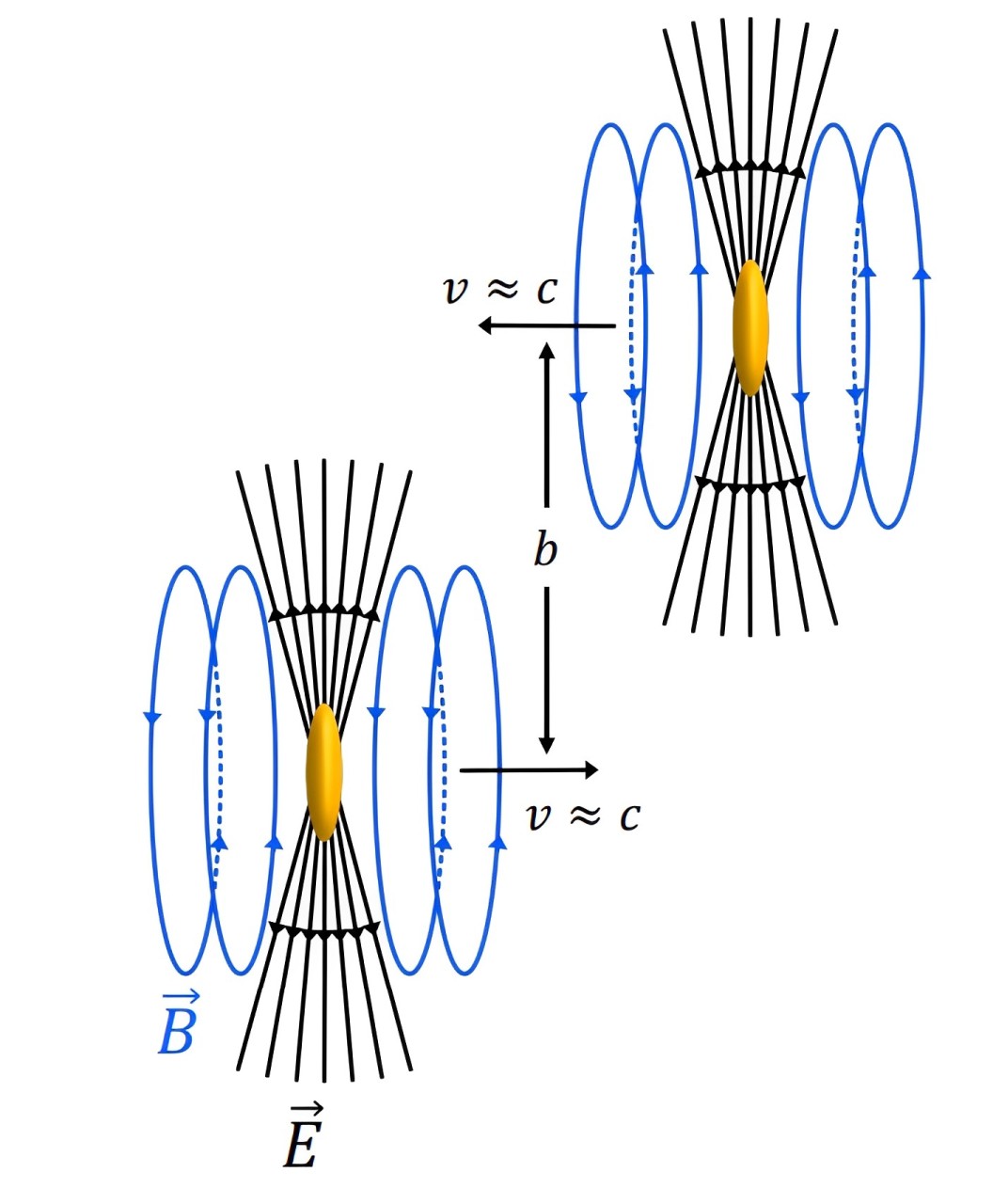} 
    \includegraphics[width=0.5\textwidth]{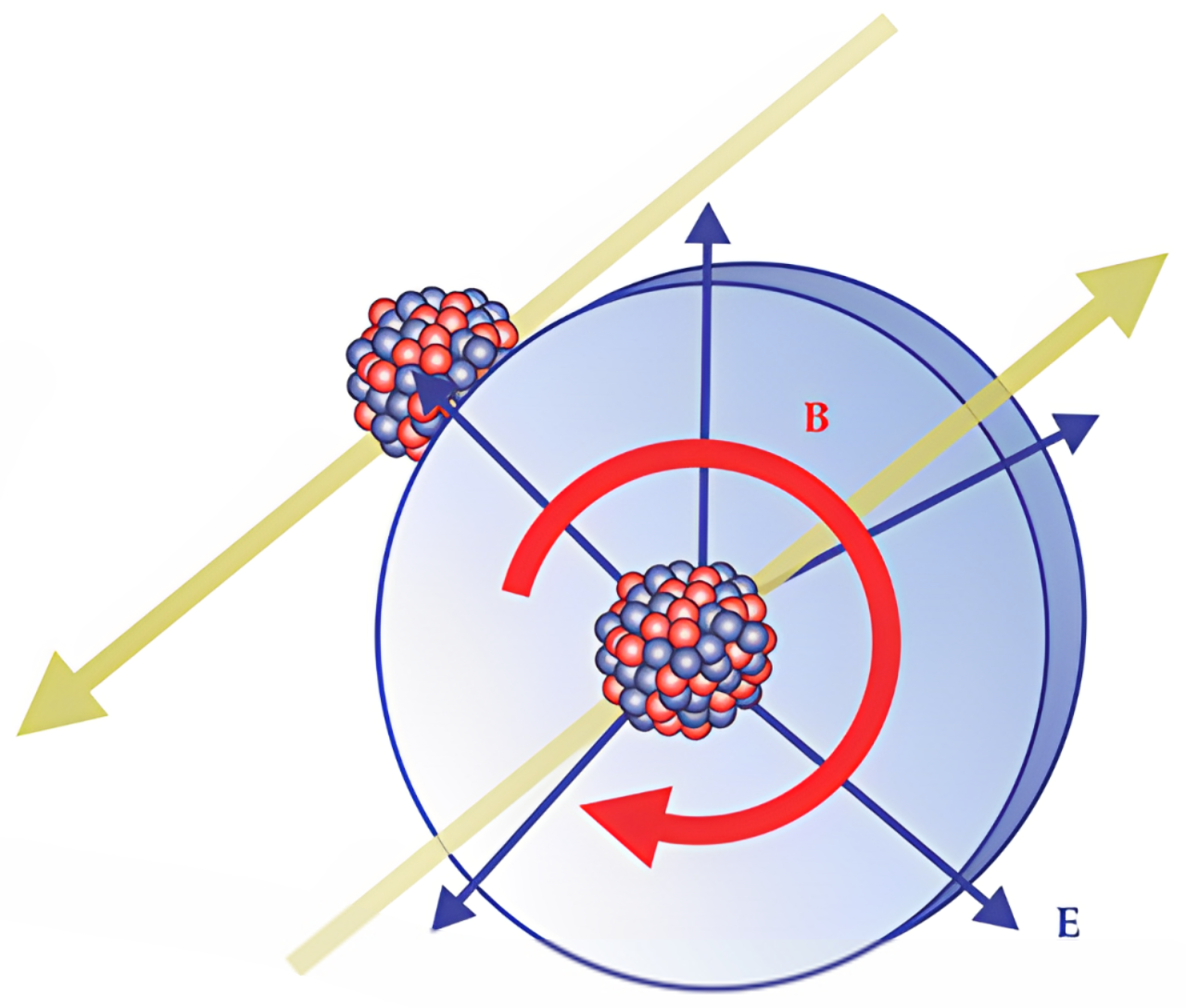} 
    \caption{Illustration of the electromagnetic field distribution in ultra-peripheral collisions ($b > R_A + R_B$, with $R_A,\ R_B$ being the radii of nucleus $A,\ B$ respectively), shown in a side view (left) and an oblique view (right). From Ref.~\cite{Klein:2017nqo}.}
    \label{fig:EPA:UPC}
\end{figure}

\hspace{2em}When a charged relativistic ion moves at high speed, the electric field around it is compressed into the transverse plane due to the relativistic effect causing the electric field vectors to extends outward radially in the transverse plane. At the same time, the moving electric field induces a magnetic field, with the direction of the magnetic field vectors arranged in concentric circles centered on the ion. As the speed becomes relativistic, the magnetic field strength grows to be approximately the same as the electric field strength (equal for $v=c$). This is illustrated in Fig.~\ref{fig:EPA:UPC} for relativistic ultra-peripheral heavy-ion collisions ($b > R_A + R_B$, with $R_A,\ R_B$ being the radii of nucleus $A,\ B$ respectively), for which no hadronic overlap takes place. In this limit, one can approximate these electromagnetic fields as a plane wave.  This approximation is the essence of the Equivalent Photon Approximation (EPA). Initially proposed by E. Fermi in 1924~\cite{1924ZPhy29315F}, this approximation was further developed to incorporate relativity by Weizsäcker and Williams~\cite{vonWeizsacker:1934nji, Williams:1934ad}, and is therefore often referred to as the ``Weizsäcker-Williams'' method.
If one considers only a small region of the electromagnetic field in the transverse plane, it can be approximated as an electromagnetic wave with a definite polarization direction {\it i.e.} as an equivalent photon with a specific polarization direction. If we Fourier transform such an electromagnetic field distribution into momentum space, the corresponding equivalent photon polarization direction is always parallel to its transverse momentum direction, that is, the equivalent photon is 100\% linearly polarized. The Breit-Wheeler process and its polarization effects have been recently observed at RHIC~\cite{STAR:2019wlg,Wang:2022ihj,brandenburgMappingElectromagneticFields2021a,Brandenburg:2022tna}. The explicit formulation of the coherent photon distribution is presented below.

 According to the uncertainty principle, the minimum wavelength of these photons at a given impact parameter is given by the thickness of the contracted field.  Then, 
\begin{equation}
\Delta t \cdot \Delta E \approx 1 \implies \frac{R}{\gamma v} \cdot \omega \approx 1 \stackrel{v \rightarrow c}{\implies} \omega_{\text{max}} \approx \frac{\gamma}{R},
\end{equation}
where $\Delta t$ is the passing time in the longitudinal direction, $\gamma$ is the Lorentz factor, and $R$ is the nucleus's radius, so $R/\gamma$ is approximately the Lorentz contracted length of the field in the beam direction. Thus, the maximum photon energy $\omega_{\text{max}}$ depends on the nucleus type $(R)$ and energy $(\gamma)$. At RHIC energy scales, the maximum $\omega_{\text{max}}$ for gold nuclei can reach 3 GeV, while at LHC energy scales, $\omega_{\text{max}}$ for lead nuclei can reach 80 GeV, in the lab frame.

In the laboratory frame, the inhomogeneous wave equation for the electromagnetic vector potential in the Lorentz gauge is ~\cite{Krauss:1997vr}:
\begin{equation}
\partial_{\mu} \partial^{\mu} A^\nu (x)=j^\nu (x),
\label{eq3-2-1-0}
\end{equation}
where $A^\nu(x)$ is the electromagnetic vector potential and $j^\nu(x)$ is the current density.  The current density can be written in momentum representation as:
\begin{equation}
j^{\nu}(k) = 2 \pi \delta(k \cdot \mu) \rho(\sqrt{-k^2}) \cdot \mu^{\nu},
\label{eq3-2-1-1}
\end{equation}
where $k^\nu = (\omega, \vec{k})$ and $\mu^\nu = \gamma(1,0,0,v)$ represents the four-velocity with a Lorentz boost along the z-axis. Fourier transforming Eq.~\ref{eq3-2-1-0} to momentum space and substituting into Eq.~\ref{eq3-2-1-1}, we obtain:
\begin{equation}
A^\nu (k) = -\frac{1}{k^2} j^\nu (k) = -2 \pi \delta(k \cdot \mu) \frac{\rho(\sqrt{-k^2})}{k^2} \mu^\nu = Z r(k) \mu^\nu,
\end{equation}
\begin{equation}
r(k) = - 2 \pi \frac{1}{Z} \delta(k \cdot \mu) \frac{\rho(\sqrt{-k^2})}{k^2} = -2 \pi e \delta(k \cdot \mu) \frac{F(-k^2)}{k^2},
\end{equation}
where $F(-k^2) = \rho(\sqrt{-k^2}) / Ze$ represents the normalized nuclear electromagnetic form factor, $Z$ is the atomic number, and the delta function ensures the validity of the Lorentz gauge condition $k_\nu A^\nu = 0$ and the transverse nature of the photon. Considering the condition $\gamma (\omega - v k_z) = 0$:
\begin{equation}
k^2 = \omega^2 - k_z^2 - \vec{k}_\perp^2 \stackrel{k_z = \frac{\omega}{v}}{=} - \left( \frac{\omega}{v \gamma} \right)^2 - \vec{k}_\perp^2 \approx - \left( \frac{\omega}{\gamma} \right)^2 - \vec{k}_\perp^2,
\end{equation}
where $\perp$ denotes the transverse component. The Fourier transform of the field strength tensor is:
\begin{equation}
F^{\mu \nu}(k) = -i (k^\mu A^\nu (k) - k^\nu A^\mu (k)).
\end{equation}

\begin{figure}[htbp]
\centering
\includegraphics[width=0.7\textwidth]{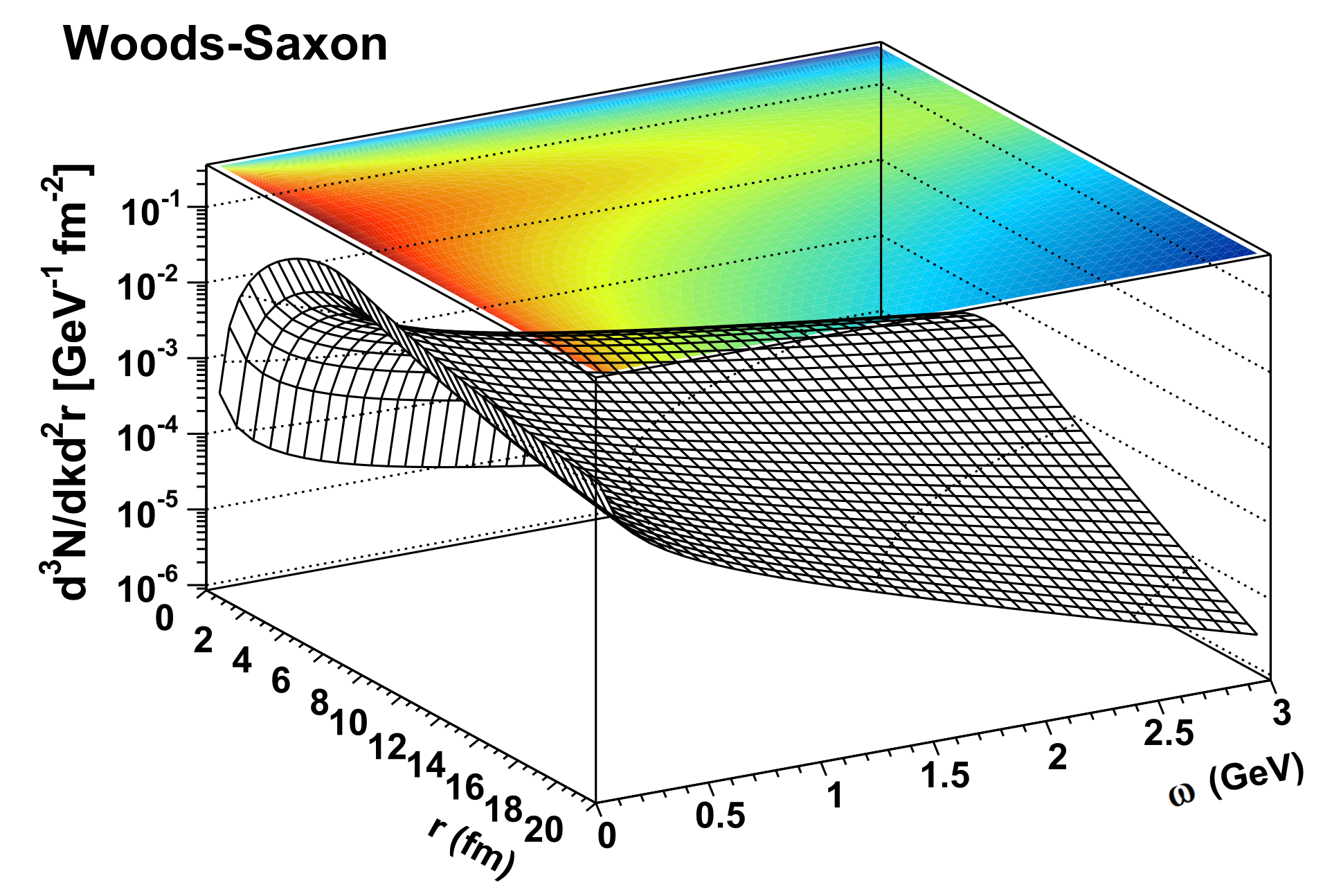}
\caption{Dependence of the equivalent photon flux on distance $r$ and photon energy $\omega$ for Au + Au collisions at $\sqrt{s_{NN}} = 200$ GeV. From Ref.~\cite{Zha:2018ywo}.}
\label{fig:flux}
\end{figure}

Since $\vec{\mu}_\perp = \vec{A}_\perp (k) = 0$, the transverse components of the electric and magnetic field strengths are:
\begin{equation}
\vec{E}_\perp (k) = -i A^0(k) \vec{k}_\perp,
\end{equation}
\begin{equation}
\vec{B}_\perp (k) = \vec{v} \times \vec{E}_\perp (k) = -iv A^0(k) (-k_y, k_x, 0)^T.
\end{equation}
As $v \rightarrow c = 1$,  the longitudinal components of both the electric and magnetic field strengths tend to zero, so:
\begin{equation}
|\vec{E}| = |\vec{B}|, \quad \vec{E} \perp \vec{B}, \quad \vec{E} \perp \vec{v}, \quad \vec{B} \perp \vec{v}.
\end{equation}
The energy flux density across the transverse section, equal to the energy flux density of the equivalent photon, is given by:
\begin{equation}
\int_{-\infty}^\infty dt \int d\vec{x}_\perp \cdot \vec{S}(\vec{r},t) \equiv \int_0^\infty d\omega \, \int d\vec{x}_\perp \omega \cdot n(\omega, \vec{x}_\perp),
\end{equation}
\begin{equation}
\vec{S}(\vec{r}, t) = \vec{E}(\vec{r}, t) \times \vec{B}(\vec{r}, t) \stackrel{v \rightarrow c}{\rightarrow} \left|\vec{E}(\vec{r}, t)\right|^2 \vec{v}.
\end{equation}
Fourier transforming $\vec{E}(\vec{k}, \omega)$ along the z-axis, we have:
\begin{align}
\vec{E}(z, \vec{k}_\perp, \omega) &= \int_{-\infty}^\infty \frac{dk_z}{2\pi} e^{ik_z z} \vec{E}(\vec{k}, \omega) \nonumber \\
&= 2\pi i \gamma (Ze) \vec{k}_\perp \int_{-\infty}^\infty \frac{dk_z}{2\pi} e^{ik_z z} \frac{F(-k^2)}{k^2} \delta(k \cdot \mu) \nonumber \\
&= (Ze) \frac{-i \vec{k}_\perp}{v} e^{i \omega_z / v} \cdot \frac{F((\frac{\omega}{v \gamma})^2 + \vec{k}_\perp^2)}{(\frac{\omega}{v \gamma})^2 + \vec{k}_\perp^2}.
\end{align}
Combining the above equations, we get:
\begin{equation}
  n(\omega, \vec{x}_\perp) = \frac{1}{\pi \omega} 
\left|\vec{E}_\perp(\vec{x}_\perp, \omega)\right|^2
= \frac{4Z^2 \alpha}{\omega} \left|\int \frac{d^2 \vec{k}_\perp}{(2\pi)^2} 
\vec{k}_\perp \frac{F((\frac{\omega}{\gamma})^2 + \vec{k}_\perp^2)}{(\frac{\omega}{\gamma})^2 + \vec{k}_\perp^2} e^{i \vec{x}_\perp \cdot \vec{k}_\perp}\right|^2.
\label{eq3-2-1-flux} 
\end{equation}

By Fourier transforming the nuclear charge density distribution, we can obtain $F(\vec{q})$. The charge density distribution for a heavy spherically symmetric nucleus is given by the Woods-Saxon distribution:
\begin{equation}
\rho_A(r) = \frac{\rho^0}{1 + \text{exp}\left[\left(r - R_{WS}\right) / d\right]},
\end{equation}
where $r$ is the distance from the nucleus center, $R_{WS}$ is the nucleus radius, $d$ is the surface diffuseness parameter describing the charge density transition rate, and $\rho^0$ is the normalization factor.  The Woods-Saxon distribution is very close to the convolution of a hard sphere with a Yukawa potential; this leads to a nice analytic expression for the form factor~\cite{Klein:1999gv}:
\begin{equation}
F(q) = \frac{4\pi\rho^0}{Aq^3}\big[\sin(qR_{WS}) - q R_{WS} \cos(qR_{WS})\big]\big[\frac{1}{1+a^2q^2}\big],
\label{eq:formfactorWS}
\end{equation}
where $a=0.7$ fm is the range of the Yukawa potential \cite{Klein:1999qj}. The radius $R_{WS}$ and surface diffuseness parameter $d$ are based on fits to electron scattering data~\cite{DeJager:1974liz}.  For protons, the form factor is better described by a dipole form factor \cite{Klein:2003vd}, while for lighter nuclei, with charge $Z\le 6$, the form factor is well fit by a Gaussian distribution \cite{Klein:2016yzr}.

Figure \ref{fig:flux} shows the dependence of the equivalent photon flux on distance $r$ and photon energy $\omega$ for Au + Au collisions at $\sqrt{s_{NN}} = 200$ GeV.  As described by Eq.~\ref{eq3-2-1-flux}, the photon flux is inversely proportional to the energy. Additionally, the photon flux initially increases with distance from the nucleus core and then decreases, reaching a maximum near the nuclear surface.

The photon transverse momentum spectrum can be extracted from Eq. \ref{eq3-2-1-flux}, but only after integrating from $b=0$ to $b=\infty$.  Then, for a fixed $\omega$,  \cite{Baltz:2009jk}:
\begin{equation}
\frac{dn}{dk_\perp} = \frac{\alpha^2Z^2F^2(k_\perp^2+(\omega^2/\gamma^2))k_\perp^3}{\pi^2(k_\perp^2+(\omega^2/\gamma^2)^2)}
\label{eq:photonpt}
\end{equation}
Very roughly, the $k_\perp$ spectrum scales with $\omega/\gamma$, with a maximum at $k_\perp\approx \sqrt{3}\omega/\gamma$ but with a large tail at large $k_\perp$.   
Since $b$ and $k_\perp$ are conjugate variables, the $k_\perp$ spectrum becomes much more complicated when putting constraints on $b$ (including requiring $b>2R_A$) \cite{Klein:2020jom,Yu:2024icm,Shi:2024gex}.  One expects $\langle k_T^2\rangle\propto \langle 1/b^2\rangle$, so that restricting $b$ leads to an increase in $k_T$.  

One key difference between photon sources in heavy-ion collisions and those from bremsstrahlung or Compton backscattering lies in the nature of the photons: heavy-ion collisions produce quasi-real photons, whereas bremsstrahlung and Compton backscattering generate real photons. For the studies discussed here, the distinction between quasi-real and real photons can be neglected since the virtuality of the quasi-real photons is negligible compared to the mass of the photoproduced particles. In bremsstrahlung and Compton scattering, photons are emitted with well-defined energy and angular distributions, and their polarization properties can be controlled using experimental setups, such as diamond radiators or polarized incident beams. These methods facilitate the generation of high-energy, polarized photon beams with straightforward kinematics. In contrast, although the photon source in heavy-ion collisions cannot be externally controlled, its energy-momentum distribution is accurately described by EPA, as demonstrated earlier. Notably, this source is 100$\%$ linearly polarized at low $x$. Additionally, the photon flux scales with the square of the nuclear charge $Z^{2}$, enabling the production of extremely intense photon beams. For quantum phenomenon studies, a critical advantage of heavy-ion collisions is the availability of coherent photon sources from both colliding nuclei, providing a dual-photon source for exploring quantum interference phenomena such as double-slit interference (discussed in detail in the next section). In contrast, bremsstrahlung and Compton backscattering provide only single-photon sources, which are limited to diffraction studies.

        \subsubsection{Accelerator and detector setups in relativistic heavy-ion collisions}
        
\hspace{2em}Due to the high flux of quasi-real photons, as described in Sec.~\ref{sec3-2-1}, relativistic heavy-ion collisions offer a unique environment to study photoproduction. The leading facilities in this field are the Relativistic Heavy Ion Collider (RHIC) at Brookhaven National Laboratory and the Large Hadron Collider (LHC) at CERN.

RHIC facilitates the exploration of the quark-gluon plasma (QGP) and various QED phenomena by colliding ions, ranging from protons to gold nuclei, at energies up to 200 GeV per nucleon pair. On the other hand, the LHC, primarily known for its high-energy proton-proton collisions, accelerates lead nuclei to 
achieve collision energies of 5.4 TeV per nucleon pair. This provides an unparalleled opportunity to probe QGP properties and study new phenomena under extreme electromagnetic conditions.

Experiments at RHIC and LHC are pivotal in these studies. At RHIC, the STAR (Solenoidal Tracker at RHIC) detector~\cite{Ludlam:2003sn,Anderson:2003ur} is notable for its comprehensive tracking and identification capabilities. STAR's Time Projection Chamber (TPC) and Time of Flight (TOF) systems enable precise momentum measurements and particle identification, while its Electromagnetic Calorimeter (EMC) measures the energy of photons and electrons. The PHENIX (Pioneering High Energy Nuclear Interaction eXperiment) detector~\cite{PHENIX:2003nhg} focused on measuring direct photons, leptons, and hadrons.

At the LHC, the ALICE (A Large Ion Collider Experiment) detector~\cite{ALICE:2008ngc} is optimized for heavy-ion physics. It excels in tracking and identifying particles at low transverse momenta with its Inner Tracking System (ITS) and TOF detector, essential for studying the dense matter and extreme electromagnetic environment produced in collisions. The CMS (Compact Muon Solenoid)~\cite{CMS:2008xjf} and ATLAS (A Toroidal LHC ApparatuS)~\cite{ATLAS:2008xda} detectors, although designed for a broad range of particle physics research, also make critical contributions to heavy-ion physics. CMS and ATLAS have robust capabilities for measuring high-energy jets, photons, and leptons, and their large acceptance allows for detailed studies of high-energy phenomena.  The LHCb experiment is a forward spectrometer designed to study $b-$hadron physics \cite{LHCb:2008vvz}, but it has also studied UPCs. 

\begin{figure}[htbp]
    \centering
    \includegraphics[width=0.7\textwidth]{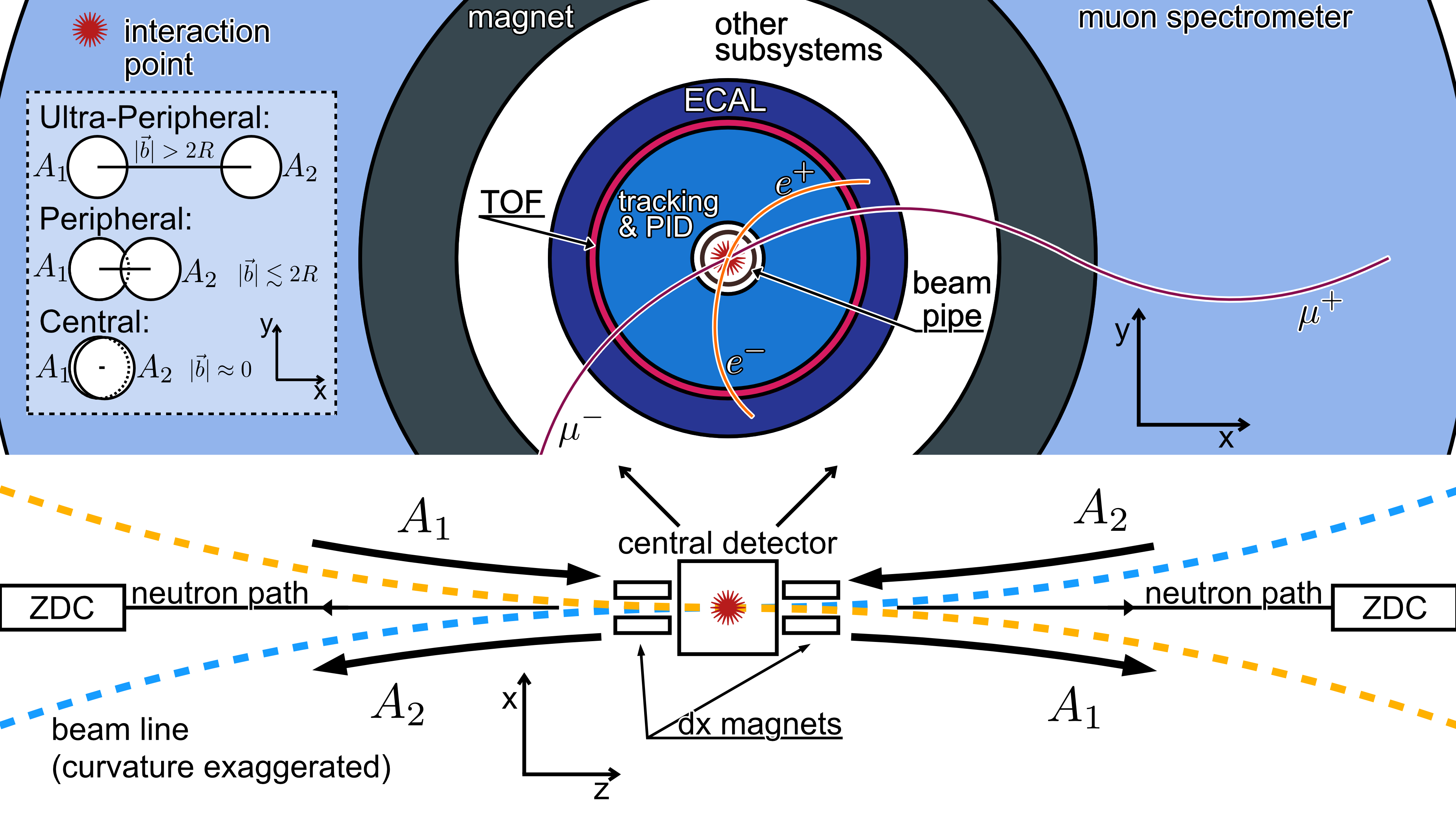}
    \caption{Diagram depicting the standard configuration of a high-energy collider experiment. Accelerated particle beams, labeled A1 and A2, converge and interact at the designated collision point. Neutrons produced during these interactions proceed directly towards Zero Degree Calorimeters (ZDCs), whereas magnetic fields deflect the charged particles resulting from the collisions. The rectangle beneath the legend of the interaction point contains representations of different types of heavy-ion collisions, including ultra-peripheral, peripheral, and central. From Ref.~\cite{Brandenburg:2022tna}.}
    \label{fig:dector_overview}
\end{figure}

The tracking and particle identification systems are crucial components of these experiments. Tracking detectors, such as the TPC in STAR and the ITS in ALICE, provide high-resolution measurements of charged particle trajectories, allowing for precise reconstruction of the collision vertex and momentum determination. Particle identification is achieved through various subsystems, including TOF detectors, calorimeters, and transition radiation detectors, which distinguish between particle species based on their mass, charge, and energy deposition patterns.

Another crucial technique is neutron tagging. This method involves detecting neutrons with near-beam energies, resulting from electromagnetic dissociation (EMD), which travel along the beam direction and are captured by Zero Degree Calorimeters (ZDCs). Neutron tagging is instrumental in identifying events where photoproduction occurs without direct hadronic interactions, enhancing the rare signal and reducing background noise, as discussed in Sec.~\ref{sec6-3}.

Figure~\ref{fig:dector_overview} illustrates the standard configuration of a high-energy collider experiment, highlighting the reconstructed particle tracks and the capabilities of the detector systems in identifying different particle types. The combination of advanced accelerator facilities and sophisticated detector systems at RHIC and LHC allows for detailed studies of photoproduction in relativistic heavy-ion collisions, enhancing our understanding of the photoproduction mechanism and related physics.

 	\subsubsection{RHIC measurements and basic UPC techniques}\label{sec3-2-3}
 \hspace{2em}To study high-energy photoproduction at ion collider required the development of a number of new techniques.  Most photonuclear interactions result in final states containing only a handful of particles. In contrast, relativistic heavy-ion collisions produce hundreds or even thousands of particles, which can completely obscure the products of a photoproduction reaction. Therefore, it is usual to focus on photoproduction events that are not accompanied by hadronic interactions - UPCs. Photoproduction in hadronic interactions will be covered later. UPCs can happen when (roughly)  $b>2R_A$.  More precisely, the
photoproduction can be calculated as
\begin{equation}
\sigma = \int d^2b P_P(b) P_{\rm nohad}(b)
\label{eq:p[ofb]}
\end{equation}
where $P_P(b)$ is the probability for having a photoproduction reaction, and $P_{\rm nohad}(b)$ is the probability of not having a hadronic interaction.

Photoproduction processes can be categorized as either coherent or incoherent. Coherent photoproduction occurs when the electromagnetic field of a relativistic nucleus interacts with the target nucleus as a whole, preserving its internal structure. In this scenario, the entire nucleus acts as a single quantum entity, leading to photon-mediated processes that produce final states with low transverse momentum due to the large spatial coherence scale. Importantly, the nucleus remains intact after the interaction. In contrast, incoherent photoproduction involves interactions between the electromagnetic field and individual nucleons within the nucleus, rather than the nucleus as a collective entity. This results in the disruption of the nuclear structure, often accompanied by nucleon dissociation or breakup. The final-state products of incoherent photoproduction typically exhibit higher transverse momentum, reflecting the smaller spatial scale of the interaction.
This review primarily focuses on coherent photoproduction, leveraging its unique characteristics to investigate the quantum phenomena underlying photoproduction processes. 

UPCs introduce several other new features.  One is the twofold ambiguity as to which nucleus emitted the photon and which is the target.  In the typical kinematic region accessible at LHC and RHIC, for coherent photoproduction, these two possibilities are indistinguishable, requiring us to sum the amplitudes rather than the square of the amplitudes.  Even without this interference, for identical nuclei, the ambiguity leads to a $d\sigma/dy$ being symmetric around $y=0$.  For slowly-rising (with $W_{\gamma p})$ cross sections, this leads to a cross-section that is almost flat in $y$, over many units in $y$.

The first experimental UPC studies came from the STAR Collaboration \cite{STAR:2002caw}, who studied $\rho$ and direct $\pi^+\pi^-$ photoproduction in 130 GeV/nucleon gold-gold collisions.  This pioneering study illustrated most of the major challenges in experimental UPC physics: triggering, and selecting UPC events.  Early large-coverage heavy-ion experiments were optimized to study events containing hundreds of particles.  Consequently, they could read out only a limited number of full events each second (100 events/second readout for the early STAR electronics \cite{Bieser:2002ah}).  Although selecting hadronic heavy-ion collisions was simple, selecting low-multiplicity was not simple, since there were significant backgrounds from very peripheral hadronic interactions, beam-gas events, and, particularly, cosmic-ray muons, which could mimic two-track events (with pair $y=0$ and $p_T=0$) if the muons passed through the center of the detector \cite{Nystrand:1998hw}.  

\begin{figure}[htbp]
    \centering
    \includegraphics[width=0.8\textwidth]{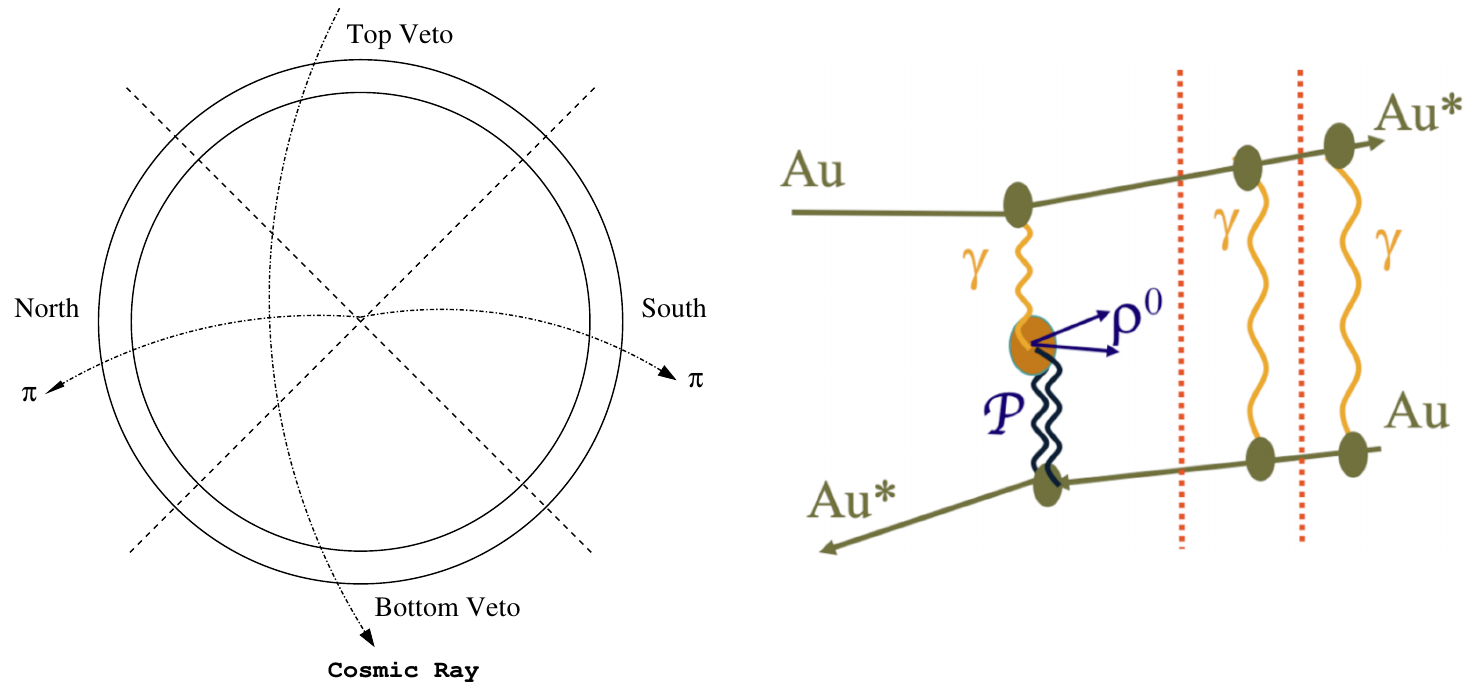}
    \caption{(left) Diagram showing how the STAR topology trigger worked.  Events with hits in the North and South quadrants of the central trigger barrel were accepted.  Events with hits on the top and bottom were likely to be from cosmic ray muons, and so were rejected.
    From Ref. \cite{Klein:2001vh}. (right) Diagram showing how multi-photon exchange factorizes, with independent photons each of which do one thing, sharing only a common impact parameter.  From Ref. \cite{Klein:2024wos}.
    \label{fig:STARtopology}
    }
\end{figure}

STAR initially used two triggers to select UPC events \cite{Klein:2001vh}. The first, `topology' trigger selected low-multiplicity events where the particles went horizontally, rather than vertically, to reject mostly-vertical cosmic-ray muons, as is shown in Fig. \ref{fig:STARtopology} (left). This trigger suffered from a rather poor selectivity, with relatively few of the events being actual UPC interactions.  

The second `minimum bias' trigger required low-multiplicity in the central detector, and also that both nuclei break up in the interactions.  The mutual dissociation  might seem to go against the requirement that there be no hadronic interaction.  However, nuclear breakup can also be electromagnetically induced.   A coherent UPC interaction can be accompanied by an additional photon exchange which can cause Coulomb excitation one of the nuclei \cite{Baltz:2002pp}.  These multi-photon exchanges involve largely independent photons (sharing only a common impact parameter), with each photon performing a single function.   Although a single photon can produce a vector meson while breaking up the nuclear target, these reactions largely occur at larger $p_T$.  At lower $p_T$, when coherent production dominates, this factorization holds quite well \cite{STAR:2007elq,ALICE:2020ugp}.  

Events where multiple photons are exchanged naturally have a lower average impact parameter $\langle b \rangle$ than single photon exchange \cite{Baur:2003ar}.  The probability of a single photon exchange goes as $1/b^2$, with the probability of having two-photon exchange being $1/b^4$, and three-photon exchange (one vector meson plus Coulomb excitation of both nuclei) scales as $1/b^6$.  These excitations almost always lead to a nuclear breakup with near-beam-energy neutron emission that would be visible in zero degree calorimeters (ZDCs).

With gold beams at RHIC, the cross-sections for vector meson photoproduction accompanied by mutual Coulomb excitation of both nuclei are about 1\% to 5\% of the cross sections for vector mesons alone, depending on the choice of meson \cite{Baltz:2002pp}.   Although this fraction is small, it is large enough to leave a robust signal for lighter mesons. A trigger that required neutrons in both ZDCs  plus a low multiplicity in the STAR central detector fired at a reasonable rate, and produced useful samples of $\rho$ and other vector mesons.

\begin{figure}[htbp]
    \centering
\includegraphics[width=0.9\textwidth]{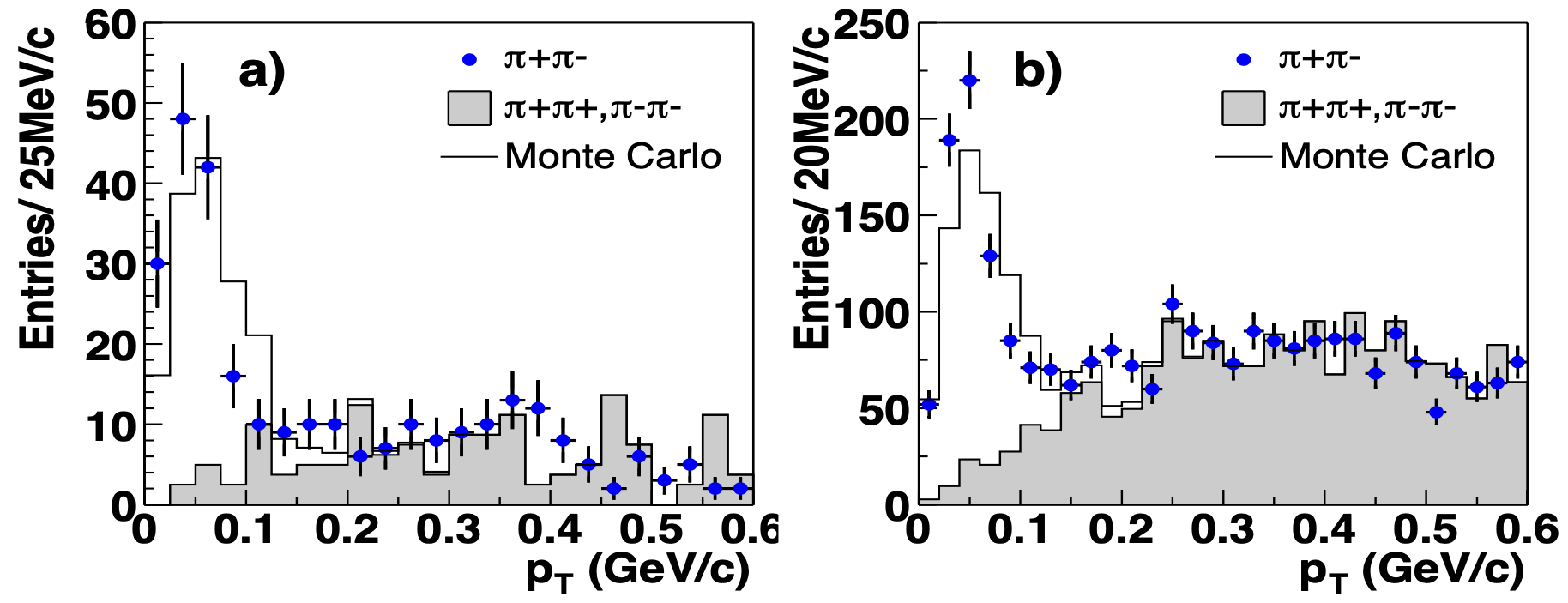}
    \caption{The $p_T$ spectra from the 2002 STAR $\rho$ analysis.  The left panel (a) shows the results from the topology trigger, while the right (b)nis from the minimum bias trigger, which required both nuclei to break up.  The blue points and error bars are the data, while the darker histogram shows the estimated background from like-sign events, scaled to match the high $p_T$ data.  The lighter histogram shows the STARlight Monte Carlo prediction.
    From Ref. \cite{STAR:2002caw}.
    \label{fig:STARrho2002}
    }
\end{figure}

With these triggers, STAR made the first observation of coherent vector meson photoproduction in UPCs \cite{STAR:2002caw} in 130 GeV/nucleon gold-gold collisions, observing $131\pm 14$ and $656\pm 36$ two-track events consistent with from the topology and minimum bias triggers respectively.  These events demonstrated many of the features that are now common in UPC studies.  
Figure \ref{fig:STARrho2002} shows a large peak for pair $p_T < 150$ MeV/c, consistent with coherent production.  Events are also present at higher $p_T$, due to a combination of background and incoherent $\rho$ photoproduction.  The $\pi^+\pi^-$ mass spectrum was dominated by the $\rho^0$, but also exhibited direct $\pi^+\pi^-$, with the ratio consistent with HERA data.  The cross section was consistent with calculations based on HERA ($\gamma p$) data and a Glauber calculation \cite{Baltz:2002pp}.  

This was the first experimental measurement, so the analysts had to make the first estimate of backgrounds.  They did so using like-sign pairs ($\pi^+\pi^+ + \pi^-\pi^-$), which can arise due to the limited detector acceptance.  Additional particles to compensate for the nonzero charge can lie outside the detector.  This background can come from peripheral hadronic interactions, beam gas, incoherent photoproduction, or coherent photoproduction of higher-mass states \cite{Nystrand:1998hw}.  In the 2002 STAR analysis, the collaboration scaled the like-sign background to match the data at high $p_T$, where coherent production was expected to be negligible.  So, their background estimate explicitly included incoherent production.   Later studies used the like-sign background directly, without scaling, leaving the incoherent emission visible.  The like-sign background is much smaller for detectors with larger acceptance, or without the requirement of ZDC hits, which signal Coulomb dissociation,  but also accompany hadronic interactions and incoherent photoproduction.

A later STAR analysis used similar techniques, but was at a higher beam energy, 200 GeV/nucleon gold-gold collisions and with a much larger data sample - about 20,000 events from the two triggers, leading to much more precise results~\cite{STAR:2007elq}.  This later analysis also measured incoherent production, by studying higher $p_T$ pion pairs (above the coherent peak), and made the first UPC measurement of the polarization of the $\rho$, finding that production was compatible with s-channel helicity conservation (SCHC).  This is expected for elastic scattering, and so critical for the interferometry discussed in this review. STAR also studied $\rho$ photoproduction in gold-gold collisions at a center of mass energy of 62.4 GeV, and used that data to measure the energy dependence of this reaction \cite{STAR:2011wtm}.  Due to the larger error bars (due to the lower luminosity and cross section) for the low-energy point, the data was compatible with multiple models. Another STAR analysis used a 200 GeV gold-gold data set to make the first observation of interference between the two photon emission directions \cite{STAR:2008llz}.  They observed a decrease in production as pair $p_T\rightarrow 0$ for events near mid-rapidity (where interference is largest). STAR also explored a more complex final state, $\pi^+\pi^-\pi^-\pi^-$, which can be produced by the decay of a $\rho'$ meson \cite{STAR:2009giy}.  STAR observed a signal compatible with a single broad resonance (mass $1540\pm 40$ MeV and width $570\pm 60$ MeV), decaying primarily to $\rho^0\pi^+\pi^-$. 

Also at RHIC, the PHENIX collaboration used a rather different strategy to make the first UPC observation of photoproduction of the $J/\psi$.  PHENIX covered a more limited solid angle than STAR, but, crucially for this study, had a high rate data acquisition system, and good particle identification.  They collected 28 events, compatible with a mixture of photoproduced $J/\psi\rightarrow e^+e^-$ and $\gamma\gamma\rightarrow e^+e^-$ \cite{PHENIX:2009xtn}, and measured a cross-section compatible with the expectations of a Glauber calculation \cite{Klein:1999qj}. 

Later STAR UPC analyses benefited greatly from the upgraded `Data Acquisition 1000 (DAQ 1000)' data acquisition system, which increased the allowable trigger rate and data acquisition speed by a factor of 10. By requiring BEMC energy deposits greater than 0.7 GeV in back-to-back azimuthal sextants, STAR collected approximately $24 \times 10^6$ UPC $J/\psi$-triggered events, corresponding to an integrated luminosity of 13.5 nb$^{-1}$. This enabled precise measurements of $J/\psi$ photoproduction in UPCs and the first observation of $\Psi(2S)$ photoproduction at RHIC~\cite{STAR:2023nos,STAR:2023vvb}. 

A 2017 study of $\rho$ photoproduction used a total of 384,000 photoproduced $\pi^+\pi^-$ \cite{STAR:2017enh}.  These statistics enabled a number of new measurements.  The resulting $\pi^+\pi^-$ mass spectrum was found to be well fit by a combination of $\rho$, direct $\pi^+\pi^-$, and $\omega\rightarrow\pi^+\pi^-$. This data also allowed the first mapping of the transverse partonic structure of the gold nucleus via Fourier transform of $d\sigma/dt$, as is discussed in Section \ref{sec:tomography}. STAR UPC analyses further took advantage of the fact that RHIC was built primarily to study heavy ion collisions, spending the bulk of the year accelerating heavier ions.  This allowed for beam time to accelerate a wider variety of heavy-ions, including copper, ruthenium, zirconium, and uranium.  These ions were important in exploring the $Z$ and $A$ scaling for photoproduction. Ruthenium-96 and zirconium-96 were of particular interest, containing the same total number of nucleons, but with zirconium having four more protons than zirconium, giving it a more intense photon field.  Another study used gold-on deuterium collisions to probe $J/\psi$ photoproduction on deuterium \cite{STAR:2021wwq}, since, for very asymmetric collisions, the heavier nucleus usually acts as the photon emitter, and the lighter nucleus as a target. 

In this time frame, the Fermilab Tevatron $p\overline p$ collider also contributed to UPC studies, with a measurement of $J/\psi$ photoproduction on proton targets \cite{CDF:2009xey}. In $p\overline p$ collisions, UPCs act similarly, except that the interference between the two photon directions is constructive as $p_T\rightarrow 0$ \cite{Klein:2003vd}.
 	\subsubsection{LHC measurements}

 \hspace{2em}
The LHC provides significant opportunities for exploring photon-induced production processes, leveraging its enhanced photon energies, wide rapidity coverage, and high luminosity. These features enable precise investigations of nuclear shadowing, gluon distributions, and collective quantum effects through coherent photoproduction, including rare processes like the production of heavier vector mesons. This makes the LHC an essential platform for advancing our understanding of high-density gluonic systems and quantum phenomena in strong interactions.

At small Bjorken-$x$, QCD predicts a high-density regime where gluon saturation occurs as the rapid growth of gluon density is balanced by recombination processes~\cite{Morreale:2021pnn}. This nonlinear dynamic fundamentally alters the gluonic structure of hadrons and nuclei, with overlapping gluon wavefunctions giving rise to coherence effects and collective behavior. UPCs at the LHC uniquely probe this regime, exploring nuclear gluonic structure in the low-$x$ region ($10^{-5} < x < 10^{-2}$). Through detailed studies of coherent photoproduction, the LHC offers critical insights into the dynamics of gluon saturation and its role in shaping high-energy nuclear interactions.

CMS conducted the measurement of exclusive $\rho$ photoproduction in ultraperipheral proton-lead collisions at $\sqrt{s_{\rm NN}} = 5.02$ TeV~\cite{CMS:2019awk}. The cross-section as a function of $W_{\gamma p}$ is consistent with fixed-target and HERA results. ALICE measured the coherent photoproduction of $\rho$ mesons in ultraperipheral PbPb and XeXe collisions~\cite{ALICE:2015nbw,ALICE:2020ugp,ALICE:2021jnv}. The XeXe data was an important intermediate point, showing that the coherent cross-section scales nearly linearly with atomic number.
The rapidity dependence of the cross-section in different neutron multiplicity classes in ultraperipheral PbPb collisions at $\sqrt{s_{\rm NN}} = 5.02$ TeV is in good agreement with predictions based on various models of nuclear shadowing~\cite{ALICE:2020ugp}. Moreover, the $A$-dependence of the coherent photoproduced $\rho$ cross-section on the nuclear mass number at mid-rapidity ($W_{\gamma \rm{N}} = 65$ GeV) is significantly smaller than expected from a purely coherent process, suggesting the importance of shadowing effects~\cite{ALICE:2021jnv}.

Compared to the $\rho$ meson, the photoproduced $J/\psi$ meson is a cleaner and more direct probe of gluon distributions in nuclei due to its higher mass, making it more suitable for precise QCD studies. 
However, there are significant theoretical complications, including large NLO calculations and high scale sensitivity
\cite{Eskola:2022vpi}. 
ALICE conducted the first measurement of exclusive $J/\psi$ photoproduction off protons in ultraperipheral proton-lead collisions~\cite{ALICE:2014eof}, with maximum $W_{\gamma p}$ around 700 GeV, corresponding to $\sim 2 \times 10^{-5}$. Together with the exclusive $J/\psi$ results from previous HERA measurements~\cite{ZEUS:2002wfj,H1:2005dtp,H1:2013okq}, the cross section of coherent $J/\psi$ follows a universal power law dependence of $W_{\gamma p}$ from $\sim$20 to $\sim$700 GeV, or equivalently, Bjorken $x$ from $\sim 2 \times 10^{-2}$ to $\sim 2 \times 10^{-5}$. More recently, LHCb has used $pp$ collisions to extend these measurements up to $W_{\gamma p} \approx 2$ TeV
\cite{LHCb:2024pcz}, probing Bjorken $x$ values below $10^{-5}$. Overall, these measurements suggest there is no significant change of the gluon PDF scaling in the proton between HERA and LHC energies. 

\begin{figure}[htbp]
    \centering
    \includegraphics[width=0.5\textwidth]{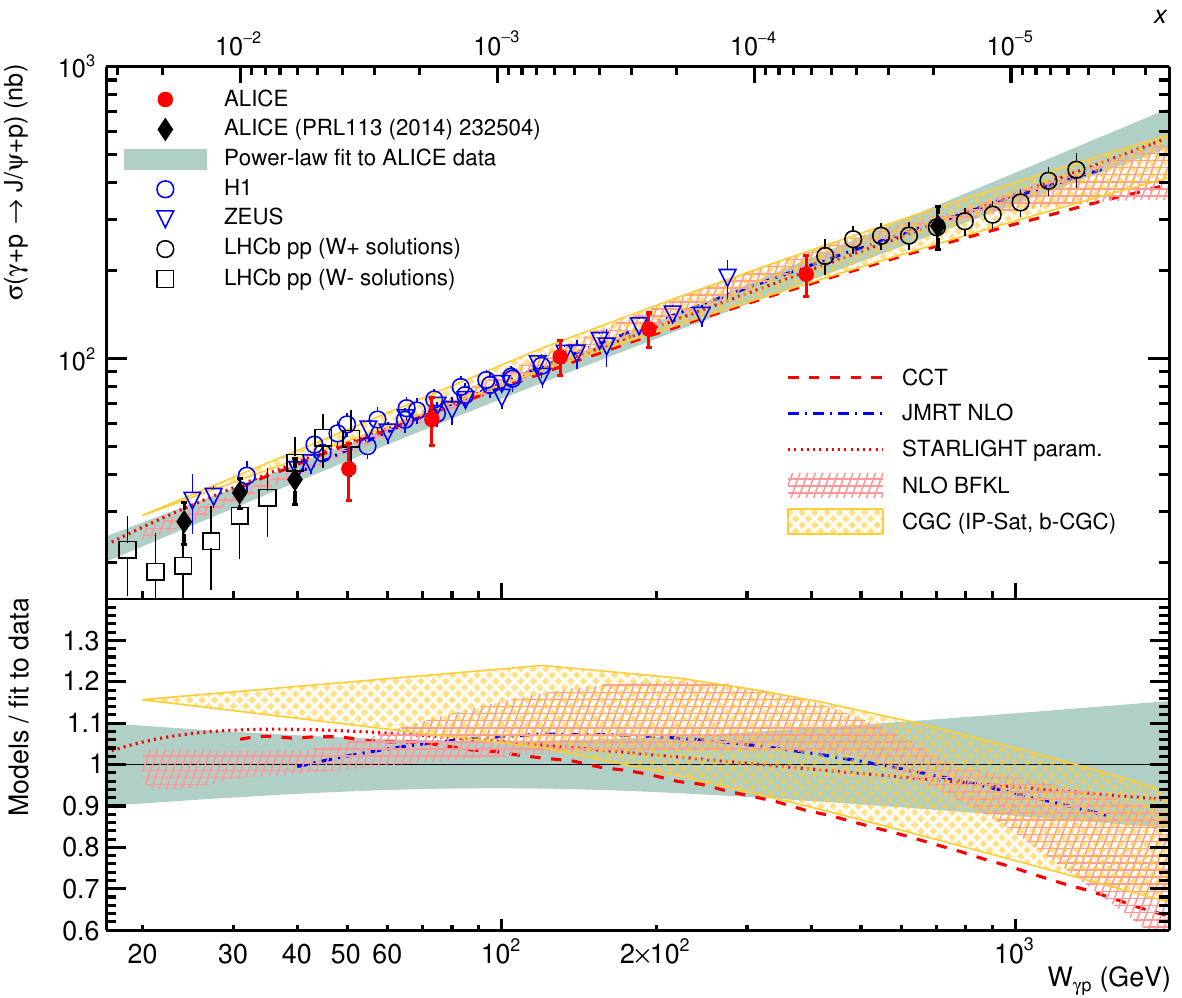}
    \caption{The cross-section for exclusive photoproduction of J/$\psi$ on protons as a function of photon-proton center-of-mass energy ($W_{\gamma\rm{p}}$).  From Ref.~\cite{ALICE:2018oyo}.}
    \label{fig:cohJpsi_pPb}
\end{figure}

\begin{figure}[!htbp]
    \centering
    \includegraphics[width=0.53\textwidth]{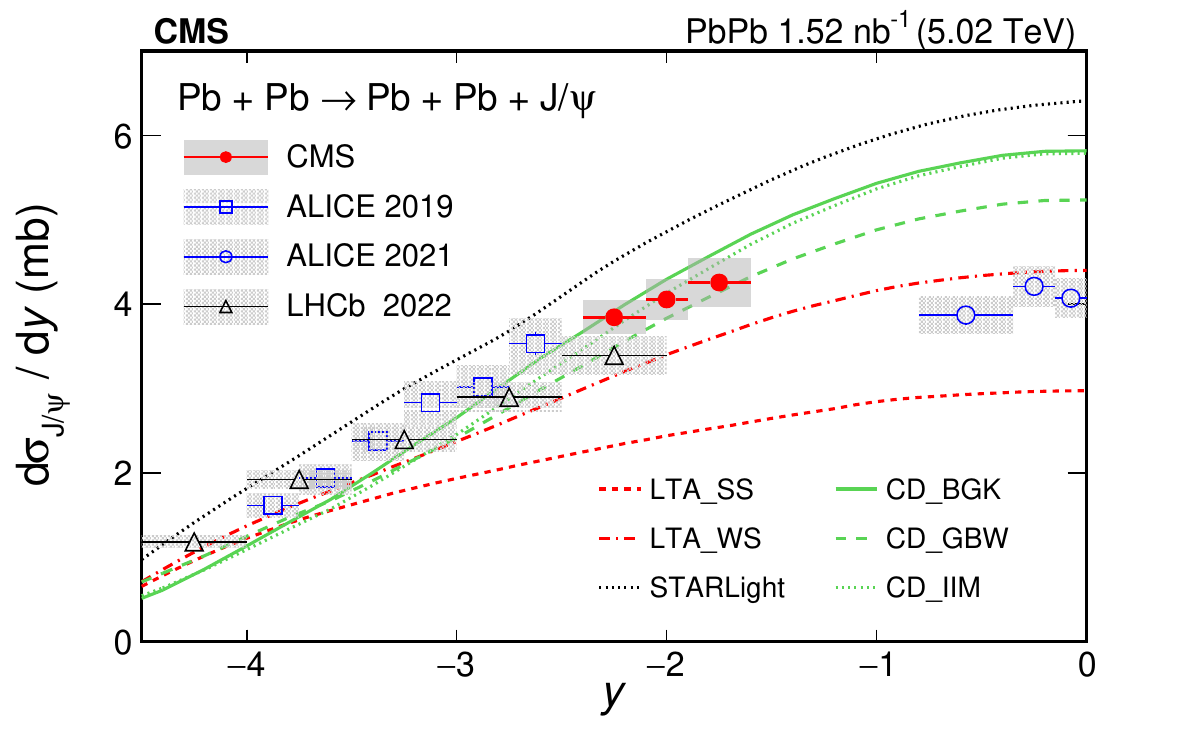} 
    \includegraphics[width=0.46\textwidth]{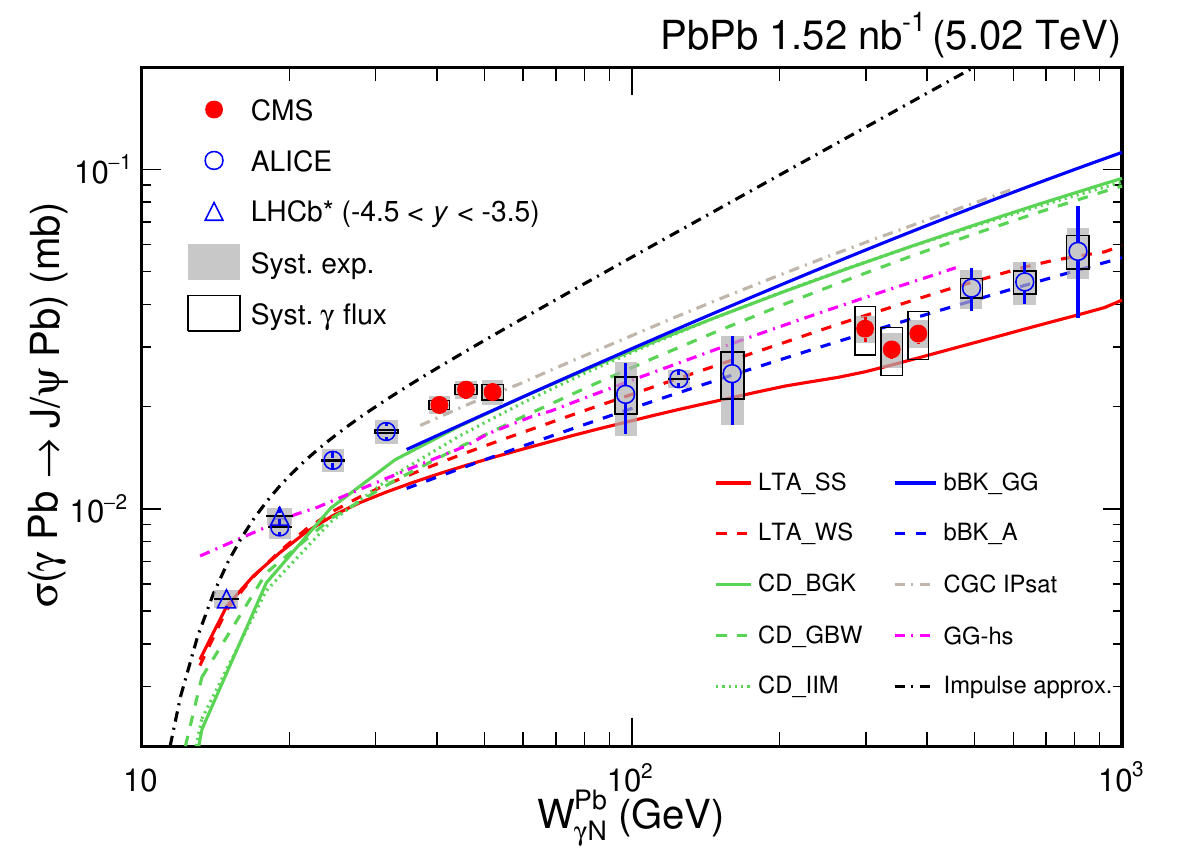} 
    \caption{(left) The differential coherent J/$\psi$ photoproduction cross section as a function of rapidity. The vertical bars and shaded boxes represent the statistical and systematic uncertainties, respectively. (right) The total cross section of coherent J/$\psi$ as a function of $W_{\gamma\rm{N}}^{\rm{Pb}}$ in Pb$+$Pb UPCs. The plots are reproduced from Refs.~\cite{CMS:2023snh,ALICE:2023jgu}.}
    \label{fig:cohJpsi_PbPb}
\end{figure}

In a nucleus with A nucleons, the parton distributions should naively be A times those in a single nucleon, but reductions, known as nuclear shadowing, are observed at small $x$~\cite{H1:2015ubc}. Similar unitarity considerations to those for a single nucleon predict that saturation should set in for large nuclei at higher $x$ values than in protons, with this behavior scaling roughly as $A^{1/3}$~\cite{Armesto:2006ph}. The LHC experiments systematically measured the coherent photoproduction of $J/\psi$ in UPCs at different collision energies and rapidity regions. The left panel of Fig.~\ref{fig:cohJpsi_PbPb} shows the coherent $J/\psi$ cross section as a function of rapidity in ultraperipheral Pb-Pb collisions at $\sqrt{s_{\rm NN}}$ = 5.02 TeV from ALICE~\cite{ALICE:2019tqa,ALICE:2021gpt}, CMS~\cite{CMS:2023snh} and LHCb~\cite{LHCb:2022ahs} Collaborations. The LHC experiments complement each other over a wide range of rapidity.

In a symmetric UPC, {\it e.g.} Pb$+$Pb, the photon-emitter nucleus and target nucleus are indistinguishable. For a given vector meson rapidity, there is a mixture of low (large) and high (small) energy photon (gluon $x$) contributions. To give further insight of gluon evolution at small-$x$, a forward neutron tagging technique proposed in Ref.~\cite{Guzey:2013jaa} is employed to obtain cross sections in the photon-nucleus frame. The forward neutrons can be used to tag certain impact parameter ranges of UPCs, which is essential to decompose the contributions of low and high energy photons. The right panel of Fig.~\ref{fig:cohJpsi_PbPb} shows the total photoproduced cross section of coherent J/$\psi$ as a function of photon-nucleus center-of-mass energies($W_{\gamma\rm{N}}^{\rm{Pb}}$)~\cite{CMS:2023snh,ALICE:2023jgu}.  
The total cross section increases rapidly at energies below around 40 GeV, resembling the behavior observed in the $\gamma$p case. However, the growth rate moderates at energies around 40-100 GeV and remains low at higher energies. This behavior aligns with the qualitative expectation that, at smaller $x$, the rapid growth slows due to nonlinear gluon evolution effects, which are closely linked to gluon saturation. These nonlinear effects arise as gluon densities reach saturation levels, leading to a suppression of further growth in the parton distribution.

None of the theoretical model calculations fully capture the trends at both high and low $x$. While the models qualitatively predict a shape change consistent with the observed data, they fail to align with the data across the entire measurement range. To confirm whether this discrepancy arises from genuine small $x$ nonlinear evolution effects tied to gluon saturation, it is crucial to continue exploring these measurements with other vector mesons. These nonlinear evolution effects are expected to be universal, as gluon saturation is a fundamental property of quantum chromodynamics at high densities, and they should not significantly depend on the specific vector meson used to probe the nuclear structure. Since different vector mesons have varying masses, corresponding to the effective size of their color dipoles, they exhibit different sensitivities to the properties of high partonic densities at small $x$. This makes them valuable tools for investigating gluon saturation and the associated nonlinear evolution dynamics.

\begin{figure}[htbp]
    \centering
    \includegraphics[width=0.5\textwidth]{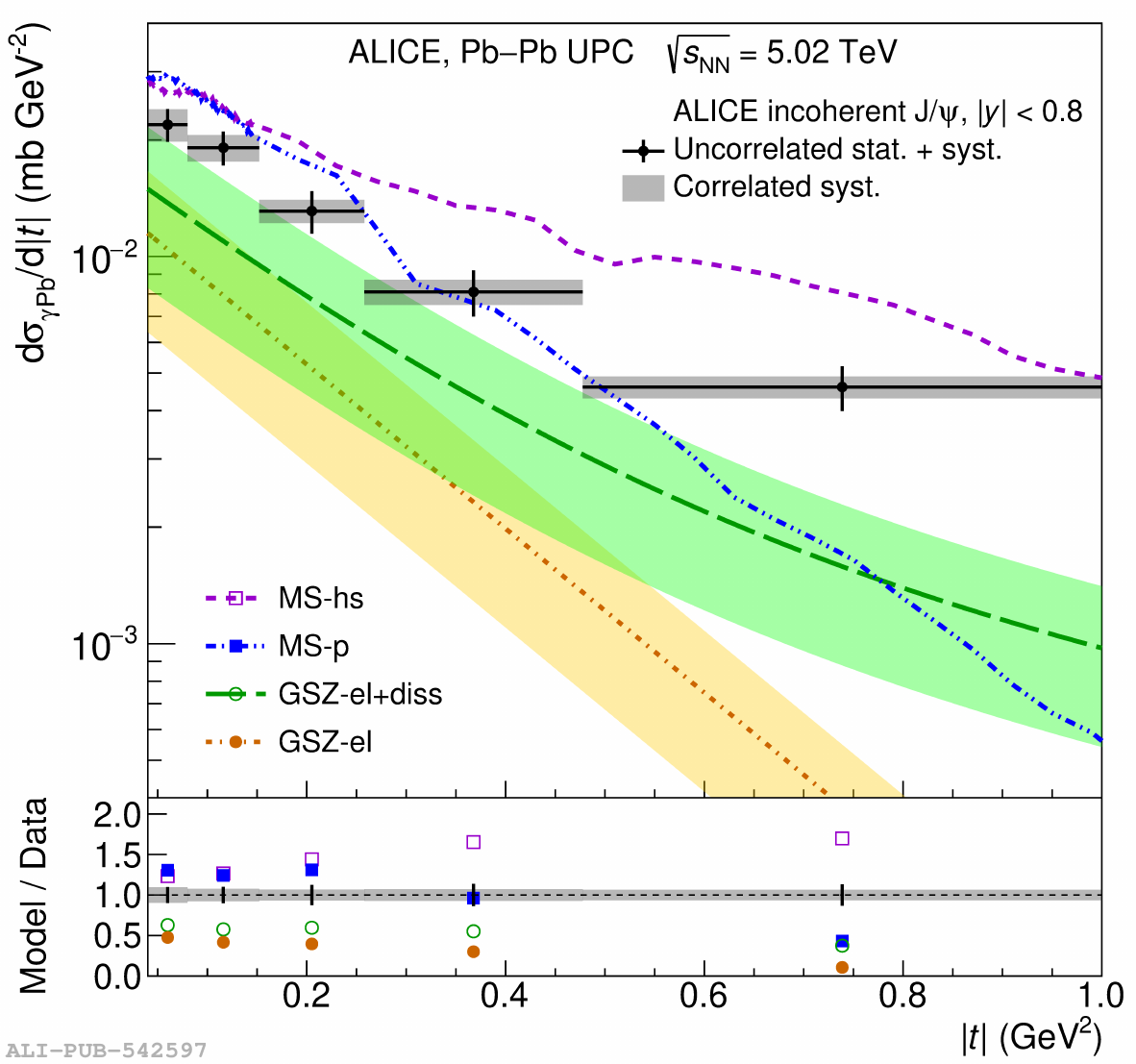}
    \caption{The cross section of incoherent J/$\psi$ as a function of $|t|$ in Pb$+$Pb UPCs. From Ref.~\cite{ALICE:2023gcs}.}
    \label{fig:incohJpsi_PbPb}
\end{figure}

Figure~\ref{fig:incohJpsi_PbPb} shows the ALICE measurement of the incoherent photoproduced J/$\psi$ cross section as a function of $|t|$ at mid-rapidity, corresponding to $0.3<x<1.4 \times 10^{-3}$, in ultraperipheral Pb$+$Pb collisions at $\sqrt{s_{\rm NN}}$ = 5.02 TeV~\cite{ALICE:2023gcs}. The measurement is compared to three groups of theoretical calculations~\cite{Mantysaari:2017dwh,Guzey:2018tlk,Mantysaari:2022sux} framed within the Good-Walker picture. Each group provides two predictions: one including only the elastic interaction with single nucleons, and another that considers the sub-nucleon fluctuation effect. None of the models describes both the absolute yield and the shape of the measured $|t|$ spectra. However, the measured $|t|$-slope is reasonably well described when the predicted dependence is moderated by incorporating sub-nucleon fluctuations.

	\section{The quantum phenomena of photoproduction in HICs}\label{third}

        \hspace{2em}In exclusive photoproduction, vector mesons are generated alongside non-resonant continuum production, wherein the photon fluctuates into pion/kaon pairs, which subsequently diffract off the target and emerge as real pion/kaon pairs—a phenomenon known as the Drell-S\"oding process \cite{Soding:1965nh,Drell:1961zz}. The indistinguishable nature of exclusive vector meson production and the Drell-S\"oding process leads to inevitable quantum interference between these contributions, resulting in significant modifications to the mass shape of the vector meson.

In relativistic heavy-ion collisions, a distinct feature arises in photoproduction: the colliding nuclei can interchangeably serve as the photon emitter and the target. This dual role results in photoproduction events originating from two distinct directions, complicating the identification of the photon source. The negative parity of the photon exacerbates this ambiguity, causing the two amplitudes to exhibit opposite signs and inducing destructive double-slit interference at the Fermi scale. This section aims to provide a concise overview of these intriguing quantum interference phenomena observed in high-energy heavy-ion collisions, shedding light on the intricate dynamics underlying these interactions.
        \subsection{The interference in the invariant mass spectrum of $
        \rho^{0}$ photoproduction}
         \hspace{2em}The invariant mass spectrum for any specific final state ({\it e. g.} $\pi^+\pi^-$ or $K^+K^-$) can include contributions from multiple intermediate resonances.  Since all of these channels are experimentally indistinguishable, they can interfere with each other.  

\begin{figure}[!htbp]
    \centering
    \includegraphics[width=0.52\textwidth]{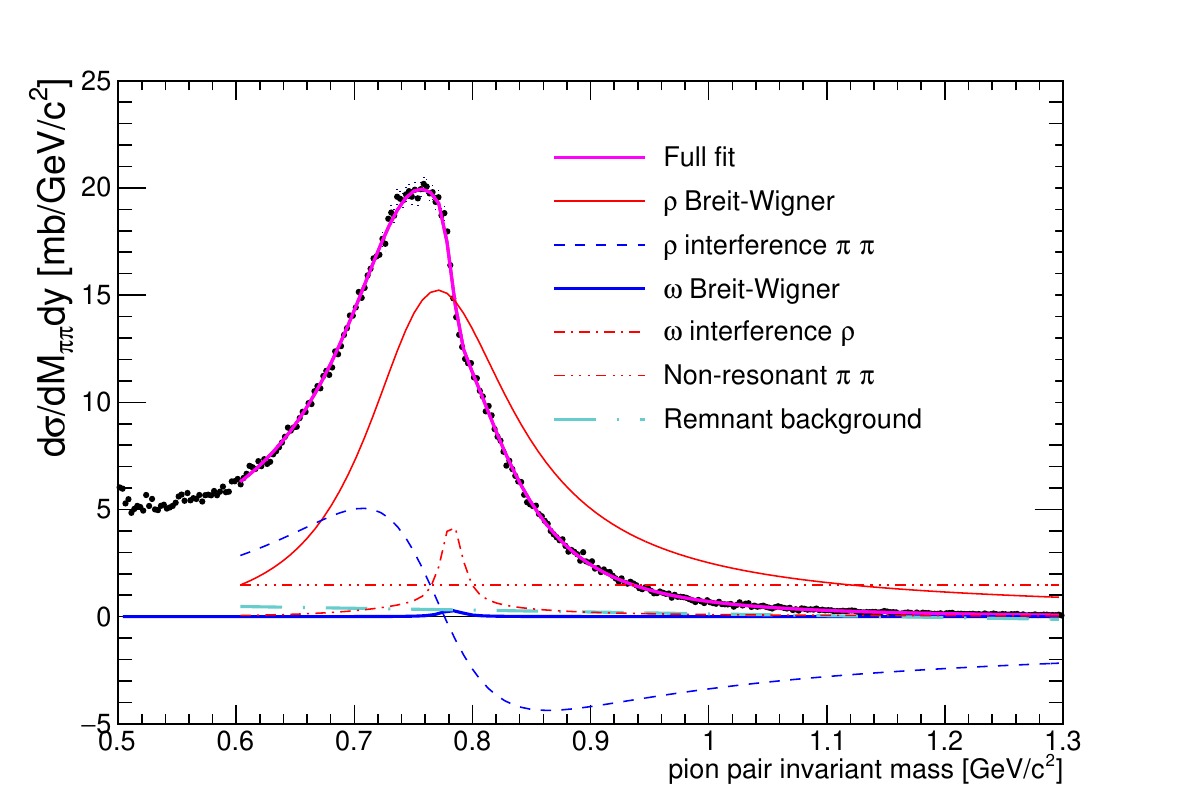} 
     \includegraphics[width=0.44\textwidth]{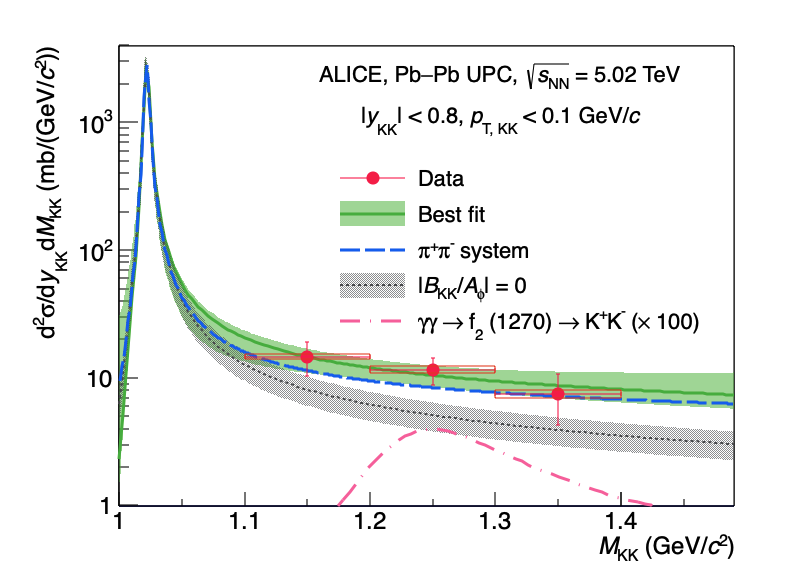} 
    \caption{(left) The $\pi^+\pi^-$ invariant mass spectrum measured by STAR in coherent photoproduction.  The spectrum is fit to the sum of amplitudes from $\rho^0$, direct $\pi^+\pi^-$ and $\omega\rightarrow\pi^+\pi^-$.  From Ref. \cite{STAR:2017enh}. (right) the $K^+K^-$ invariant mass spectrum measured by ALICE in coherent photoproduction. The data is fit to the sum of amplitudes from $\phi\rightarrow K^+K^-$ and direct $K^+K^-$ production, with the phase angle fixed.  The data is compatible with a $\phi:{\rm direct}\ K^+K^-$ ratio that is measured for the $\rho$ and direct $\pi^+\pi^-$.  From Ref. \cite{ALICE:2023kgv}. 
    }
\label{fig:STARdipionmassspectrum}
\end{figure}

The most precise study of the $\pi^+\pi^-$ invariant mass spectrum, from the STAR Collaboration, is shown in Fig. \ref{fig:STARdipionmassspectrum} (left).  The spectrum includes 384,000 entries, and is fit to a sum of $\rho^0$, direct $\pi^+\pi^-$ and $\omega\rightarrow\pi^+\pi^-$ amplitudes.  The fit was done to a modified S\"oding functional form \cite{Soding:1965nh}, with the $\omega$ added in:
\begin{equation}
 \frac{d\sigma}{dM_{\pi^{+}\pi^{-}}}\! \propto\! \left | A_{\rho}\frac{\sqrt{M_{\pi\pi}M_{\rho}\Gamma_{\rho}}}{M_{\pi\pi}^{2}-M_{\rho}^{2}+iM_{\rho}\Gamma_{\rho}}+B_{\pi\pi}+C_{\omega}e^{i\phi_{\omega}}\frac{\sqrt{M_{\pi\pi}M_{\omega}\Gamma_{\omega\rightarrow\pi\pi}}}{M_{\pi\pi}^{2}-M_{\omega}^{2}+iM_{\omega}\Gamma_{\omega}} \right |^{2} + f_{p}
\label{eq:massFitFunction}
\end{equation}
where $A_{\rho}$, $B$ and $C_{\omega}$ are the three amplitudes, $\phi_{\omega}$ is the phase difference between the $\rho$ and $\omega$,
$M_{\pi\pi}$ is the dipion invariant mass, and $M_x$ and $\Gamma_x$ are the resonance masses and widths respectively.   Here, $\Gamma_{\omega\rightarrow\pi\pi}={\rm Br}(\omega\rightarrow\pi^+\pi^-) \cdot\Gamma_{\omega}$ is the partial width for the $\omega$ to decay to two pions.  

The widths in Eq. \ref{eq:massFitFunction} depend on the available phase space, leading to momentum-dependent widths:
\begin{equation}
\Gamma_{\rho} = \Gamma_{0} \frac{M_{\rho}}{M_{\pi\pi}}\left(\frac{M_{\pi\pi}^{2}-4m_{\pi}^{2}}{M_{\rho}^{2}-4m_{\pi}^{2}}\right)^{3/2}
\label{eq:momentumdependewntwidth}
\end{equation}
where $\Gamma_0$ is the $\rho$ pole width.   The $\omega$ is similar, but with one difference: the most common decay, with an 89.2\% branching fraction, is to $\pi^+\pi^-\pi^0$.  This alters how the phase space varies with energy.  So, the $4m_{\pi}^2$ is replaced with $9m_{\pi}^2$.  This effectively assumes that phase space acts as if the $\omega$ branching ratios to 3 pions is 100\%, which is reasonably accurate, but not perfect. 

The fit to the data has a $\chi^2/DOF=255/270$; even with this level of precision, two resonances plus a direct component do a good job of describing the data. One caveat is that the data ends at a mass of 1.3 GeV.  At higher masses, additional resonance structure is visible.  Both STAR and ALICE have reported higher mass peaks in the $\pi^+\pi^-$ mass spectrum.  STAR has observed a peak with mass $1653\pm 10$ MeV and a width of $164\pm 15$ MeV \cite{Klein:2016dtn}, and noted that it might be compatible with the production and decay of the $\rho_3(1690)$.   ALICE has also observed a high-mass $\pi^+\pi^-$ peak, with a mass of 1700 MeV and a width of 140 MeV \cite{ALICE:2020ugp}.  Although one might initially expect that spin-3 mesons would not be produced in photon-Pomeron interactions, detailed calculations indicate that, even though production is off-diagonal, a $k_T$ factorization scheme predicts that it may be produced with appreciable rates, comparable to the $\rho (1700)$ \cite{Caporale:2005me}. It would be very interesting to look for other final states from $\rho_3(1690)$ decays. 

Only recently have we had similar measurements for the $K^+K^-$ system.   ALICE has observed photoproduction of kaon pairs, in the mass range $1.1 < M_{KK} < 1.4$ GeV \cite{ALICE:2023kgv}.  Unfortunately, the study was insensitive to the $\phi$ peak, because the charged kaons produced in $\phi$ decay are soft, with a momentum of 135 MeV/c in the kaon rest frame.  So, $\beta\approx 0.2$; not only do the kaons have very little kinetic energy, they lose energy rapidly.  
        
        \label{sub:chater4:first}
    \subsection{The theoretical framework for photoproduction in heavy-ion collisions}\label{sub:chater4:first2}
    \hspace{2em}Before delving deeper into the quantum phenomena observed in photoproduction, it is essential to introduce several theoretical models that elucidate the underlying mechanisms of this process. These include the Vector Meson Dominance (VMD) model and the color dipole approach. The VMD model provides a comprehensive framework for understanding the interaction between photons and vector mesons, while the color dipole approach elucidates the formation and evolution of quark-antiquark pairs during the photoproduction process.

The double slit-like interference effect can be naturally formulated in these models by taking into account the impact parameter dependence of the amplitude. They also account for significant effects such as gluon saturation and nuclear shadowing, which play a critical role in high-energy collisions. 

     \subsubsection{Vector Meson Dominance}

     \hspace{2em}Vector Meson Dominance (VMD) is commonly applied to photoproduction. In it, a photon can fluctuate into a vector meson, such as $\rho$, $\omega$, or $J/\psi$, which subsequently interacts with the nuclei by exchanging a Pomeron. This conversion process is governed by the vector meson propagator and the coupling constants derived from experimental data~\cite{PhysRevC.60.014903}. The scattering amplitude ${\Gamma}_{{\gamma}A{\rightarrow}VA}$ with the shadowing effect can be obtained by the quantum Glauber~\cite{Miller:2007ri} + VMD~\cite{Bauer:1977iq} approach:
\begin{equation}
	\Gamma_{{\gamma}A{\rightarrow}VA}(\vec{x}_\perp)=\frac{f_{{\gamma}N{\rightarrow}VN}(0)}{\sigma_{VN}}2\left[1-\mathrm{exp}\left(-\frac{\sigma_{VN}}{2}T^{\prime}\left(\vec{x}_\perp\right)\right)\right] ,\label{5}
\end{equation}
where $f_{{\gamma}N{\rightarrow}VN}(0)$ is the forward-scattering amplitude for $\gamma+N\,{\rightarrow}\,V+N$ and $\sigma_{VN}$ is the total $VN$ cross section. $T^{\prime}(\vec{x}_\perp)$ is the modified thickness function accounting for the coherent length effect:
\begin{equation}
	T^{\prime}(\vec{x}_\perp)=\int_{-\infty}^{+\infty}\rho(\sqrt{{\vec{x}_\perp}^2+z^2})e^{iq_Lz}dz,\quad q_L=\frac{M_Ve^y}{2\gamma}, \label{6}
\end{equation}
where $q_L$ is the longitudinal momentum transfer required to produce a real vector meson. The $f_{{\gamma}N{\rightarrow}VN}(0)$ can be determined from the measurements of the forward-scattering cross section $\frac{d\sigma_{{\gamma}N{\rightarrow}VN}}{dt}|_{t=0}$, which is well parameterized in Ref.~\cite{Klein:2016yzr}. Using the optical theorem and VMD relation, the total cross section for VN scattering is
\begin{equation}
	\sigma_{VN}=\frac{f_V}{4\sqrt{\alpha}C}f_{{\gamma}N{\rightarrow}VN}(0),\label{7}
\end{equation}
where $f_V$ is the $V$-photon coupling and $C$ is a correction factor for the off-diagonal diffractive interaction~\cite{Hufner:1997jg}.

The amplitude for vector meson photoproduction in heavy-ion collisions can then be expressed as:

\begin{equation}
A\left( \vec{x}_\perp \right) = \Gamma_{\gamma A \to V A}\left( \vec{x}_\perp \right) \sqrt{\frac{d^2N_\gamma\left(\vec{x}_\perp\right)}{d^2\vec{x}_\perp}},
\end{equation}
where $\Gamma_{\gamma A \to v A}(\vec{x}_\perp)$ is the scattering amplitude incorporating the shadowing effect, and $d^2N_\gamma/d^2\vec{x}_\perp$ is the spatial photon flux induced by the ions, given by EPA approach described in Sec~\ref{sec3-2-1}.

The total cross-section is determined by the square of the amplitude, which naturally incorporates the effects of the nuclear form factor and coherent length effect, which accounts for the gradual loss of coherence with increasing $k_\perp$.   For an optically thin target (where vector mesons do not interact multiple times, {\i. e.} for heavy vector mesons), this is just the square of the sum of the amplitudes for scattering from the individual nucleons, and the typical $k_\perp$ is $\hbar/R_A$.

     \subsubsection{Color dipole model}     
   
     \hspace{2em}The color dipole model has been widely applied to compute the cross section of the diffractive production of vector mesons~\cite{Brodsky:1994kf}. In this framework, the evolution of the dipole-proton interaction can be classified into two regimes: linear evolution, which applies at relatively moderate gluon densities, and nonlinear evolution, which becomes significant at high gluon densities where saturation effects play a role. Here, we use diffractive $\rho^0$ production as an example to introduce the most basic formulation of the calculation within the dipole model framework.
The diffractive photo-production can be divided into  three steps: quasi-real photon splitting into quark and anti-quark pair, the color dipole scattering off nucleus, and subsequently recombining  to form  a vector meson after penetrating the nucleus target. The production amplitude $ {\cal A}(Y,\Delta_\perp)$ can be  expressed as the convolution of the dipole scattering amplitude and the overlap between the
vector meson and photon wave functions in transverse position space, 
 \begin{eqnarray}
 {\cal A}(Y,{\bold \Delta_\perp})=i \!\int \! d^2
b e^{i  {\bold \Delta_\perp} \cdot b} \!\int \! \frac{d^2
r_\perp}{4\pi} \! \int_0^1 \!\! dz \  \Psi^{\gamma\rightarrow q\bar q} (r_\perp,z,\epsilon_{\perp}^{\gamma})N(r_\perp,b)
\Psi^{ V\rightarrow q \bar q *} (r_\perp,z,\epsilon_{\perp}^{V}),
 \end{eqnarray}
where ${\bold \Delta_\perp}$ is the nucleus recoil transverse
momentum. $\epsilon_{\perp}^{\gamma}$ and $\epsilon_{\perp}^{V}$ are
the transverse polarization vectors for the incident
quasi-real photon and final outgoing vector meson, respectively. The
polarization dependent wave function $\Psi^{\gamma\rightarrow q\bar
q}$ ($\Psi^{ V\rightarrow q \bar q}$) of the quasi-real photon
(vector meson) is determined from light cone perturbation theory. $z$ denotes the fraction of the
photon's light-cone momentum carried by the quark.
$N(r_\perp,b)$ is the elementary amplitude for the scattering
of a $q\bar q$ dipole of size $r_\perp$ on a target nucleus at the
impact parameter $b$ of the $\gamma A$ collision.
The overlap of photon and meson wave functions  can be cast into the following form after
integrating out the azimuthal angle of $r_\perp$, 
\begin{eqnarray}
\sum_{a,a',\sigma,\sigma'} \int \! \frac{d^2
r_\perp}{4\pi} \Psi^{\gamma \rightarrow q\bar q}
\Psi^{V\rightarrow q \bar q  *} N(r_\perp,b)= & \!\! (\epsilon_{\perp}^{V*}  \cdot
\epsilon_{\perp}^\gamma ) \displaystyle \frac{e e_q}{2\pi} 2N_c  \int \frac{d^2
r_\perp}{4\pi}  \ N(r_\perp,b) \Bigg \{ \left [ z^2+(1-z)^2
\right ]
\nonumber \\
 \times & \!\!\!
\displaystyle \frac{\partial\Phi^*(|r_\perp|,z)}{\partial |r_\perp|}
 \frac{\partial  K_0(|r_\perp| e_f)}{\partial |r_\perp|}
  +m_q^2 \Phi^*(|r_\perp|,z)K_0(|r_\perp| e_f) \Bigg \},
 \label{eq-wave-combine}
\end{eqnarray}
where $\Phi^*(|r_\perp|,z)$ is the scalar part of the meason wave function~\cite{Kowalski:2003hm}. The photon's polarization vector $\epsilon_{\perp}^\gamma$ manifestly couples to meson's transverse polarization vector $\epsilon_{\perp}^{V*}$. Most importantly, one should notice  that $\epsilon_{\perp}^\gamma$ is parallel to the incident photon's transverse momentum as it is fully linearly polarized. 

All the nontrivial dynamics is encoded in the dipole scattering amplitude $N(r_\perp,b_\perp)$.  The formal operator definition of  the dipole scattering amplitude  is given by,
\begin{eqnarray}
N(b, r_\perp)=1-  \frac{1}{N_c}\langle
{\rm Tr} U(b+r_\perp/2) U^\dag(b-r_\perp/2)  \rangle
\end{eqnarray}
The dipole amplitude is typically computed using the color glass condensate(CGC) effective theory. More specifically, the $x$ dependence of the dipole amplitude  is determined by  solving the  Balitsky-Kovchegov(BK) equation~\cite{Balitsky:1995ub,Kovchegov:1999yj}, where the initial condition is either fitted to experimental data or derived from the McLerran-Venugopalan(MV) model~\cite{McLerran:1993ni,McLerran:1993ka}. However, implementing the impact parameter-dependent BK equation numerically poses significant challenges. For the sake of simplicity, researchers often resort to a phenomenological parametrization to characterize the $b_\perp$ dependence of the dipole amplitude for a nuclear target, as described in Refs.~\cite{Kowalski:2003hm,Kowalski:2006hc},
\begin{eqnarray}
N(b, r_\perp)
=1-e^{-2\pi B_p A T_A(b) {\cal N}(r_\perp)}
\end{eqnarray}
where  ${\cal N}(r_\perp)$ is the dipole-nucleon scattering amplitude.
For heavy nuclei, the nuclear thickness function $T_A(b)$ is conventionally determined with the Woods-Saxon distribution. Numerous parameterizations exist for the dipole-proton cross section, including models like the GBW model~\cite{Golec-Biernat:1998zce} and the Bartels, Golec-Biernat, and Kowalski(BGBK) parametrization~\cite{Bartels:2002cj}.

        \subsection{The double-slit interference in UPCs}\label{sub:chater4:second}
   
        \begin{figure}[htbp]
    \centering
    \includegraphics[width=0.4\textwidth]{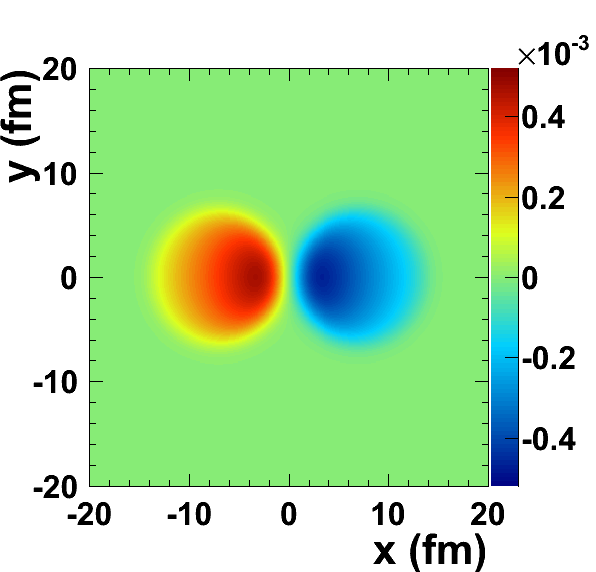}
    \includegraphics[width=0.4\textwidth]{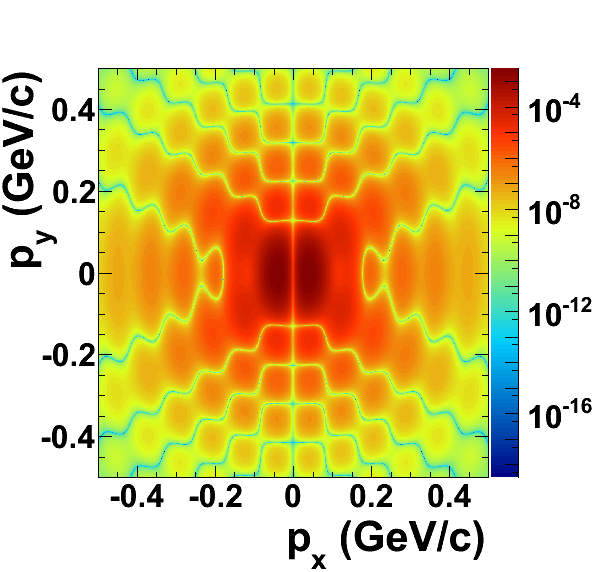}
    \caption{Amplitude (left) and momentum (right) distribution profiles for coherent J/$\psi$ photoproduction in Au+Au collisions at $\sqrt{s_{\rm{NN}}} = 200$ GeV with an impact parameter of 15 fm at midrapidity ($y = 0$) as predicted by the VMD model. From Ref.~\cite{Zha:2018jin}.}
    \label{fig:amp:mom}
\end{figure}

\hspace{2em}In extending from a single photon hitting a single target to UPCs, an important interference effect emerges. For coherent photoproduction, the kinematics of photon emission are akin to those of dipole-nucleus scattering, making it indiscernible which nucleus is the photon emitter and which is the target. Consequently, the amplitudes are added, forming a Young's double-slit interference experiment with nuclei acting as the slits \cite{Klein:1999gv}.  The cross sections can be expressed as:
\begin{equation}
\sigma(p_\perp,y,b) = \big| A_1(p_\perp,y,b) - A_2(p_\perp,y,b)e^{i(k\cdot b)} \big|^2
\label{eq:interfere}
\end{equation}
The negative sign between the amplitudes arises because vector mesons have negative parity, and swapping the photon emitter and target is equivalent to a parity inversion. The total $p_\perp$ is the vector sum of the photon and Pomeron transverse momenta, with the Pomeron component tending to dominate, especially at LHC energies. 

Interestingly, for photoproduction in $p\overline p$ collisions (as observed at the Fermilab Tevatron), the sign is positive because swapping the photon emitter and target equates to a charge-parity inversion, and vector mesons exhibit positive CP \cite{Klein:2003vd}.

\begin{figure}[htbp]
    \centering
    \includegraphics[width=0.5\textwidth]{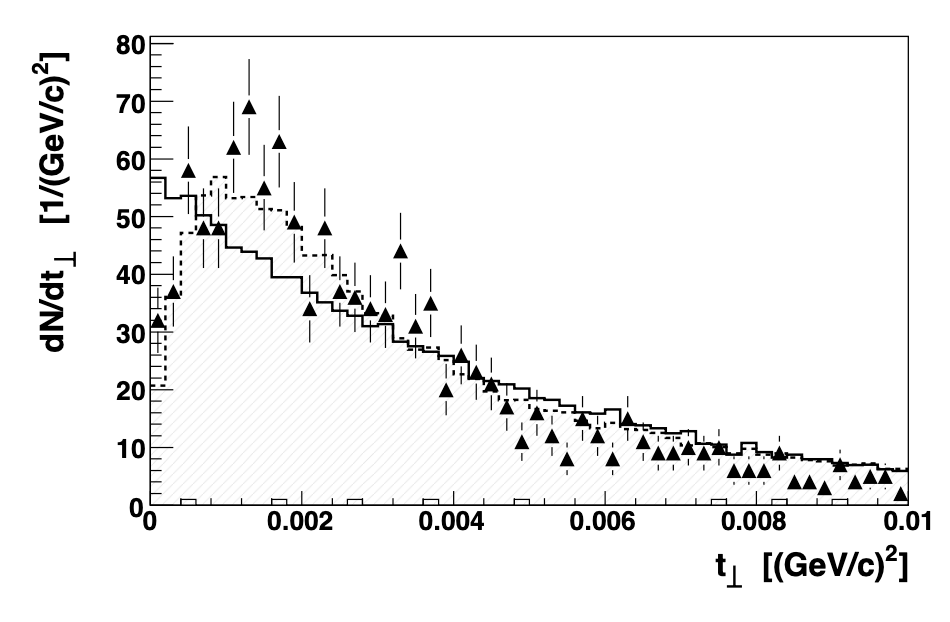}
    \includegraphics[width=0.4\textwidth]{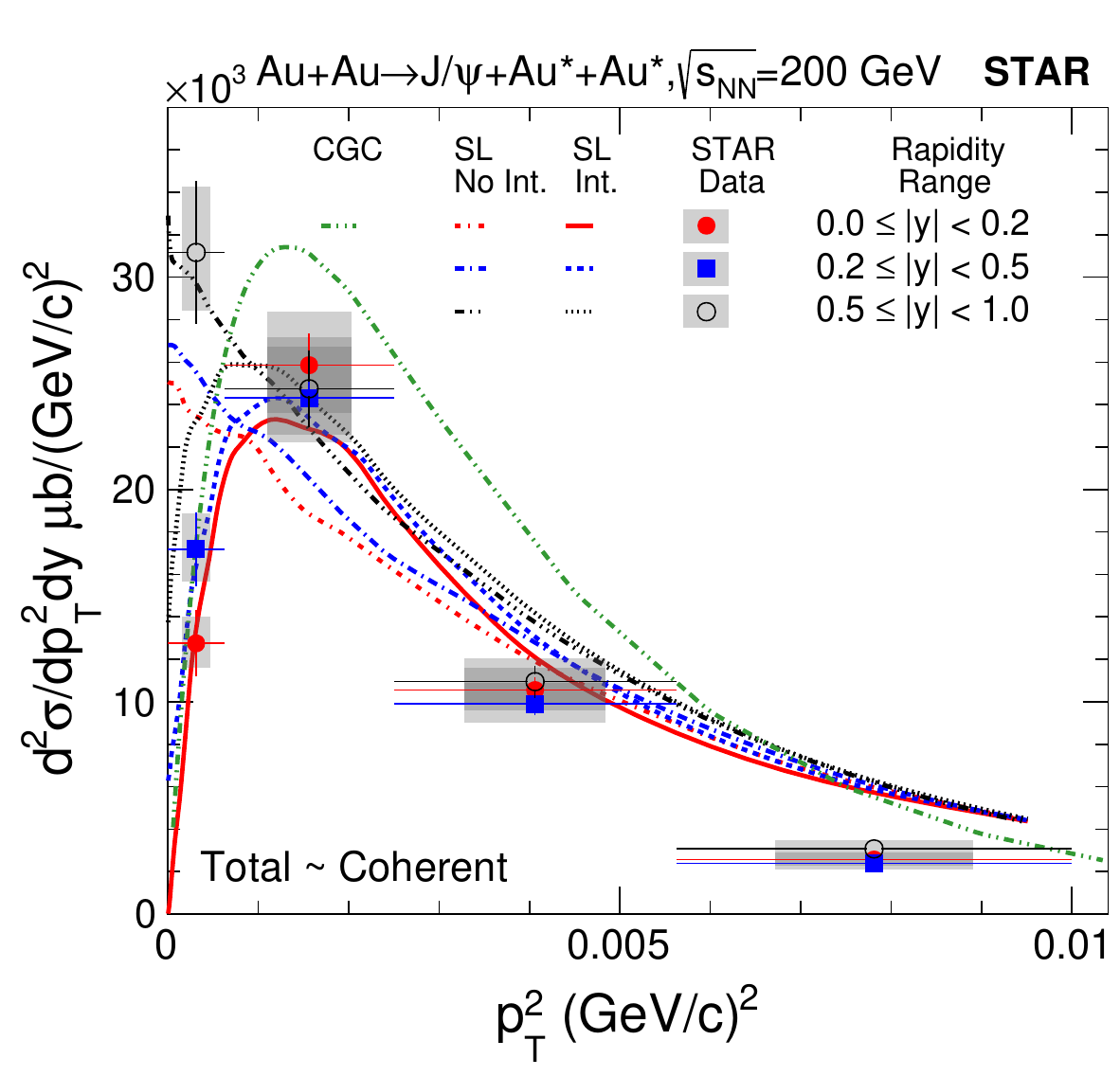}
    \caption{Left Panel: STAR data on $\rho$ photoproduction in the region $|y|<0.5$ (triangles and error bars) compared with calculations including (solid histogram) and excluding (dashed histogram) interference effects. From Ref. \cite{STAR:2008llz}. Right Panel: Differential cross section measurements from STAR, $\frac{d^2\sigma}{dp_T^2 dy}$, as a function of $p_T^2$ at very low $p_T$, with different rapidity $|y|$ bins in Au+Au UPCs at $\sqrt{s_{\rm{NN}}} = 200$ GeV. STARlight~\cite{Klein:2016yzr} events and color glass condensate (CGC)~\cite{Mantysaari:2022sux} calculations are compared with the data. A systematic uncertainty of 10\% from the integrated luminosity is not shown. From Ref. \cite{STAR:2023vvb}.}
    \label{fig:STARinterferencemeasurment}
\end{figure}

Figure \ref{fig:amp:mom} (left panel) illustrates the amplitude distributions for asymmetric slits (where the amplitude from the two slits have opposite signs) in Au+Au collisions at  a per-nucleon center of mass energy $\sqrt{s_{\rm{NN}}} = 200$ GeV at midrapidity ($y = 0$) as estimated by the VMD model~\cite{Zha:2018jin}. The right panel shows the resulting interference pattern, where the typical interference fringes exhibit sinusoidal variation. However, this variation is not apparent due to the logarithmic scale of the z-axis. Additionally, due to the non-spherically symmetric slit profile, the diffraction pattern does not display typical symmetric diffraction rings, causing the interference fringes to curve.

Due to the averaging of different orientations in UPCs, it is impossible to determine a specific direction for each event. Consequently, a two-dimensional momentum distribution cannot be observed, and only the one-dimensional (integrated over angle) transverse momentum spectrum can be measured. A significant feature of the one-dimensional transverse momentum spectrum is the substantial suppression of yield as the transverse momentum approaches zero.  This can  be intuitively understood from Eq.~\ref{eq:interfere}. The exponential term in Eq. \ref{eq:interfere} represents the phase difference for the vector meson to move between the two nuclei, setting the $p_T$ scale for the problem. When $p_T < \hbar/b$, the exponential is near 1, and the two amplitudes directly subtract. For larger $p_T$, the amplitudes combine with a random phase, and after integration over $b$, the interference averages out. Therefore, the $b$-integrated cross-section is reduced when $p_T < \hbar/\langle b\rangle$, where $\langle b\rangle$ is the mean impact parameter. Interference is largest for distributions with smaller $\langle b\rangle$, {\it i. e.} collisions with multiple photon exchanges.

The left panel of Fig.~\ref{fig:STARinterferencemeasurment} presents STAR data showing this interference pattern \cite{STAR:2008llz}. The collaboration plotted $d\sigma/dt_\perp = d\sigma/dp_T^2$ because the interference is more visible using $t_\perp$. Without interference, $d\sigma/dp_T$ is roughly exponential (following the nuclear form factor), with a maximum at $t_\perp=0$ while with interference $d\sigma/dt_\perp \rightarrow 0$ as $t_\perp\rightarrow 0$. Due to the finite rapidity range, the two amplitudes in Eq. \ref{eq:interfere} are not identical everywhere, and thus $d\sigma/dp_T$ does not reach 0 at $p_T = 0$. The curves in this panel
are based on Monte Carlo simulations that include this incomplete interference. The data shows a distinct downturn at low $t_\perp$, consistent with the prediction that includes interference. After studying data using both topological and minimum bias UPC triggers across two rapidity ranges ($|y|<0.5$ and $0.5 < |y| <0.9$), the collaboration found that the interference was $87\pm 5 ({\rm stat.}) \pm 8 ({\rm syst.})$\% of the expected strength.

The STAR experiment also measured exclusive J/$\psi$ photoproduction in Au+Au UPCs at $\sqrt{s_{\rm{NN}}} = 200$ GeV~\cite{STAR:2023nos,STAR:2023vvb}. The right panel of Fig. \ref{fig:STARinterferencemeasurment} displays the differential cross section of J/$\psi$ photoproduction at very low $p_T^2$ for three rapidity bins. In this region, the cross-section is predominantly from  coherent photoproduction, with approximately 1\% contamination from incoherent processes. The cross section at the lowest $p_T^2$ bin shows more than 50\% suppression at the lowest rapidity compared to the highest rapidity bin, with intermediate rapidity showing less suppression. At higher $p_T^2$ bins, the cross sections are roughly equal across all rapidities. This pattern is attributed to quantum interference in symmetric Au+Au UPCs, where the ambiguity of the photon source nucleus leads to destructive interference. STARlight calculations, both with and without interference, quantitatively demonstrate this effect. When turning off the interference in STARlight simulation, no suppression is observed at the lowest $p_T^2$. Accordingly, STARlight calculations incorporating interference predict the observed trend: increased suppression at the lowest $p_T^2$ as rapidity approaches zero.

Although this interference resembles a two-slit system, it possesses several unique aspects~\cite{Klein:2002gc} that will be elaborated upon in subsequent sections. First, unlike a conventional double-slit interferometer, the two paths (photon-Pomeron interactions) that produce a vector meson do not share any common space-time history; there is no overlap. Instead, the fixed relative phase, which typically arises from a shared history, is derived from the inherent symmetry of the problem. Second,  the decay $\rho \rightarrow \pi^+\pi^-$ occurs almost instantaneously, with $c\tau \approx 1.2$ fm. This distance is significantly smaller than the separation between the two nuclei, which ranges from 20 to 300 fm depending on the meson, beam energy, and the presence of additional photon exchanges \cite{Baltz:2002pp}. Consequently, the two $\rho$ mesons cannot directly interfere; any observed interference must involve the $\pi^+\pi^-$ pairs at a time after the amplitudes from the two nuclear sources have overlapped, i.e., after the $\pi^+$ and $\pi^-$ are well separated. This indicates that particle decay does not induce wave function collapse—the wave function retains amplitudes for all possible decay channels. This scenario exemplifies the Einstein-Podolsky-Rosen paradox, which will be discussed in detail in Sec.~\ref{Sec5-3}. Finally, unlike the standard double-slit experiment, which involves the measurement of a single particle, the interference effect observed in photonuclear interactions involves the simultaneous measurement of two particles—the daughters of the produced vector meson. In this regard, the measurement bears similarities to intensity interferometry, which is sensitive to interference effects arising from higher-order coherence.

        \subsection{The double-slit interference in hadronic heavy-ion collisions}\label{sub:chater4:third}
   
        \hspace{2em}Conventionally, coherent photoproduction processes are expected only in UPCs, where the quantum-mechanical coherence criterion - that the nucleus does not break up - can be met. In hadronic heavy-ion collisions, the colliding nuclei are shattered, so  coherent photoproduction was expected to be prohibited in such collisions. The shift in this traditional perspective began with ALICE $J/\psi$ measurements conducted in peripheral collisions at very low transverse momenta.
\begin{figure}[!htbp]
    \centering
    \includegraphics[width=0.4\textwidth]
    {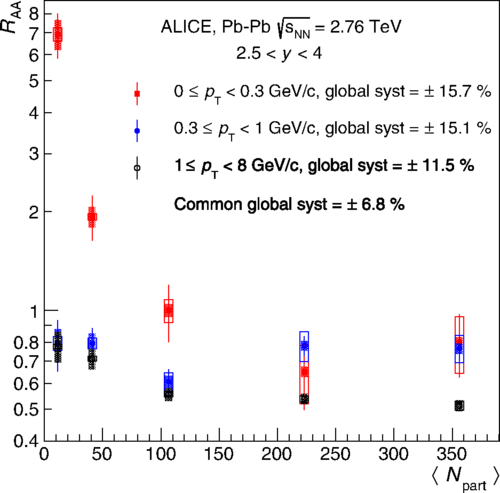}
     \includegraphics[width=0.5\textwidth]{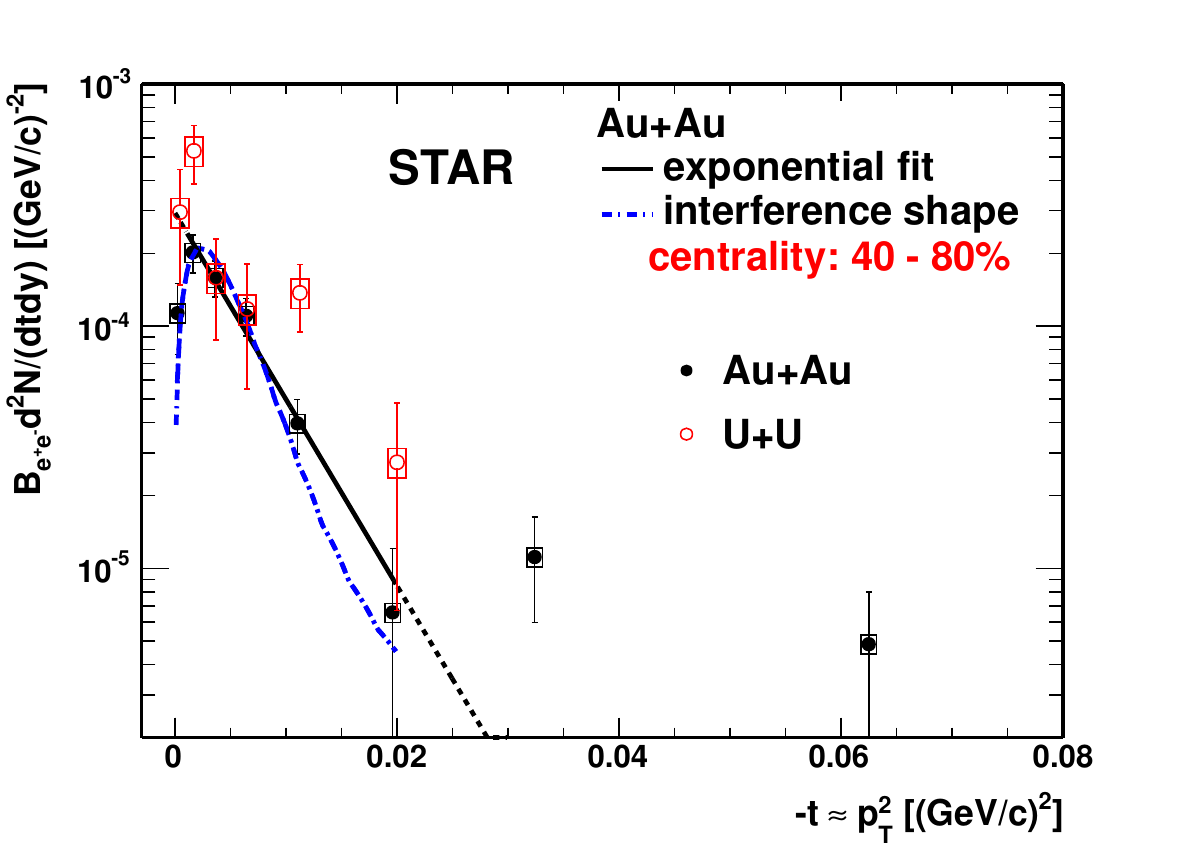} 
    \caption{(Left Panel) The nuclear modification factor (\(R_{\text{AA}}\)) of \(J/\psi\) as a function of number of participant nucleons in Pb+Pb collisions at \(\sqrt{s_{\text{NN}}} = 2.76\) TeV. (Right Panel) The \(J/\psi\) yield, after subtracting the anticipated hadronic contributions, plotted against the negative squared momentum transfer \((-t \sim p_{\text{T}}^2)\) for the 40-80\% centrality class in both Au+Au and U+U collisions. Error bars indicate statistical uncertainties, while boxes denote systematic uncertainties.  From Ref.~\cite{ALICE:2015mzu,STAR:2019yox}. 
    }
\label{fig:STAR:HHIC:RAA-T}
\end{figure}
In 2016, the ALICE Collaboration~\cite{ALICE:2015mzu} focused on measuring the yield of $J/\psi$ production at extremely low transverse momentum via the dimuon decay channel in peripheral Pb + Pb collisions at $\sqrt{s_{\rm{NN}}}$ = 2.76 TeV. The analysis involved approximately 17 million dimuon minimum bias events, detected using a sophisticated triggering system that identified opposite sign (OS) tracks in the forward muon spectrometer ($2.5 < y < 4.0$). The results revealed a significant excess compared to hadronic production, which may originate from coherent photoproduction. Subsequently, STAR conducted similar measurements using the Au + Au data collected during the 2010-2011 RHIC runs at $\sqrt{s_{\rm{NN}}}$ = 200 GeV and the U + U data collected in 2012 at $\sqrt{s_{\rm{NN}}}$ = 193 GeV~\cite{STAR:2019yox}. The total number of events used in Au + Au collisions and U + U collisions were 720 million and 270 million, respectively. As shown in the left panel of Fig.~\ref{fig:STAR:HHIC:RAA-T}, a significant enhancement was also observed in peripheral collisions at very low transverse momentum ($p_{T} < 0.1$ GeV/c). Within this $p_{T}$ range, suppression of $J/\psi$ production due to the color screening effect and cold nuclear matter effect is expected. Conversely, the regeneration effect, which would typically elevate $J/\psi$ production, is deemed negligible~\cite{Yan:2006ve, Rapp:2008tf}. Consequently, the prevailing expectation is a suppression of $J/\psi$ production for hadronic processes. Hence, the anomalous surplus observed at extremely low $p_{T}$ indicates the existence of alternative mechanisms governing $J/\psi$ production. The anomalous enhancement, primarily concentrated at extremely low transverse momenta and peripheral collisions, closely resembles the characteristics of coherent photoproduction, suggesting the presence of coherent photoproduction even in non-UPCs. Models incorporating coherent photoproduction mechanisms~\cite{Zha:2017jch,Klusek-Gawenda:2015hja} can reasonably describe the observed excesses reported by the ALICE and STAR Collaborations. In addition to the measurements conducted by ALICE and STAR, further experimental measurements~\cite{LHCb:2021hoq, Mallick:2024yre,Li:2022ncj,ALICE:2022zso} have also observed the anomalous enhancement of $J/\psi$. These experimental results, along with theoretical explorations, corroborate the existence of coherent photoproduction in hadronic heavy-ion collisions.

The presence of coherent photoproduction in hadronic heavy-ion collisions naturally provokes some questions: What are its diffraction properties? Does interference manifest? And if so, does the interference behavior differ from that observed in UPCs? Pursuing these questions, STAR conducted additional measurements, specifically focusing on the distribution of the negative momentum transfer square ($-t \sim p_{T}^{2}$) for the photoproduction of $J/\psi$ in peripheral Au+Au and U+U collisions. The measured $-t$ distributions are shown in the right panel of Fig.~\ref{fig:STAR:HHIC:RAA-T}, where the expected hadronic contribution has been excluded. The shape of the $-t$ distributions closely resembles that observed in UPC collisions~\cite{STAR:2002caw}, further confirming that the excess originates from coherent photoproduction. An exponential fit has been performed on the measurements over the range $0.001 < -t < 0.015\ {\rm GeV}/c^2$,  with the slope parameter sensitive to the effective size of the photoproduction target. The fitted slope parameter is $177 \pm 23$ GeV/c$^2$, which aligns within uncertainties with expectations for an Au nucleus (199 GeV/c$^2$), as well as with measurements in UPCs~\cite{ALICE:2021tyx}. This alignment suggests that the diffraction pattern in hadronic collisions mirrors that observed in UPCs, attesting to the precision of the measurements. Intriguingly, the first data point at $-t < 0.001$ GeV/$c^2$ is more than 3$\sigma$ lower than the extrapolation of the exponential fit, hinting at the presence of interference in hadronic collisions. Theoretical calculations~\cite{Zha:2017jch} incorporating interference, depicted as the blue curve, well describe the $-t$ distributions for Au+Au collisions. Given the smaller impact parameter for hadronic heavy-ion collisions compared to UPCs, suppression due to interference is expected at larger $p_T$.  Although the data is suggestive, its current precision of data prevents us from drawing definitive conclusions.

        \subsection{The spin interference in HICs }\label{sub:chater4:fourth}
        \subsubsection{The linearly polarized photons in HICs }
        \label{sub:4-6-1}

\hspace{2em}As explained earlier, the coherent photons are highly linearly polarized. This phenomenon can also be analyzed in the context of Transverse Momentum Dependent(TMD) factorization. In the covariant gauge, the dominant component of the gauge potential $A$ is the light-cone component along the direction of the charged particle movement: the plus component. The dominant component of the electromagnetic field strength tensor can therefore be written as:
$F_{+ }^\mu=\partial_+ A^\mu-\partial^\mu A_+$
where $\mu$ is a transverse index. By taking the partial derivative, one obtains,
$F_{+ }^\mu \propto k_+ A^\mu-k_\perp^\mu A_+$
where $k_\perp$ and $k_+$ are the transverse component and the plus component of 
 small $x$ photons, the photon field strength can be further approximated as,
$F_{+ }^\mu \propto -k_\perp^\mu A_+$. 
 Therefore, the transverse component of the field strength tensor is parallel to the photon transverse momentum direction. That is, in the small $x$ region, the photons are 100\% linearly polarized. In the TMD factorization framework, this polarization phenomenon is described by the linearly polarized photon TMD. Its operator definition is given by the following formula ~\cite{Mulders:2000sh,Pisano:2013cya},
\begin{eqnarray}
\int \frac{2dy^- d^2y_\perp}{xP^+(2\pi)^3} e^{i\bold{k} \cdot y} \langle P
|  F_{+\perp}^\mu(0) F_{+\perp}^\nu(y)  |P \rangle \big|_{y^+=0}=
{\bold \delta_\perp}^{\mu \nu} f_1^\gamma(x,\bold{k}_\perp^2)+ \left (\frac{2\bold{k}_\perp^\mu
\bold{k}_\perp^\nu}{\bold{k}_\perp^2}-{\bold \delta_\perp}^{\mu\nu} \right )
 h_1^{\perp \gamma}(x,\bold{k}_\perp^2) , \label{gmat}
\end{eqnarray}
where $f_1^\gamma(x,\bold{k}_\perp^2) $ and  $h_1^{\perp \gamma}(x,\bold{k}_\perp^2)$ are the photon unpolarized TMD and linearly polarized TMD respectively.  And the second rank  tensor is given by $\delta_\perp^{\mu \nu}=-g^{\mu nu}+\frac{P^\mu \bar P^\nu+P^\nu \bar P^\mu}{2 P \cdot \bar P} $ with $P^\mu=(P^+,0,0,P^+) $ and $\bar P^\mu=(P^-,0,0,-P^-) $ being the commonly defined light cone vectors.

\begin{figure}[htbp]
    \centering
    \includegraphics[width=0.6\textwidth]{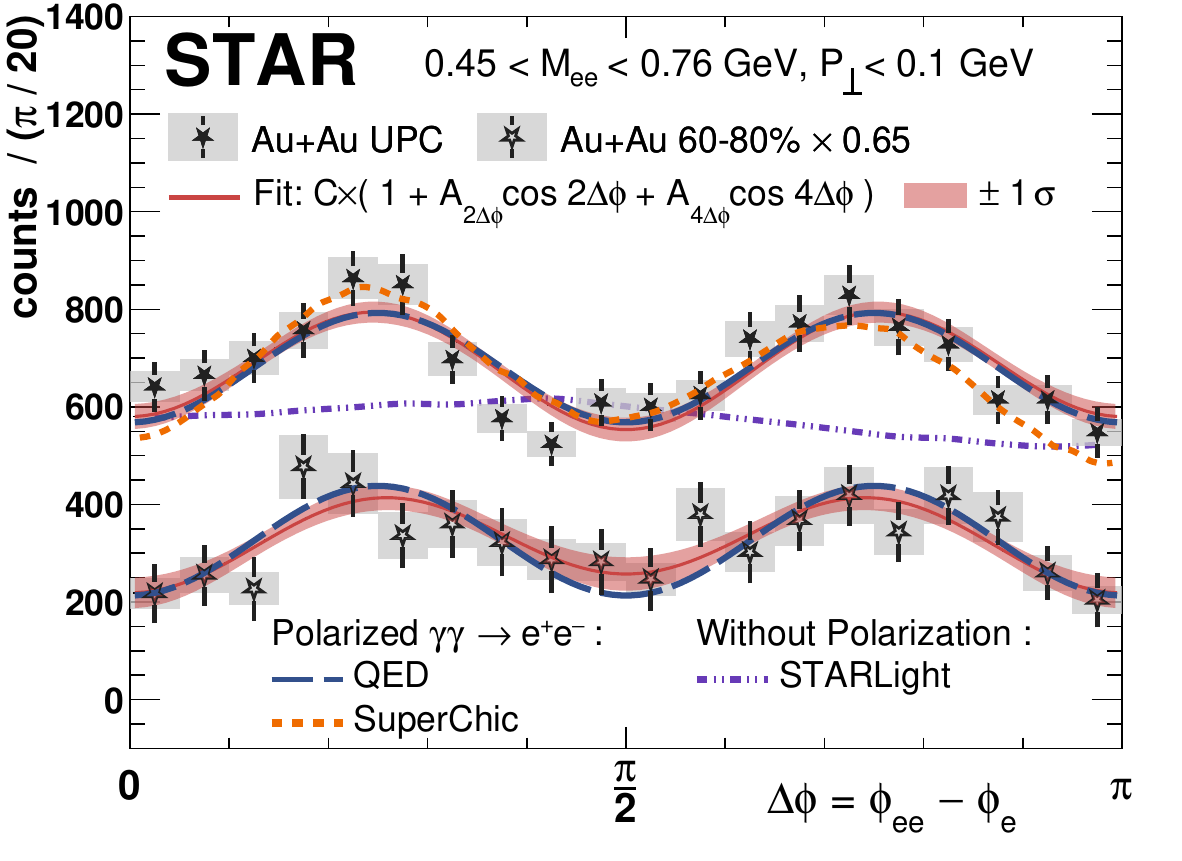}
    \caption{The $\Delta \phi = \phi_{ee} - \phi_{e}$ distribution from ultra-peripheral and 60\%–80\% central Au +Au collisions at $\sqrt{s_{\rm{NN}}}$ = 200 GeV for $M_{ee} > 0.45 \ \mathrm{GeV}$, along with calculations from QED~\cite{Li:2019yzy}, STARLight~\cite{Klein:2016yzr}, and the publicly available SuperChic3 code~\cite{Harland-Lang:2018iur}. From Ref.~\cite{STAR:2019wlg}.
}
    \label{fig:dielectron:cos4phi}
\end{figure}

The linear polarization in the small $x$ region is  a universal feature of gauge theory.  The gauge particles in QCD, gluons,  are also highly linearly polarized in the small $x$ region~\cite{Metz:2011wb,Dominguez:2011br}. 
The linearly polarized gluon TMD can be measured via the azimuthal asymmetries in various high energy scattering processes. 
In close analogy to the QCD case, the linearly polarized photon distribution  can be probed via $\cos4\phi$ azimuthal asymmetry in lepton pair production in UPCs~\cite{Pisano:2013cya,Li:2019sin,Li:2019yzy}, where $\phi $ is the angle between $\bold{k}_\perp$  and individual lepton's transverse momentum $\bold{p}_\perp$.  Observations from the STAR experiment have demonstrated a significant fourth-order cosine oscillation in the azimuthal angle distribution of electron-positron pairs. This phenomenon, depicted in Fig.~\ref{fig:dielectron:cos4phi} , which shows the $\Delta \phi = \phi_{ee} - \phi_{e}$ distribution from ultra-peripheral and 60–80$\%$ central Au + Au collisions at $\sqrt{s_{\rm{NN}}}$ = 200 GeV for $M_{ee} > 0.45 \ \mathrm{GeV}$. Note that the azimuthal asymmetries are not affected by the Lorentz boost along the beam direction, as they arise from the interference between two different helicity states of coherent photons.

This azimuthal asymmetry in lepton pair production can be intuitively understood as the following.  Linearly polarized photons can be regarded as being in a superposition state of spins $+1$ and $-1$. The lepton pair produced through the fusion of two photons with spin $+1$ in the amplitude carries an orbital angular momentum of $+2$, while the lepton pair produced by two photons with spin $-1$ in the conjugate amplitude carries an orbital angular momentum of $-2$. The interference of the two lepton pair wave functions with different orbital angular momenta naturally leads to the $\cos 4\phi$ azimuthal asymmetry.  Apart from this $\cos 4\phi$ azimuthal asymmetry, the phenomenology resulting from the linear polarization of coherent photons has been investigated in various  contexts~\cite{Dai:2024imb,Shao:2023bga,Zhao:2023nbl,Wu:2022exl,Niu:2022cug,Brandenburg:2022tna,Wang:2022gkd,Shi:2024gex,Lin:2024mnj,Zhang:2024mql,Yu:2024icm,Jia:2024xzx}.

        \subsubsection{The linearly polarized photoproduction in HICs }
 
        \begin{figure}[htbp]
    \centering
    \includegraphics[width=0.4\textwidth]{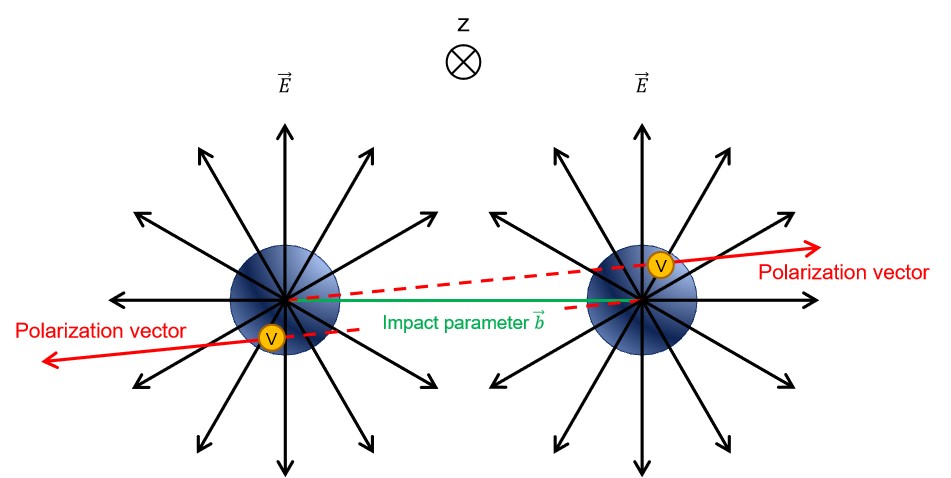}
    \includegraphics[width=0.25\textwidth]{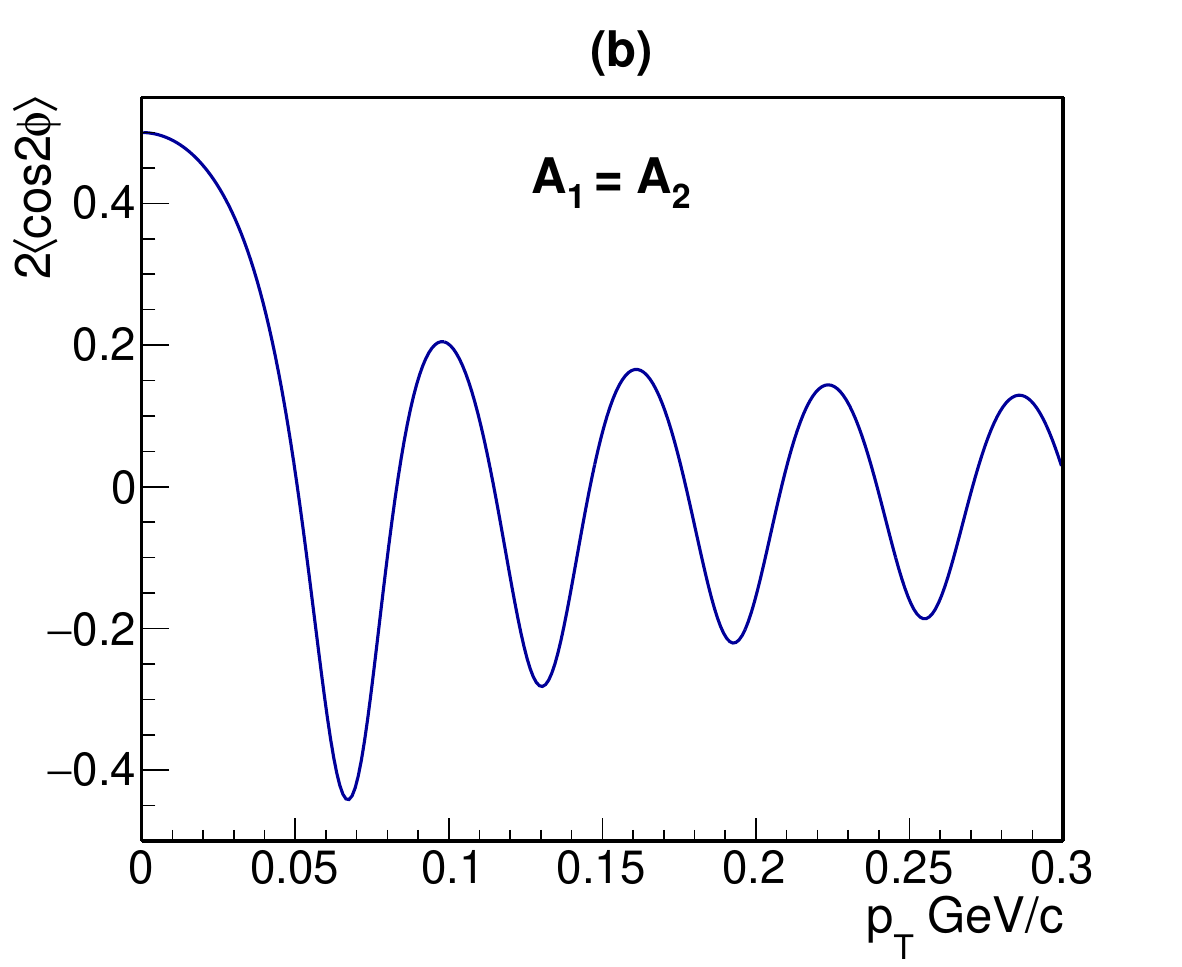}
    \includegraphics[width=0.25\textwidth]{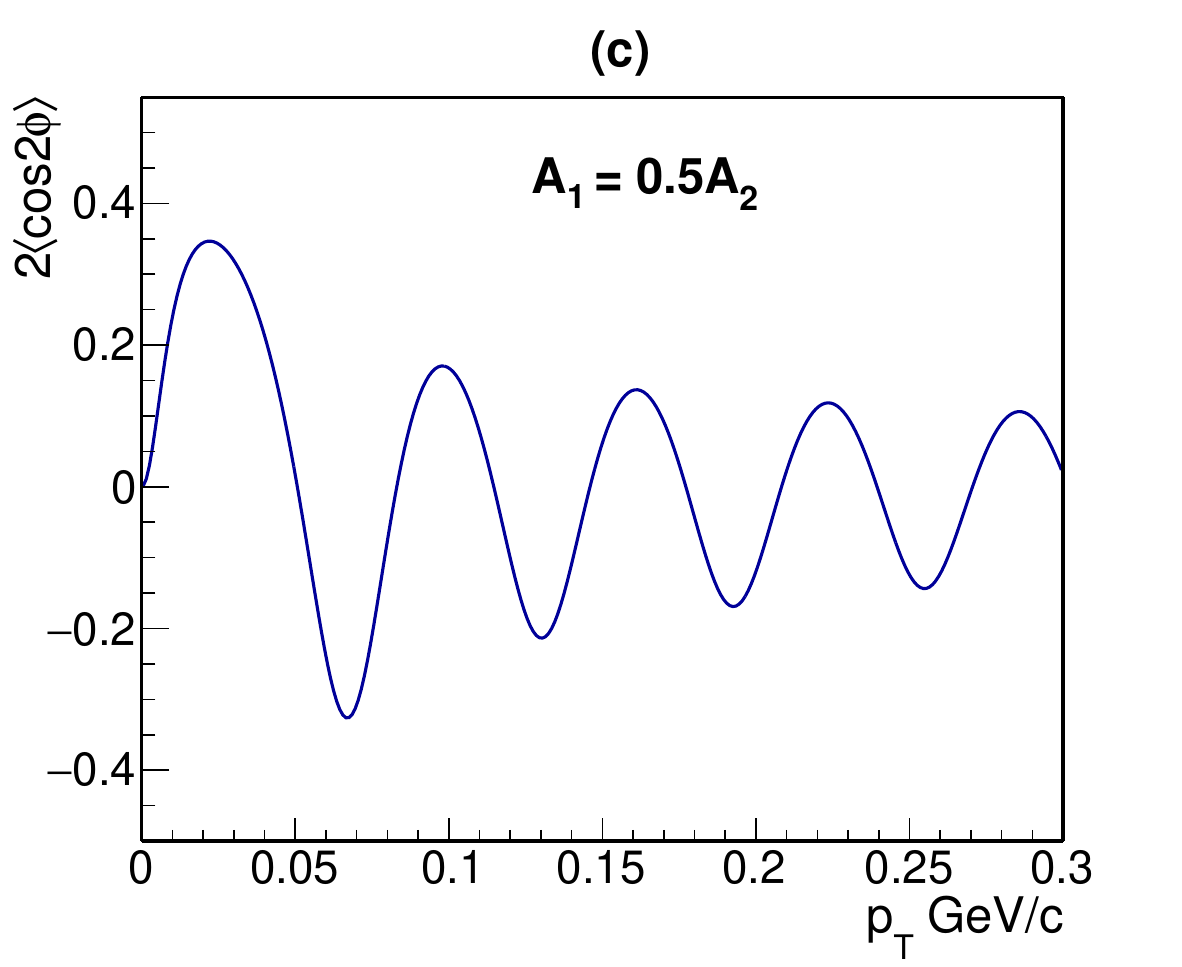}
    \caption{(Left panel) The schematic diagram for the direction of the polarization direction of the photons for vector meson photoproduction. (Middle panel) The modulation strength, 2$\langle \cos\phi \rangle$, as a function of transverse momentum at mid-rapidity. (Right panel) 2$\langle \cos\phi \rangle$ as a function of transverse momentum at forward rapidity ($A_{1} \neq A_{2}$).  From Refs.~\cite{Zha:2020cst,Wu:2022exl}.
}
    \label{fig:rho:spin-interference:schematic}
\end{figure}

\begin{figure}[htbp]
    \centering
    \includegraphics[width=0.75\textwidth]{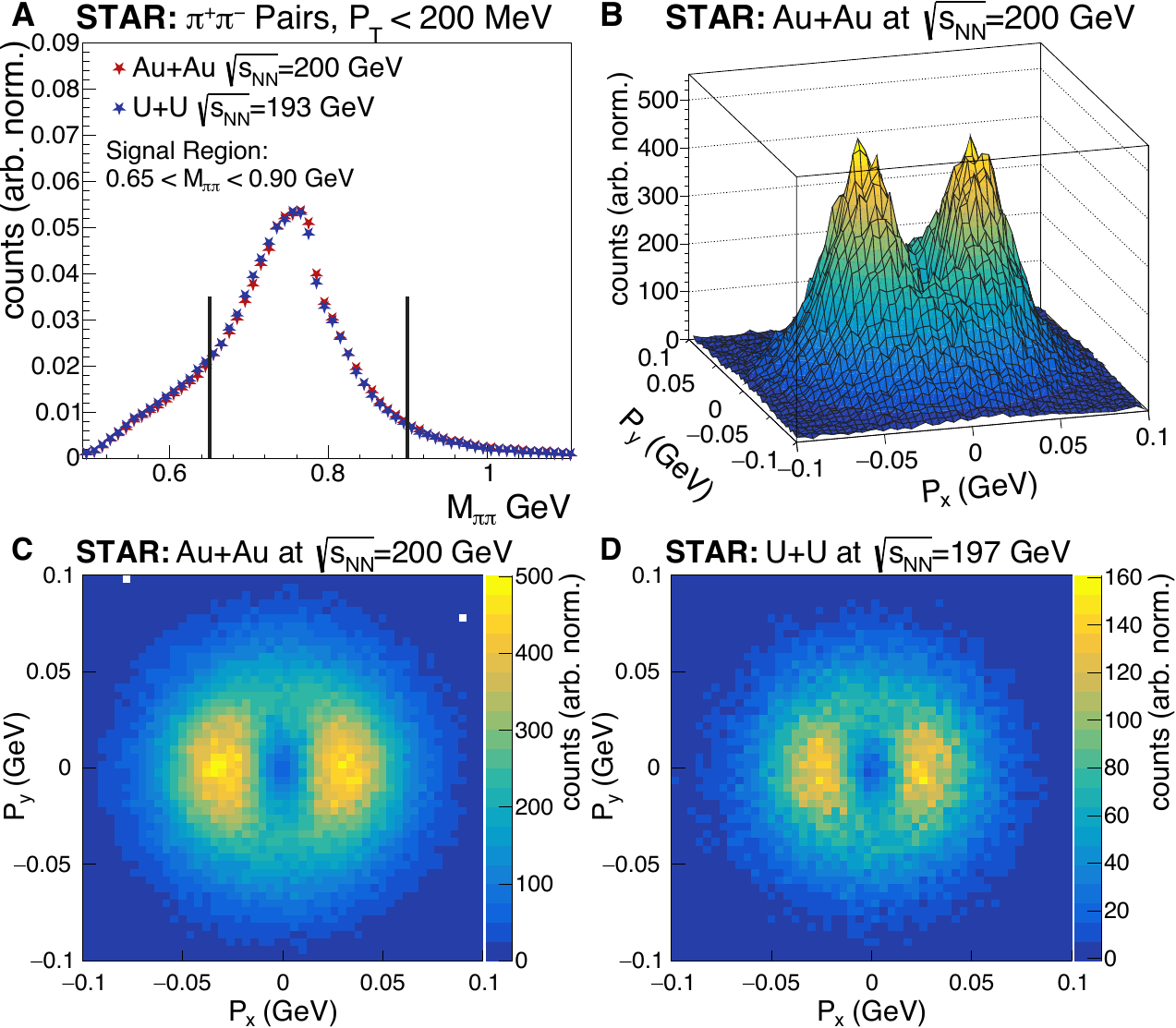}
    \caption{(A) The invariant mass distribution of \(\pi^+\pi^-\) pairs obtained from Au+Au and U+U collisions. The vertical black lines denote the selected mass range, which maintains uniform detector acceptance and efficiency in \(\phi\). Panels (B) to (D) illustrate the two-dimensional distribution of the \(\rho^0\) transverse momentum, with \(P_x = p_T \cos\phi\) and \(P_y = p_T \sin\phi\). Panel (B) presents the Au+Au collision data as a continuous surface, while Panel (C) shows it as a 2D image. For comparison, Panel (D) displays the U+U collision data as a 2D image. From Ref.~\cite{STAR:2022wfe}.
}
    \label{fig:rho:spin-interference:figure1}
\end{figure}

\begin{figure}[htbp]
    \centering
    \includegraphics[width=0.95\textwidth]{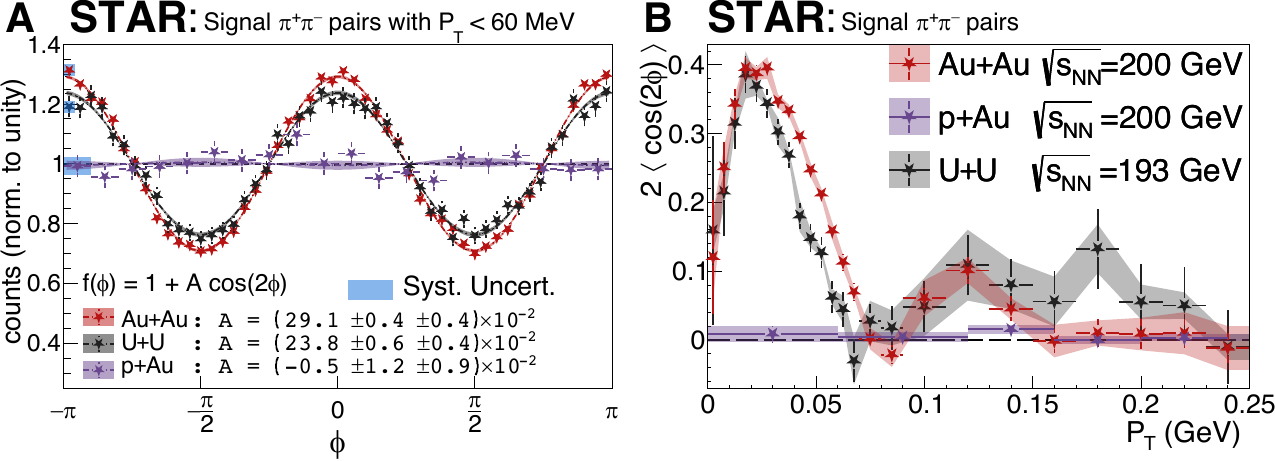}
    \caption{(A) The \(\phi\) distribution for \(\pi^+\pi^-\) pairs collected from Au+Au and U+U collisions with a pair $p_T$
    %transverse momentum (\(p_T\)) 
    less than 60 MeV and an invariant mass between 650 and 900 MeV. Statistical uncertainties are represented by vertical bars on all points, while systematic uncertainties are indicated as filled boxes on the leftmost points for better clarity. (B) The fully corrected \(\langle \cos 2\phi \rangle\) modulation as a function of $p_T$ for Au+Au and U+U collisions. Statistical uncertainties for each data point are shown as vertical bars, and systematic uncertainties are displayed in shaded bands. From Ref.~\cite{STAR:2022wfe}.
}
    \label{fig:rho:spin-interference:figure2}
\end{figure}

Under the helicity no-flip assumption, the vector meson produced in photoproduction inherits the polarization state of the photon, which is completely linearly polarized in heavy-ion collisions. This helicity conservation assumption has been validated by numerous experimental measurements~\cite{Criegee:1968fxl,Ballam:1970ex,Eisenberg:1975jr,ZEUS:1996esk}. Following Ref.~\cite{Schilling:1969um}, 
the decay angular distribution of the vector meson can be expressed as:
\[
W(\theta, \phi) \propto \sum_{l=0,\pm1}\sum_{m,m'} D_{ml}^{1}(\phi, \theta, -\phi)D_{m'l}^{1*}(\phi, \theta, -\phi)\rho_{mm'},
\]
where $D_{ml}^{1}(\theta, \phi)$ are the complex rotation matrix elements, and $\rho_{mm'}$ is the spin density matrix. The complex rotation matrix elements \(D^1_{ml}(\phi, \theta, -\phi)\) are defined as:
\[
D^1_{ml}(\phi, \theta, -\phi) = e^{-i(m - l)\phi} d^1_{ml}(\theta), \tag{A3}
\]
where the coefficient $d^1_{ml}(\theta)$ can be found in the Clebsch-Gordan coefficient tables. The spin density matrix $\rho_{mm'}$ can be derived from the polarization state of vector meson, it reads:
\[
|V\rangle = -\frac{1}{\sqrt{2}} e^{-i\Phi} |+1\rangle + \frac{1}{\sqrt{2}} e^{i\Phi} |-1\rangle \tag{A6}
\]
where \(\Phi\) is the angle between the linear polarization vector and the production plane of the vector meson.
The decay angular distribution of a vector meson into two spinless products (e.g., \(\rho^{0} \rightarrow \pi^{+} + \pi^{-}\)) is then given by:
\begin{equation}
\frac{d^{2}N}{d\cos\theta d\phi} = \frac{3}{8\pi}\sin^{2}\theta [1 + \cos2(\phi-\Phi)],
\label{equation001}
\end{equation}
while for spin-1/2 products (e.g., \(\rho^{0} \rightarrow e^{+} + e^{-}\)), it is expressed as:
\begin{equation}
\begin{aligned}
\label{equation_Jpsi_0}
\frac{d^{2}N}{d\cos\theta d\phi} = \frac{3}{16\pi}(1+\cos^{2}\theta) \left[1-\frac{\sin^{2}\theta}{1+\cos^{2}\theta} \cos2(\phi-\Phi)\right],
\end{aligned}
\end{equation}
where \(\theta\) and \(\phi\) are the polar and azimuthal angles of the decay particles in the rest frame of the vector meson, and \(\Phi\) is the azimuthal direction of the linear polarization. The right-handed coordinate system for decay is comprehensively discussed in Ref.~\cite{Zha:2020cst}. As shown in Eqs.~\ref{equation001} and \ref{equation_Jpsi_0}, linearly polarized states lead to second-order modulations in azimuth, with the strength of the modulation given by:

\begin{equation}
2\langle \cos(2\phi) \rangle = \cos(2\Phi)
\label{equation002}
\end{equation}

and 

\begin{equation}
2\langle \cos(2\phi) \rangle = -\frac{\sin^{2}\theta}{1+\cos^{2}\theta}\cos(2\Phi),
\label{equation003}
\end{equation}
for the \(\rho^{0} \rightarrow \pi^{+} + \pi^{-}\) and \(\rho^{0} \rightarrow e^{+} + e^{-}\) cases, respectively. The modulation strength is dictated by the direction of the linear polarization, which is aligned with the electric field vector. As illustrated in the left panel of Fig.~\ref{fig:rho:spin-interference:schematic}, the electric field vector is parallel to the impact parameter at leading order. This implies that the modulations in the decay distribution of the vector meson are determined by the anisotropy in the two-dimensional transverse momentum distribution with respect to the impact parameter at leading order. The variation in polarization direction due to the finite size of nuclei should be minimal in UPCs, which can be explored within the theoretical framework mentioned in Sect.~\ref{sub:chater4:first2}.

To simplify the physical picture, the production over the two nuclei can be approximated as two point sources: \(A_{1,2}(x_{\perp}) = \pm A_{1,2}\delta(x_{\perp} \pm \frac{1}{2}b)\), where \(A_{1}\) and \(A_{2}\) are the production amplitudes for the two colliding nuclei, and the minus sign in \(A_{2}(x_{\perp})\) arises from the negative parity of the vector meson. With the interference effect, as demonstrated in Sect.~\ref{sub:chater4:fourth} and the left panel of Fig.~\ref{fig:amp:mom} for the case at mid-rapidity (\(A_{1} = A_{2}\)), the transverse momentum reveals a typical Young’s double-slit interference pattern with alternating light and dark fringes, corresponding to significant asymmetries in azimuth. According to Eq.~\ref{equation002}, the modulation strength factor for vector meson decay to two spinless products, \(2 \langle \cos(2\phi) \rangle\), as a function of transverse momentum (\(p_{T}\)) can be extracted from the two-dimensional distribution. The result, shown in the middle panel of Fig.~\ref{fig:rho:spin-interference:schematic}, exhibits periodic oscillations akin to modified (hyperbolic) Bessel functions with respect to transverse momentum. The oscillation cycle is determined by the distance between the two sources (impact parameter). For production at forward rapidity in heavy-ion collisions, the relative amplitudes from the two nuclei are not equal. The right panel of Fig.~\ref{fig:rho:spin-interference:schematic} shows the modulation strength as a function of transverse momentum with unequal amplitudes from the two sources. Since the interference is not complete in this case, the oscillation amplitude is weaker than at mid-rapidity. Furthermore, the modulation strength is zero at \(p_{T} = 0\), while it is maximal at mid-rapidity. This is because the interference contribution is negligible at \(p_{T} = 0\).

For detailed calculations in the dipole framework, one must first formulate the calculation in impact parameter space. It is crucial to take into account the $b$ dependence in the calculation, where  $b$ refers to the transverse distance between two colliding nuclei. By Fourier transform to transverse momentum space, one eventually ends up with the joint $b$ and $\rho^0$ transverse momentum $q_\perp$ dependent cross section~\cite{Xing:2020hwh},
 \begin{eqnarray}
\frac{d \sigma}{d^2 p_T dY d^2 b}\!\!
&=&\!\! \frac{1}{(2\pi)^4} \int d^2 {\bold \Delta_\perp} d^2k_\perp d^2 k_\perp'\delta^2(k_\perp+{\bold \Delta_\perp}-p_T)
 (\epsilon_{\perp}^{V*}  \cdot \hat k_\perp )(\epsilon_{\perp}^{V} \cdot \hat k_\perp' )
 \nonumber \\
 && \bigg \{ \left [  e^{i  b \cdot (k_\perp' -k_\perp)}
{\cal A}(Y,{\bold \Delta_\perp}) {\cal A}^*(Y, {\bold \Delta_\perp}')
{\cal F}(Y,k_\perp){\cal F}(Y,k_\perp')
 \right ]  
  \nonumber \\
 &&+ \left [  e^{i  b \cdot ({\bold \Delta_\perp}'-{\bold \Delta_\perp})}
{\cal A}(-Y,{\bold \Delta_\perp}) {\cal A}^*(-Y, {\bold \Delta_\perp}')
{\cal F}(-Y,k_\perp){\cal F}(-Y,k_\perp')
 \right ]
\nonumber \\
&& + \left [ e^{i  b \cdot ({\bold \Delta_\perp}' -k_\perp)}
 {\cal A}(Y,{\bold \Delta_\perp}) {\cal A}^*(-Y, {\bold \Delta_\perp}'){\cal F}(Y,k_\perp){\cal F}(-Y,k_\perp')
 \right ]
    \nonumber \\
   && + \left [ e^{i  b \cdot (k_\perp' -{\bold \Delta_\perp})}
 {\cal A}(-Y,{\bold \Delta_\perp}) {\cal A}^*(Y, {\bold \Delta_\perp}'){\cal F}(-Y,k_\perp){\cal F}(Y,k_\perp')
 \right ]
\bigg \},
\label{fcs}
 \end{eqnarray}
where ${\cal F}(Y,k_\perp)$ is linked to the transverse momentum-dependent (TMD) distribution of coherent photons via the relation ${\cal F}^2(Y,k_\perp) =x
f(x,k_\perp)$, and its specific expression can be found in Ref.~\cite{Xing:2020hwh}. ${\bold \Delta_\perp}'$  adheres to transverse momentum conservation:
$k_\perp+{\bold \Delta_\perp}=k_\perp'+{\bold \Delta_\perp}'$.  Notably, $k_\perp$ and $k_\perp'$ correspond to the incident photon’s transverse momenta in the amplitude and the conjugate amplitude, respectively. Their non-coincidence arises from $ b$ dependence.   The presence of different phase factors, namely $e^{-i b\cdot k_\perp} $ and $e^{-i b \cdot {\bold \Delta_\perp}}$, can lead to substantial destructive interference which is particularly prominent  at mid-rapidity and low transverse momentum~\cite{Klein:1999gv,STAR:2008llz}.

The STAR Collaboration conducted the first spin-interference measurements for the linearly polarized photoproduction process of \(\rho^0\) mesons in UPCs~\cite{STAR:2022wfe,ma2023new}. The data were collected from three different collision systems: Au+Au collisions at \(\sqrt{s_{NN}} = 200\) GeV (2010-2011), U+U collisions at \(\sqrt{s_{NN}} = 193\) GeV (2012), and p+Au collisions at \(\sqrt{s_{NN}} = 200\) GeV (2015). 
The UPC events were selected using a specialized trigger system that relied on mutual Coulomb dissociation signals. This was achieved by requiring coincidences in the Zero Degree Calorimeters (ZDCs) located on both sides of the interaction point. These events were also required to have signals in the mid-rapidity detectors, indicating the presence of at least 2 but not more than 6 charged particles, to effectively select exclusive events. 

Figure~\ref{fig:rho:spin-interference:figure1}(A) presents the measured invariant mass distribution for selected $\pi^+\pi^-$ candidates with pair $p_T < 200$ MeV from Au+Au and U+U collisions
A broad, prominent peak around the $\rho^0$ mass of approximately 770 MeV is evident, confirming that the trigger system and track selection criteria effectively select exclusive $\rho^0 \rightarrow \pi^+\pi^-$ processes. For further analysis, candidate $\pi^+\pi^-$ pairs with invariant masses ($M_{\pi\pi}$) between 650 and 900 MeV are chosen. This selection range primarily includes $\pi^+\pi^-$ pairs from the decay of $\rho^0$ vector mesons produced in diffractive photonuclear interactions, ensuring a high signal-to-background ratio and roughly uniform acceptance.

The transverse momentum distribution of the parent $\rho^0$ meson can be decomposed into two components: one parallel ($p_x$) and one perpendicular ($p_y$) to the decay azimuthal angle:

\begin{equation}
\label{eq:sub4-6-2:eq1}
p_x = p_T \cos\phi, \quad p_y = p_T \sin\phi.
\end{equation}
Given the preferential alignment of the decay angle with the direction of linear polarization, and the high alignment of the polarization direction with the reaction plane, $p_x$ and $p_y$ are the transverse momentum components relative to the reaction plane. The determination of this plane has an accuracy of approximately 0.5, as discussed in Sect.~\ref{sec:collectivity}. This setup enables the observation of the 2D transverse momentum distribution in UPCs for the first time. The resulting 2D representation of the $\rho^0$ transverse momentum is shown in Fig.~\ref{fig:rho:spin-interference:figure1}(B) and (C) for Au+Au collisions and in Fig.~2(D) for U+U collisions. In all three cases, a significant interference pattern is observed, closely matching the theoretical prediction shown in Fig.~\ref{fig:amp:mom} Sect.~\ref{sub:chater4:second}. The differences observed between Au+Au and U+U collisions arise from variations in nuclear density distributions and the impact parameter distributions of the different collisions.

The $\phi$ distribution for Au+Au and U+U events is displayed in the left panel of Fig.~\ref{fig:rho:spin-interference:figure2} for pairs with $650 < M_{\pi\pi} < 900$ MeV and $p_T < 60$ MeV. Each distribution is normalized such that the average yield (integrated over $\phi$) is unity to facilitate comparison. The low $p_T$ range is chosen to emphasize the region where coherent production dominates, minimizing contributions from incoherent photonuclear interactions. Both Au+Au and U+U results exhibit a clear and pronounced $\cos 2\phi$ modulation, consistent with the expected spin-interference effect. The modulation amplitude is quantitatively determined by fitting the distribution to the function

\begin{equation}
f(\phi) = 1 + A \cos 2\phi,
\end{equation}
where $A$ represents the amplitude of the $\cos 2\phi$ modulation. The extracted amplitude is larger for Au+Au collisions due to the smaller nuclear size and larger average impact parameter, which leads to a stricter alignment along the reaction plane. To further validate the origin of the $\cos 2\phi$ modulation, the $\phi$ distribution for p+Au collisions is also shown for comparison. As expected, the p+Au data exhibit an isotropic distribution without modulation, originating from negligible interference between the production amplitudes from p and Au.

The fully corrected $2\langle \cos 2\phi \rangle$ distributions as a function of the $\rho^0$ transverse momentum are shown in the right panel of Fig.~\ref{fig:rho:spin-interference:figure2} for Au+Au, U+U, and p+Au collisions. Both Au+Au and U+U cases display a prominent maximum at low $p_T$, which rapidly decreases towards zero, followed by a smaller peak at higher $p_T$. This structure closely resembles the simple estimation shown in the right panel of Fig.~\ref{fig:rho:spin-interference:schematic}, demonstrating that the spin-interference assumption captures the main features. In contrast, the data from p+Au collisions show no significant structure, remaining consistent with zero across all $p_T$. At the highest reported $p_T$ (~240 MeV), the modulation strength is consistent with zero in all three datasets, indicating dominance by incoherent contributions, which do not produce spin-interference. Model calculations incorporating spin-interference effects within the VMD~\cite{Zha:2020cst} and dipole frameworks~\cite{Xing:2020hwh,Brandenburg:2022jgr} (Model I and II in the right panel of Fig.~\ref{fig:rho:spin-interference:figure4} Sect.~\ref{sec:tomography}) qualitatively describe the Au+Au measurements, further confirming the observation of spin-interference for $\rho^0$ photoproduction in heavy-ion collisions. Model I successfully predicts the peak modulation strength and the position of the first peak, aligning well with the experimental data, whereas Model II fails to accurately replicate the measured magnitude or the observed structure. Model II incorporated gluon saturation, utilizing an effective nuclear radius of 6.9 fm to replicate the $|t|$ spectra. Although neither model accurately captures the precise location and amplitude of the second peak, both theoretical frameworks suggest that the fine details of the modulation pattern are highly sensitive to the gluon distribution within the nucleus. A detailed discussion of the application to nuclear tomography will be presented in Sect.~\ref{sec:tomography}.

Extending the analysis to diffractive $J/\psi$ production in UPCs is straightforward. Here, the mass of the charm quark serves as a hard scale, thus justifying a perturbative treatment.  Similarly, the linear polarized photon distribution leads to a comparable azimuthal asymmetry in $J/\psi$ production.  Additionally, the final state soft photon emitted from the produced dilepton pair can also give rise to significant azimuthal asymmetries due to the mechanism discovered in Refs.~\cite{Catani:2014qha,Catani:2017tuc,Hatta:2020bgy,Hatta:2021jcd}. The azimuthal asymmetries in dileptons produced via the photon fusion process $\gamma\gamma\rightarrow l^+l^-$ in UPCs are also influenced by this effect, particularly at relatively large pair transverse momentum~\cite{Shao:2022stc,Shao:2023zge}.  The detailed calculation for the same observable  in $J/\psi$ production including the aforementioned effects was carried out  in Ref.~\cite{Brandenburg:2022jgr}.  The observation of linear polarization of $J/\psi$ production through the S-channel helicity conservation conservation in peripheral collisions has been reported in ~\cite{Massacrier:2024fgx}.

Furthermore, the $\cos 4\phi$ asymmetry in coherent photo-production of $\rho^0$ in UPCs could potentially offer insights into the elusive elliptic gluon Wigner distribution~\cite{Hatta:2016dxp,Zhou:2016rnt,Boussarie:2018zwg,Mantysaari:2020lhf,Dumitru:2021mab,Hagiwara:2021xkf}. In summary, due to the large flux of quasi-real linearly polarized photons, photo-nuclear production  in UPCs provide us unique chances to investigate partonic structure of nucleus before the advent of the EIC era.  Leveraging the linear polarization of interacting photons, as evidenced by STAR measurements, provides a fresh avenue for delving into gluon tomography within the nucleus.

	\section{Theoretical concerns and paradigms regarding interference}\label{fourth}

        \hspace{2em}In typical thought experiments exploring quantum phenomena, establishing a controlled single-particle environment is straightforward. However, achieving such an environment in real-world experiments presents significant challenges. In heavy-ion collisions, photoproduction typically yields only one vector meson, inherently framing an experimental setup that proceeds in a singular, sequential fashion. Unlike conventional interference and diffraction experiments involving electromagnetic interactions, interactions between photo-produced particles and nuclei are governed by the strong nuclear force. In heavy-ion collisions, the impact parameters operate on the fermi scale, which also encompasses the distance between the two 'slits,' rendering this the smallest interference and diffraction experiment documented to date. Within this configuration, quantum hypothesis testing utilizes unstable particles like the $\rho^0$, which has a lifespan of 1.2 fm/$c$, shorter than the slit-to-slit distance. Thus, the wave function from the two 'slits' usually fails to overlap before decay, implying that interference and diffraction phenomena are manifested by the decay products, exemplifying the Einstein-Podolsky-Rosen paradox \cite{Klein:2002gc}. Additionally, photoproduction in heavy-ion collisions exhibits linear polarization, resulting in unique interference phenomena in polarization space. In the following, we will delve into the distinctive features of using photo-induced production in heavy-ion collisions to scrutinize various quantum hypotheses.

        \subsection{Coherence requirements for interference}
        \label{sec:GoodWalker}

\hspace{2em}For a two-slit interferometer to work, the two slits must not be able to tell if a particle passed through them, {\it i. e.} they must not be disturbed by the reaction.  In UPC interferometry, this is like saying that the two ions must not be excited in the process.  This is encapsulated in the Good-Walker paradigm for diffractive interactions~\cite{Good:1960ba}.  It equates coherent production with the target remaining in the ground state. The cross section for a coherent interaction may be written
\begin{equation}
\frac{d\sigma_{\rm coherent}}{dt} = \frac{1}{16\pi} \big\|\langle A(K,\Omega)\rangle\big\|^2
\label{eq:coherent}
\end{equation}
where $A(K,\Omega)$ is the amplitude for a photoproduction reaction; $K$ are the kinematic factors in the reaction (outgoing particle momenta, etc.) while $\Omega$ is the target configuration - the positions of the individual nucleons, the presence or absence of gluonic hot spots,etc.   This configuration varies with time, but on a time scale that is long compared to the time scale of the interaction.

Unfortunately, this definition fails when applied to coherent photoproduction accompanied by single or mutual Coulomb excitation, or to coherent photoproduction in hadronic heavy ion collisions.  Another, classical or semi-classical definition of coherence is when the amplitudes add in-phase, so the cross section scales as the square of the number of targets.  In other words, the cross section for producing a vector meson on a target consisting of $N$ nucleons, each at positions $\vec{b}_i$
\begin{equation}
\frac{d\sigma_{\rm coherent}}{dt} \propto \big\|\Sigma_i A_i \exp(i\vec{\Delta}\cdot\vec{b}_i)\big\|^2
\label{eq:semiclassical}
\end{equation}
where $A_i$ are the amplitudes for production on the individual nucleons.   When the argument of the exponential is small, then the phase angles are all near zero, and the amplitudes add coherently.  So, for identical $A_i$, $d\sigma_{\rm coherent}/dt \propto N^2 A_i^2$.   When the argument of the exponential is large, the phase angles are random, and the amplitudes add incoherently.   So, at small $|\Delta|$, the cross-section is almost entirely coherent, while at large $|\Delta|$, the cross-section becomes almost entirely incoherent.  This is very different from the requirement that the target remain in the ground state.  For UPC interferometry, one could extend the sum to run over all of the nucleons in both nuclei, with an appropriate minus sign to account for the photon direction. 

However, this definition of coherence does not always match our observations.  There are at least two cases where coherent in-phase addition has been observed, but when the target nucleus does not remain intact \cite{Klein:2023zlf}.  The first is in UPC reactions where the final state consists of a vector meson plus two excited nuclei, with the excitation observed via the emission of neutrons, as was depicted in Fig. \ref{fig:STARtopology}.   This signature was central to much of the STAR UPC program, discussed in Sec. \ref{sec3-2-3}.  The second is the observation of coherent $J/\psi$ photoproduction in peripheral heavy ion collisions.  In these cases, the $J/\psi$ may be accompanied by tens or hundreds of particles from the hadronic part of the collision.  In both case, the coherence clearly manifests itself as a large increase in the cross-section for $p_T < {\rm few}\ \hbar/R_A$, as is shown in Fig. \ref{fig:STARrho2002}.  The resolution of this paradox is not yet known.

An extension of the Good-Walker paradigm, by Miettinen and Pumplin included incoherent production \cite{Miettinen:1978jb}.  The incoherent cross section is just the total cross section, minus the coherent component.  The total cross section is 
\begin{equation}
\frac{d\sigma_{\rm total}}{dt} 
 = \frac{1}{16\pi} \langle \big\| A(K,\Omega)\big\|^2 \rangle
 \label{eq:sigmatotal}
\end{equation}
and the incoherent cross section is 
\begin{equation}
  \frac{d\sigma_{\rm incoherent }}{dt}  = \frac{1}{16\pi} 
  \langle \big\| A(K,\Omega)\big\|^2 \rangle - 
  \big\|\langle A(K,\Omega)\rangle\big\|^2.
  \label{eq:sigmaincoherent}
\end{equation}
The incoherent cross section is the difference between the sum of squares and the square of sums.   Miettinen and Pumplin pointed out that the incoherent cross section is related to event-by-event fluctuations in the nuclear configuration, including the presence of gluonic hot spots.  These differences lead to a difference between the sum of squares and the square of sums.  Without these fluctuations, the incoherent cross-section is zero. This is a very different prediction than from Eq. \ref{eq:semiclassical}, where, at large $|t|$, the incoherent cross-section scales as $N$ times the single-nucleon cross-section.   

There are tests that could differentiate between these two scenarios \cite{Klein:2023zlf,Klein:2023tqv}.  The first is to compare $d\sigma/dt$ for photoproduction or electroproduction with and without nuclear breakup.  This can be done in a fairly straightforward manner at the EIC , where exclusive vector meson production may be detected in a central detector \cite{Arrington:2021yeb}, and the products of nuclear breakup may be seen in a ZDC \cite{AbdulKhalek:2021gbh}.  It may also be possible to study this at lower energies, with using light meson photoproduction with 12 GeV electrons at Jefferson Lab \cite{Klein:2024cck}.

A second test is to compare lead and gold targets \cite{Klein:2023zlf}.  Lead and gold are both heavy nuclei, with nuclear distributions that are well described by Woods-Saxon distributions, with slightly different radii.  They should have similar nuclear shadowing, and similar nucleonic and subnucleonic fluctuations.  So, in the Good-Walker picture, they should have similar probabilities for incoherent photoproduction.  However, they have very different nuclear level structures.  The differences in required energy transfer to excite these nuclei lead to very different nuclear breakup probabilities at low $|t|$.   Lead-208 is doubly magic, so has a lowest excited level at $E_{\rm min}=2.6$ MeV.  No excitation is possible at smaller energy transfers.  In contrast, gold-197 has a lowest lying excited state at $E_{\rm min}=77$ keV.  If one assumes that the energy transfer to the target nucleus is to a single nucleon, then the equivalent momentum transfer is $p=\sqrt{2E_{\rm min} m_p}$, where $m_p$ is the proton mass.  For lead and gold respectively, this is $p=69$ MeV/c and 12 MeV/c.  So, in an exclusive production reaction, for a vector meson with $12 < p_T < 69$ MeV/c, one expects incoherent production in gold but not in lead.  This is dramatically different from the Good-Walker picture, where nuclear energy levels do not play a role.   It should be noted that the single-nucleon recoil picture is supported by high-statistics STAR data, where incoherent $\rho$ photoproduction at larger $|t|$; the data is well fit by a dipole form factor, but not by an exponential~\cite{Klein:2021mgd}.  

	\subsection{The quantum entanglement paradox for photoproduction in HICs  }\label{Sec5-3}

\newcommand{\upcphase}[3]{e^{{#3}i\left( \phi_#2 + k^{#1} \cdot r_{#2}^{#1} \right)}}
\newcommand{\sphase}[3]{e^{{#3}i \psi_{#2}^{#1}}}
\newcommand{\afa}[3]{ f_{#1#2}^{#3}}
\newcommand{\phase}[2]{e^{i \left( \phi_{#2}^{#1} + k r_{#2}^{#1} \right)}}
\newcommand{\iphase}[2]{e^{-i \left( \phi_{#2}^{#1} + k r_{#2}^{#1} \right)}}
\newcommand{\pibra}[2]{ f_{#2#1}^* \bra{ \pi^{#1}_{#2} } \sphase{#1}{#2}{-} }
\newcommand{\piket}[2]{ f_{#2#1} \ket{ \pi^{#1}_{#2} } \sphase{#1}{#2}{} }
\newcommand{\pibk}[4]{ f_{#2#1}^{*} f_{#4#3} \braket{ \pi^{#1}_{#2} | \pi^{#3}_{#4} } \sphase{#1}{#2}{-} \sphase{#3}{#4}{} } 
\newcommand{\sbk}[3]{ f_{#2#1}^{*} f_{#3#1} \braket{ \pi^{#1}_{#2} | \pi^{#1}_{#3} } \sphase{#1}{#2}{-} \sphase{#1}{#3}{} } 
\newcommand{\bk}[2]{ |f_{#2#1}|^{2} \braket{ \pi^{#1}_{#2} | \pi^{#1}_{#2} } } 

\newcommand{\upcbk}[3]{ f_{#2#1}^{*} f_{#3#1} \braket{ \pi^{#1}_{#2} | \pi^{#1}_{#3} } \upcphase{#1}{#2}{-} \upcphase{#1}{#3}{} } 
\newcommand{\upcpsi}[2]{k^{#1} \cdot r_{#2}^{#1}}

\newcommand{\gbra}[2]{ f_{#1^{#2}}^* \bra{ #1^{#2} } \sphase{#2}{#1}{} }
\newcommand{\gket}[2]{ f_{#1^{#2}} \ket{ #1^{#2} } \sphase{#2}{#1}{} }
\newcommand{\gbk}[2]{ |f_{#1^{#2}}|^{2} \braket{ #1^{#2} | #1^{#2} } } 
\newcommand{\gbkp}[4]{ f_{#1^{#2}}f_{#3^{#4}}^{*} \braket{ #1^{#2} | #3^{#4} } \sphase{#2}{#1}{} \sphase{#4}{#3}{} } 

\hspace{2em}Unfortunately, the classical or semi-classical definitions of coherence discussed earlier fail to capture the full depth of quantum mechanical phenomena. In quantum mechanics, one of the most striking features, beyond coherent superposition, is quantum entanglement. While coherence describes the ability of wavefunctions to interfere constructively or destructively, entanglement delves deeper into the fundamental correlations between quantum systems, even when they are spatially separated. Quantum entanglement, while necessarily ubiquitous in high energy particle and nuclear physics, has proven exceedingly elusive to experimental scrutiny. The parton model is an example. Since its introduction, the parton model has become a cornerstone of high-energy physics, supported by extensive deep inelastic scattering (DIS) measurements~\cite{PhysRev.179.1547,PhysRevLett.23.1415,PhysRevLett.30.1343}. The basic tenet of the parton model is that nucleons are composed of quasi-free point-like `partons' frozen in the infinite momentum frame due to Lorentz dilation. Despite the stupendous success of the parton model, at its core lies a conundrum - how can a zero (von Neumann) entropy pure state (i.e the proton) be described by a collection of quasi-free objects~\cite{Kharzeev:2017qzs}. In other words, how can a maximally entangled state be described by quasi-free constituents absent of correlations? Therefore, the search for experimental signatures of entanglement has persisted as a topic of intense interest across diverse areas of high-energy physics~\cite{PhysRevD.105.014002,Hentschinski:2023izh,Hentschinski:2023izh,Kharzeev:2021yyf,H1:2020zpd,ATLAS:2023fsd}

The recent observation by the STAR experiment of spin interference in photonuclear production claims to be a novel experimental signature of entanglement. However, entanglement is not an obvious consequence of the analogy to a double-slit interference experiment. The entanglement and interference of continuous variables (momentum in this case) is not easily subject to a traditional Bell-type inequality, making the signatures of entanglement more subtle \cite{Klein:2002gc}. In the following, a verbose formalism is used where we retain the often-implicit phase term which is commonly dropped in amplitude level discussions. Additionally, the decay of the vector meson is included in a general way. This somewhat verbose formalism is utilized to discuss the role of entanglement in this process, making a clear parallel to the $E^2I^2$ phenomena discussed in Ref.~\cite{COTLER2021168346}.

In A+A collisions, photonuclear interactions occur when a photon from the Lorentz-contracted electromagnetic field of one nucleus interacts with the target nucleus. 
As discussed, the situation is somewhat analogous to the famous double slit experiment, which is central to the history of quantum mechanics~\cite{Einstein:1905cc,feynman1963feynman}. Taking this view, the photoproduction process can be thought of as potentially proceeding through two distinct (but macroscopically indistinguishable) amplitudes:
\begin{align}
    A_{\gamma\mathbb{P}} = \gamma^A + \mathbb{P}^B \rightarrow \rho^0 \\
    A_{\mathbb{P}\gamma} = \gamma^B + \mathbb{P}^A \rightarrow \rho^0 
\end{align}
where nucleus A acts as the photon emitter with nucleus B acting as the target, and vice versa.
As discussed, one can identify the coherent contribution resulting from the sum of amplitudes (Eq.~\ref{eq:semiclassical}) squared and the incoherent as the sum of squared amplitudes (Eq.~\ref{eq:sigmaincoherent}).

In the case of the quantum double slit experiment, a single quantum object is thought to traverse two indistinguishable paths, resulting in wavefunction interference. As noted previously, a crucial ingredient is the shared space-time history of the two paths which fixes the relative phase of the interfering waves. Such a ``double-slit'' interference effect in photonuclear production could be represented as described in Eq.~\ref{eq:interfere}:
\begin{align}
    \sigma_{\rm total} = \sigma_{\gamma \mathbb{P} + \mathbb{P} \gamma} &\propto |A_{\gamma\mathbb{P}}e^{i\phi_{\gamma\mathbb{P}}} + A_{\mathbb{P}\gamma} e^{i\phi_{\gamma\mathbb{P}}}|^2 \\
    &= |A_{\gamma\mathbb{P}}|^2 + |A_{\mathbb{P}\gamma}|^2 +  A_{\gamma\mathbb{P}}^* A_{\mathbb{P}\gamma} e^{i(\phi_{\mathbb{P}\gamma} - \phi_{\gamma\mathbb{P}})} + A_{\mathbb{P}\gamma}^*A_{\gamma\mathbb{P}} e^{i(\phi_{\gamma\mathbb{P}} - \phi_{\mathbb{P}\gamma})}
    \label{eq:e2i2_4terms}
\end{align}
Keeping the phase terms allows the total cross section to be written concisely. The additional cross terms mix the amplitudes between the two configurations (paths). In the standard double-slit experiment the phase term of the two amplitudes ($e^{i\phi_{\gamma\mathbb{P}}}$ and $e^{i\phi_{\mathbb{P}\gamma}}$ in this case) must be related by a fixed phase difference $\delta_\phi$, otherwise there is no observable expectation value from the cross terms. In the ``standard'' quantum interference of amplitudes from two-slits, this requirement is trivially satisfied when a single quantum object (like a photon or an electron) takes part in two indistinguishable processes (or paths). This is generally what has been assumed in the literature for the photonuclear process in the past. However, in UPC photoproduction, the photon emitted by nucleus A which animates one amplitude and the photon emitted by nucleus B which animates the other amplitude are separate, distinguishable objects (due to their opposite %and observable spin 
directions). It is only after interacting with the target nucleus that the two channels become indistinguishable. 

We can express the full case involving the ``double-slit'' effect resulting from the two indistinguishable processes at a distance ($\rho^0$ decay at nucleus $A$ and $B$) and the decay into a pair of distinguishable daughters, $\rho^0\rightarrow \pi^+\pi^-$
In our case we can think of the $\pi^+$ being measured at detector 1 and $\pi^-$ being measured at detector 2, and vice versa. The full process including the decay to entangled daughters can be expressed as:
\begin{align}
    \langle \gamma\mathbb{P} | \pi^+ \pi^- \rangle &= \langle \gamma^A\mathbb{P}^B | \rho^B \rangle \langle \rho^B | \pi^+ \pi^-, B \rangle \\ 
    & \times \langle \pi^+ \pi^-, B | \left( |\pi^+, 1 \rangle | \pi^-, 2 \rangle + |\pi^+, 2 \rangle | \pi^-, 1 \rangle \right) \\
    &+ \langle \gamma^B\mathbb{P}^A | \rho^A \rangle \langle \rho^A | \pi^+ \pi^-, A \rangle \\ 
    & \times \langle \pi^+ \pi^-, A | \left( |\pi^+, 1 \rangle | \pi^-, 2 \rangle + |\pi^+, 2 \rangle | \pi^-, 1 \rangle \right)
\end{align}
where $\langle \gamma^A\mathbb{P}^B | \rho^B \rangle$ represents the transition from $\gamma^{A,B} + \mathbb{P}^{A,B}\rightarrow \rho^0$ ($A_{\gamma\mathbb{P}}$ and $A_{\mathbb{P}\gamma}$, respectively) and $\langle \rho^{A,B} | \pi^+ \pi^-, (A,B) \rangle$ represents the $\rho^0 \rightarrow \pi^+\pi^-$ decay. Finally, the $\langle \pi^+ \pi^-, (A,B) | \left( |\pi^+, 1 \rangle | \pi^-, 2 \rangle + |\pi^+, 2 \rangle | \pi^-, 1 \rangle \right)$ represents the probability for the detection of a $\pi^+$ and $\pi^-$ at detectors 1 and 2. Another important note is that the above notation expresses the $\rho$ wavefunction as transitioning to a $\pi^+$ and a $\pi^-$ such that the $\pi^+$ and $\pi^-$ cannot be written as a simple product of individual wavefunctions, {\it i.e} they are entangled. 

Furthermore, unlike the traditional double-slit experiment the experimental observation in UPCs is always of {\it two} particles. Indeed, in certain ways, the situation in photonuclear interaction is more similar to intensity interferometry observed in two photon measurements, i.e. the Hanbury Brown and Twiss effect (HBT)~\cite{brown_correlation_1956,brown1956twiss,aspect2020hanburry} than the double slit interference effect; see Ref.~\cite{Baym:1997ce} for a review of HBT and intensity interferometry in nuclear collisions.  
Intensity interferometry differs from the standard double-slit experiment in at least two important aspects: 1) intensity interferometry involves two simultaneous measurements instead of only a single measurement, and 2) intensity interferometry can result from incoherent sources due to higher-order coherence~\cite{PhysRevA.37.1607}. Crucially, even in intensity interferometry the interfering states must be indistinguishable or no interference pattern is observed. Therefore, the STAR observation in diffractive photonuclear production involves another complication. The observed spin interference pattern is found in the simultaneous measurement of two particles, a but they are a $\pi^+$ and $\pi^-$ that are clearly distinguishable according to their charge. 

Through the above exercise we have expressed the same basic concept of the double slit interference effect in various notations. Only after introducing the transition (decay) of the vector meson to its daughter particles do we confront the essential problems with the ``double-slit'' analogy, {\it i.e.} that the measurement is of two particles instead of a single particle. However, the situation is also incompatible with the conventional formulation of intensity interferometry since the two particles measured in the final state are clearly distinguishable.

 Quickly after the first photonuclear measurements from nucleus-nucleus collisions, it  was suggested that the observation of interference in photonuclear production demonstrates that particle decay does not cause wavefunction collapse~\cite{Klein:2002gc}. This concept motivated a formal description illustrating the necessary conditions for interference, especially when involving final states of distinguishable particles. The authors of \cite{Duan:2024} claim that the $E^2I^2$ framework provides a unified formalism\footnote{Importantly the framework is model independent in the sense that it uses only fundamental tenets of quantum mechanics to describe the phenomenon} for understanding the analogies between photonuclear production and the double-slit and/or intensity interferometry cases.

Following the $E^2I^2$ formalism\cite{COTLER2021168346}, we illustrate here the origin and persistence of the interference terms, even with distinguishable particles. Specifying the result in terms of the measurement of the final state $\pi^+$ and $\pi^-$ gives:
 \begin{equation}
     \begin{split}
         \ket{\pi^+} = \piket{+}{A} + \piket{+}{B}    \\
         \ket{\pi^-} = \piket{-}{A} + \piket{-}{B},
     \end{split}
 \end{equation}
where again the $A,B$ indicates the nucleus at which the $\rho^0$ is produced and $f_{A,B\pm}$ represent the (complex) elements related to the photon form factor, Pomeron form factor, and $\rho$ decay matrix elements for each case.

 We can express the amplitude for a correlation of the $\pi^+$ and the $\pi^-$ (at the wavefunction level) as:
 \begin{equation}
     \begin{split}
         \ket{\pi^+}\ket{\pi^-} & = \\
         & \left( \piket{+}{A} + \piket{+}{B} \right) \left( \piket{-}{A} + \piket{-}{B} \right) = \\
         & \piket{+}{A}\piket{-}{A} + \piket{+}{A}\piket{-}{B} + \\
         & \piket{+}{B}\piket{-}{A} + \piket{+}{B}\piket{-}{B}
     \end{split}
 \end{equation}
 and the simultaneous \textit{observation} of the $\pi^+$ and $\pi^-$ as:
 \begin{equation}
     \begin{split}
         \bra{\pi^+}\braket{\pi^- | \pi^-} \ket{\pi^+} & = \\
         & \left( \pibra{+}{A}\pibra{-}{A} + \pibra{+}{A}\pibra{-}{B} \right. + \\
         & \left. \pibra{+}{B}\pibra{-}{A} + \pibra{+}{B}\pibra{-}{B} \right) \\
         & \left( \piket{+}{A}\piket{-}{A} + \piket{+}{A}\piket{-}{B} \right. + \\
         & \left. \piket{+}{B}\piket{-}{A} + \piket{+}{B}\piket{-}{B} \right)
     \end{split}
 \end{equation}
 When expanded it yields the following {\it 16} terms (compared to the 4 terms in previous expressions, e.g. Eq.~\ref{eq:e2i2_4terms}) :
\begin{equation}
\begin{split}
         \bra{\pi^+}\braket{\pi^- | \pi^-} \ket{\pi^+} & = \\
         & \bk{+}{A} \bk{-}{A} +\\
         & \bk{+}{A} \sbk{-}{A}{B} +\\
         & \sbk{+}{A}{B} \bk{-}{A} +\\
         & \sbk{+}{A}{B} \sbk{-}{A}{B} \\
         & + \\
         & \bk{+}{A} \sbk{-}{B}{A} +\\
         & \bk{+}{A} \bk{-}{B} + \\
         & \sbk{+}{A}{B} \sbk{-}{B}{A} +\\
         & \sbk{+}{A}{B} \bk{-}{B} \\
         & + \\
         & \sbk{+}{B}{A} \bk{-}{A} + \\
         & \sbk{+}{B}{A} \sbk{-}{A}{B} + \\
         & \bk{+}{B} \bk{-}{A} + \\
         & \bk{+}{B} \sbk{-}{A}{B} \\
         & + \\
         & \sbk{+}{B}{A} \sbk{-}{B}{A} + \\
         & \sbk{+}{B}{A} \bk{-}{B} + \\
         & \bk{+}{B} \sbk{-}{B}{A} + \\
         & \bk{+}{B} \bk{-}{B}  \\
        \label{eq:e2i2a}
\end{split}
\end{equation} 
In the above, the phase terms have been retained to represent the most general case. Now, instead of assuming that the phases $\sphase{+}{A}{}$ and $\sphase{+}{B}{}$ are unrelated, we can consider the case where the $\pi^+$ and $\pi^-$ daughters of a $\rho^0$ decay are entangled, and include the geometry of the UPC configuration. Therefore:
 \begin{equation}
     \begin{split}
         \psi^{+}_a & = \phi_a + k^{+} r_a^+ \\
         \psi^{-}_a & = \phi_a + k^{-} r_a^- \\
     \end{split}
 \end{equation}
where $\phi_A$ is the initial phase of $\rho^0$ produced at nucleus $A$ and $k^{\pm}$ , $r_a^{\pm}$ are momentum and position of $\pi^{\pm}$ in their corresponding wavefunction while ignoring $\rho^0$ internal structure and spin; 
and similarly, from source $B$ we have entangled daughters:
 \begin{equation}
     \begin{split}
         \psi^{+}_b & = \phi_b + k^{+} r_b^+ \\
         \psi^{-}_b & = \phi_b + k^{-} r_b^- \\
     \end{split}
 \end{equation}

With these phases related by the entanglement of the daughters, we can express the observation (e.g. the real part) as:
\begin{equation}
     \begin{split}
         [\bra{\pi^+}\braket{\pi^- | \pi^-} \ket{\pi^+}] & = \\
%         & [\braket{\pi^+ | \pi^+}][\braket{\pi^- | \pi^-}] + \\
         & \bk{+}{A} \bk{-}{A} +\\
         & \bk{+}{B} \bk{-}{B} +\\
         & \bk{+}{A} \bk{-}{B} + \\
         & \bk{+}{B} \bk{-}{A} + \\
         & \mathbf{Re}~
         f_{A} f_{B}^{*} (|f_{A}|^2+ |f_{B}|^2)  \braket{ \pi^{+}_{A} | \pi^{+}_{B} } \braket{ \pi^{-}_{B} | \pi^{-}_{A} } [\cos{(k^+\cdot{B})}+\cos{(k^-\cdot{B})}]+\\
         & 2 \mathbf{Re}~ 
         f_{A-} f_{B+} f_{B-}^{*} f_{A+}^{*}  \braket{ \pi^{+}_{A} | \pi^{+}_{B} } \braket{ \pi^{-}_{B} | \pi^{-}_{A} } \cos{(P_{\perp}\cdot{B})}
         \label{eq:upcavg4}
     \end{split}
 \end{equation}
which results in four additional terms absent from uncorrelated emission of $\pi^{\pm}$ from sources. Note that this result persists even if the two original sources are not necessarily indistinguishable (e.g. a fixed phase between $\phi_a$ and $\phi_b$).
The last cross term is analogous to the $E^2I^2$ result of Eq.4 in Cotler-Wilczek~\cite{COTLER2021168346} formalism with a correlation measurement of two distinguishable particles from two different sources. In previous sections, the analog to the double-slit experiment was performed as if the $\rho^0$ did not decay. This is, of course, not the case. The experimental detection is of the $\pi^{\pm}$ pair as illustrated in this section. However, how this illustration could be quantitatively related to the experimental measurements in  Sec.\ref{sec:tomography} and the formula discussed in Sec.~\ref{sub:chater4:second} with a realistic decay $\rho^0\rightarrow\pi^+\pi^-$ of $J^{PC}=1^{--}$  requires further numerical investigations~\cite{Duan:2024}. 

Even without quantitative predictions for comparisons to data, the $E^2I^2$ formalism offers several key insights. Firstly, it provides a formal description of the essential insight from Ref.~\cite{Klein:2002gc}, i.e. that particle decay does not cause wavefunction collapse - by describing the daughter particles as an entangled state. In addition to this, the $E^2I^2$ formalism describes how interference (of any kind) could persist in measurements of distinguishable particles.

Another key takeaway from the framework is that the interference effects observed in the exclusive photoproduction of $\rho^0-$mesons are not merely a consequence of the geometry of the collision but also encapsulate rich information about the nonperturbative dynamics of QCD. By analyzing the phase information carried by entangled states, the framework allows one to extract detailed information about the spatial distribution and spin structure of the underlying color singlet (Pomeron) configurations within nuclei. While one may integrate elements from the Color Glass Condensate effective field theory (or other models) with nonperturbative techniques to compute the production and decay amplitudes, the characteristic interference effect results from fundamental aspects of quantum mechanics independent of models.  

Crucially, the derivation is not specific to the $\rho^0-$meson or the $\pi^+\pi^-$ final state and should therefore apply equally to other production processes (e.g. $\phi$, $J/\psi$, etc.) and decay channels. Since the exact quantitative and kinematic dependence of the azimuthal asymmetry depends on nonperturbative physics, comparison of measurements which access various different nonperturbative matrix elements will be useful for validating models.  Specifically, the interference patterns observed in various processes offers a model-independent means to extract the effective Pomeron flux and its spatial variation within the nucleus. Furthermore, the formalism provides sensitivity to the spin structure of the Pomeron couplings, suggesting that such interferometric techniques can reveal subtle aspects of the strong force that have so far been challenging to isolate in high-energy experiments. Again, due to the generality of the effect, it is expected to be present in processes involving Odderon exchange as well. Since the order of the azimuthal asymmetries induced by the $E^2I^2$ effect depend on the spin quantum numbers in the interaction, such measurements involving a $C$-odd exchange (e.g. in the production of pseudoscalar states such as the $\eta,\ \eta^\prime,\ \eta_c$ etc.) would provide a clear signature of this long predicted partner to the C-even Pomeron. 

Looking ahead, future measurements could further test and refine the $E^2I^2$ framework. Experiments at current facilities like RHIC and the LHC should aim to gather high-statistics data on exclusive vector meson decays, with particular emphasis on measuring the azimuthal modulations in different kinematic regimes. Additionally, exploring other decay channels and heavier quarkonium states could help extend the applicability of the technique and test its predictions in various perturbative and nonperturbative domains. Finally, the upcoming Electron-Ion Collider offers a promising environment to apply $E^2I^2$ in deeply inelastic scattering experiments\footnote{While $e+p/A$ collisions are not expected to result in a ``double-slit'' configuration, $E^2I^2$-like interference may still take place in $e-A$ collisions if the large nuclear target acts as an incoherent source. Analogously, the astronomical HBT technique employs intensity interferometry utilizing photons from a single star (incoherent source). }, potentially allowing for the isolation of rare Odderon contributions and a more precise mapping of the spatial gluon distributions in nuclei~\cite{Duan:2024}.

        \subsection{The observer effect for photoproduction in HICs  }

        \begin{figure}[htbp]
    \centering
    \includegraphics[width=0.4\textwidth]{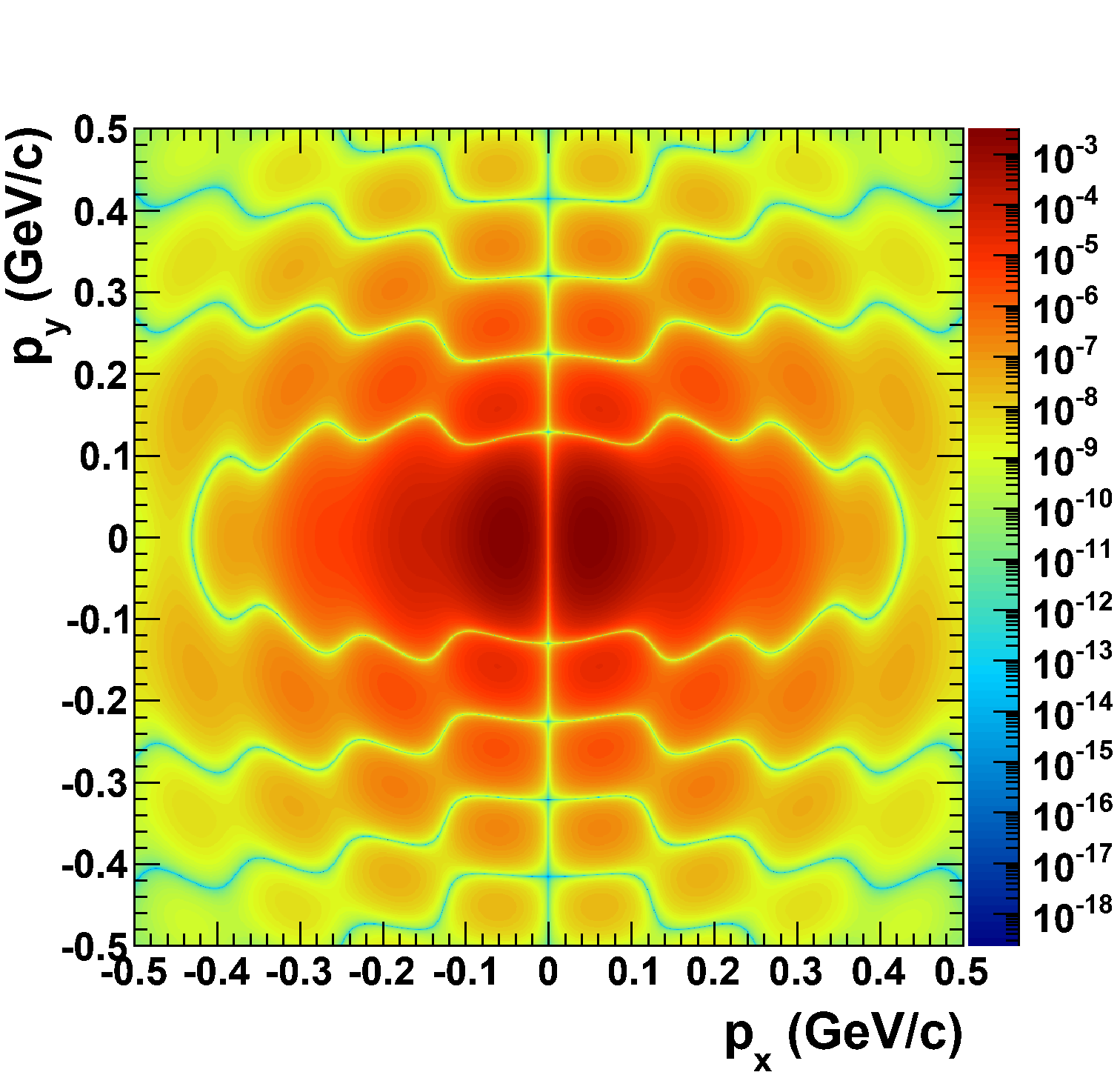}
    \includegraphics[width=0.4\textwidth]{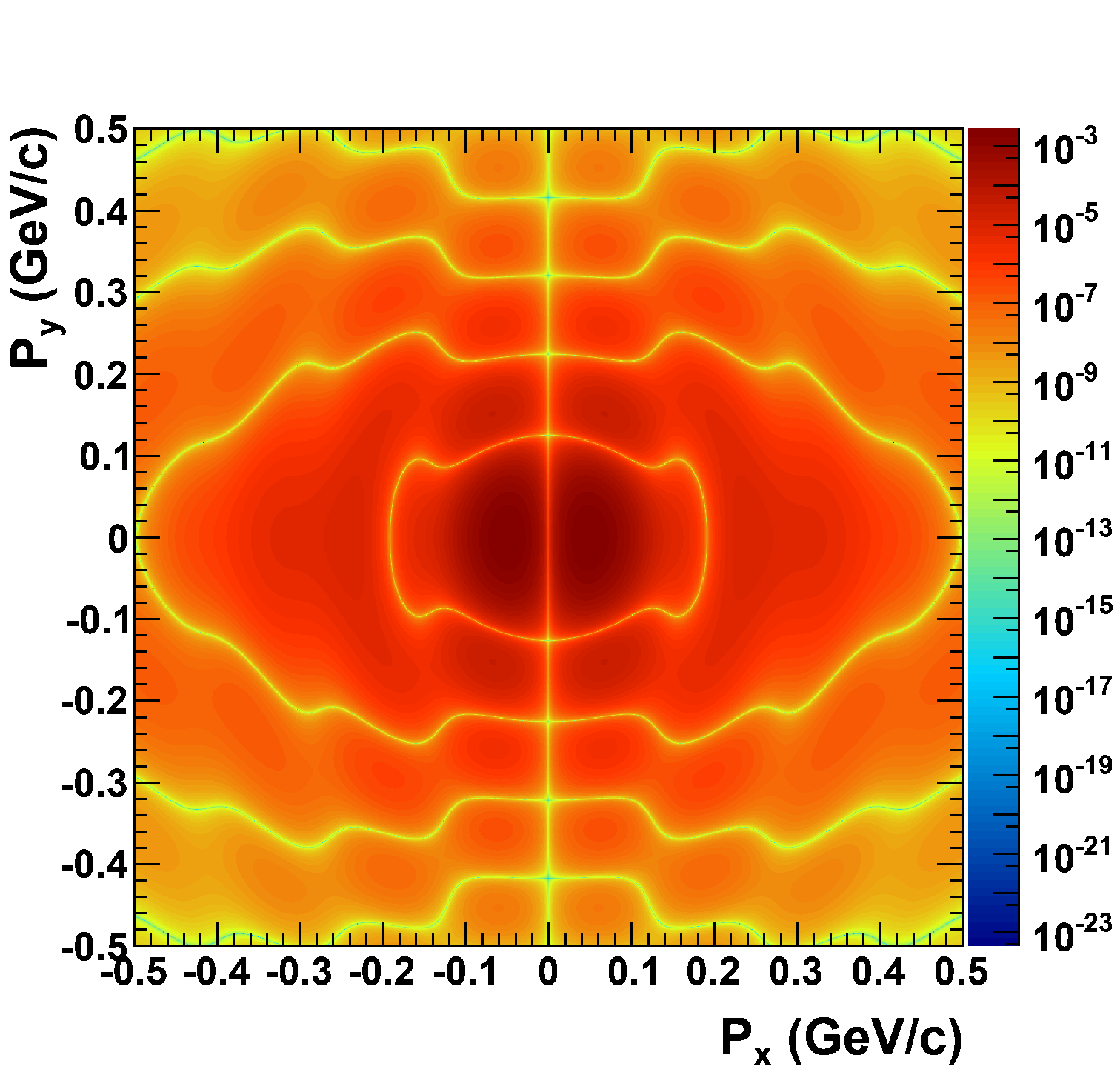}
    \caption{2D momentum distribution patterns of coherent \(J/\psi\) photoproduction from VMD calculations at mid-rapidity (\(y=0\)) in Au+Au collisions at \(\sqrt{s_{\rm{NN}}} = 200\) GeV without (left panel) and with (right panel) the observation effect for a fixed impact parameter (\(b=10\) fm). The nuclei are positioned at $x = \pm \ 5$ fm in the transverse plane. From Ref.~\cite{Zha:2018jin}.
}
    \label{fig:observation:effect:momentum}
\end{figure}

The which-way double-slit experiment serves as a quintessential demonstration of the perplexing nature of quantum mechanics, showcasing the phenomenon of wave-particle duality and the collapse of the wave function upon measurement. The crux lies in the notion that merely knowing which path a particle takes collapses its wave function, thus eliminating the interference pattern characteristic of wave-like behavior. In this thought experiment, detectors placed to determine through which slit a particle passes cause the wave function of the particle to collapse. This measurement transforms the particle's behavior from a wave-like interference pattern to a particle-like distribution, with distinct clusters corresponding to each slit. This "observation effect" underscores the fundamental principle that the act of measurement alters the quantum system, leading to the loss of coherence. Drawing from this foundational concept, the intricate dynamics of hadronic heavy-ion collisions reveal a parallel narrative. In these collisions, the intense strong interactions and the potential presence of quark-gluon plasma or hadronic matter within the overlap region should act analogously to the detectors in the double-slit experiment. These interactions effectively "observe" the paths taken by coherent J/$\psi$ particles. As a result, the strong interactions cause the wave function of these particles to collapse, leading to decoherence and impeding coherent photoproduction. Despite the expected decoherence, observations by the ALICE~\cite{ALICE:2015mzu,ALICE:2022zso,ALICE:2024whv} and STAR~\cite{STAR:2019yox,Li:2022ncj} collaborations have unveiled a significant surplus in J/$\psi$ production, particularly at very low transverse momenta. This anomaly finds a plausible qualitative explanation in the framework of coherent photoproduction mechanisms. Moreover, the STAR collaboration~\cite{STAR:2019yox,Li:2022ncj} has even documented evidence suggestive of interference akin to the double-slit experiment, in the distribution of four momentum transfer. The detection of coherent photoproduction and the emergence of interference phenomena in hadronic heavy-ion collisions compel a more nuanced examination of the observational effects within the overlap region. 

Although strong interactions are inherently short-range, sensitive only to the region where the photoproduction takes place,
the 'which slit' problem is only partially resolved. For the sake of simplicity, 
it is reasonable to assume that 'which-way' information comes primarily from the initial overlap region. Nevertheless, questions linger regarding whether the hadronic interactions within the overlap region impact photon emission. Considering the retarded (delayed) time at which the photon is created \cite{jackson_classical_1999}, it is likely that photon emission occurs prior to the onset of hadronic collisions, thereby remaining largely unaffected by them. However, photoproduction and hadroproduction occur on similar time scales, so it seems necessary to consider both possible time orderings in calculating the cross-section.   If the target undergoes a hadronic interaction first, the photon-nucleon center of mass energy will be reduced. 

\begin{figure}[htbp]
    \centering
    \includegraphics[width=0.9\textwidth]{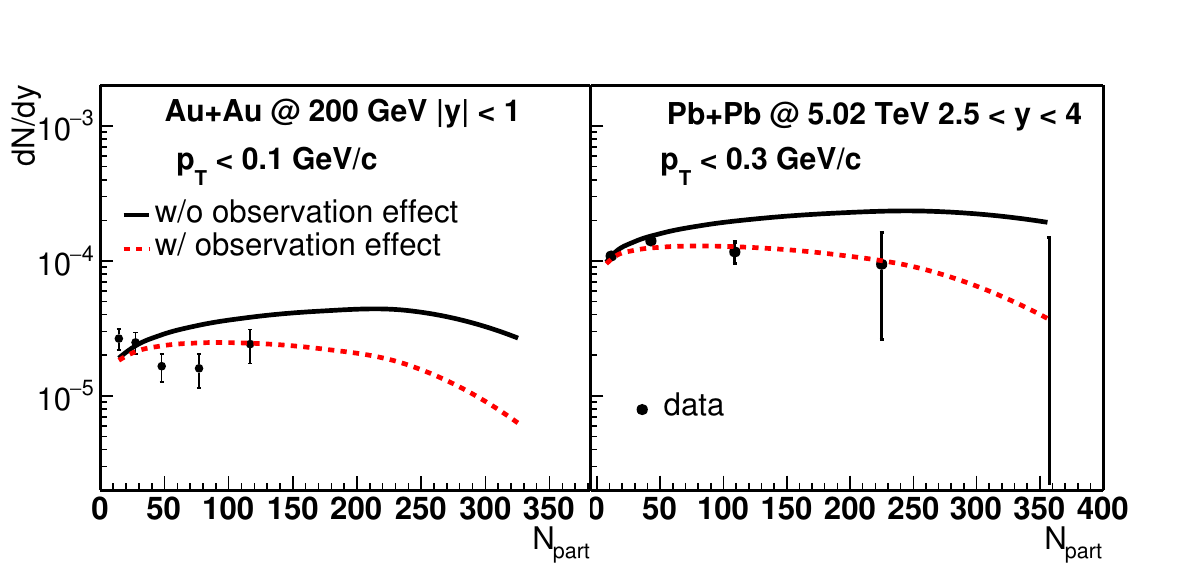}
    \caption{Theoretical calculations~\cite{Zha:2018jin} of coherent \(J/\psi\) production yield in the VMD framework with and without “observation” effect as a function of \(N_{\text{part}}\) in (left panel) Au+Au collisions at \(\sqrt{s_{\rm{NN}}} = 200\) GeV and (right panel) Pb+Pb collisions at \(\sqrt{s_{\rm{NN}}} = 5.02\) TeV. The experimental measurements~\cite{STAR:2019yox,ALICE:2022zso} are shown for comparison.
}
    \label{fig:observation:effect}
\end{figure}

Similarly, concerns about the effects of fragmentation of nuclei remnants (spectators) should be negligible, given the significantly longer timescales associated with fragmentation compared to vector meson photoproduction. Given these assumptions, the depiction in Fig. \ref{fig:observation:effect:momentum} demonstrates the alterations in momentum patterns resulting from the partial obstruction of slits in Au+Au collisions at $\sqrt{s_{\rm{NN}}} =$ 200 GeV for $J/\psi$ photoproduction. Observing these modifications in transverse momentum distributions requires high-precision data and a reliable strategy to exclude incoherent contribution. Fortunately, the process of decoherence within the overlap region reduces the cross-section, thereby alleviating stringent statistical requirements. Figure \ref{fig:observation:effect} provides insight into the integrated yields of coherent J/$\psi$ production as a function of $N_{\rm{part}}$, comparing scenarios with whole slits versus those with partially blocked slits ('observation' effect) for Au+Au and Pb+Pb collisions at different centralities. Notably, the 'observation' effect significantly diminishes the integrated production rate, particularly in semi-central and central collisions. While experimental data from ALICE and STAR prefer a scenario incorporating the 'observation' effect, the limited statistical power prevents a definitive distinction between the two scenarios.

 	\section{Related topics and future opportunities}\label{fifth}
        \subsection{The application to nuclear tomography}
    
        \hspace{2em}The Good-Walker paradigm, discussed in Section \ref{sec:GoodWalker}, relates $d\sigma_{\rm coherent}/dt$ to the transverse size of the target \cite{Klein:2019qfb}.  The two are related by a Fourier-Bessel (Hankel) transform:
\begin{equation}
F(b) \propto \frac{1}{2\pi}\int_0^{\infty} p_T dp_T J_o(bp_T) \sqrt{\frac{d\sigma}{dt}}
\end{equation}
where $J_0$ is a Bessel function.  The square root converts the cross-section to amplitude.  Because there is a sign ambiguity in the square root, it is necessary to flip the sign of the integral when integrating past each diffractive minimum, to account for the phase shift.  There are several other complications to this conversion \cite{Klein:2021mgd}.  The relationship is exact when the $p_T$ integration goes to infinity, but real data is cut off at  finite $p_T$, with the limit coming at the $p_T$ where the incoherent contribution to the cross-section is large enough to preclude measurement of the coherent contribution.   Any cutoff introduces windowing artifacts.  The collaboration explored these factors by varying the maximum $p_T$ used in the integral~\cite{Klein:2021mgd}. 

There are also two other contributors to the measured $p_T$, due to the experimental resolution, and the photon $p_T$.  These factors can in principle be removed via deconvolution \cite{ALICE:2021tyx}.  However, the photon $p_T$ spectrum is complex, since $b$ and $p_T$ are conjugate variables.  The $p_T$ spectrum is known if collisions are integrated over all $b$ \cite{Vidovic:1992ik}.  However, when constraints are put on $b$ (including $b>2R_A$) or the $b-$spectrum  is modified due to the exchange of additional photons, no simple solution is possible \cite{Klein:2020jom}.

\begin{figure}[htbp]
    \centering
    \includegraphics[width=0.75\textwidth]{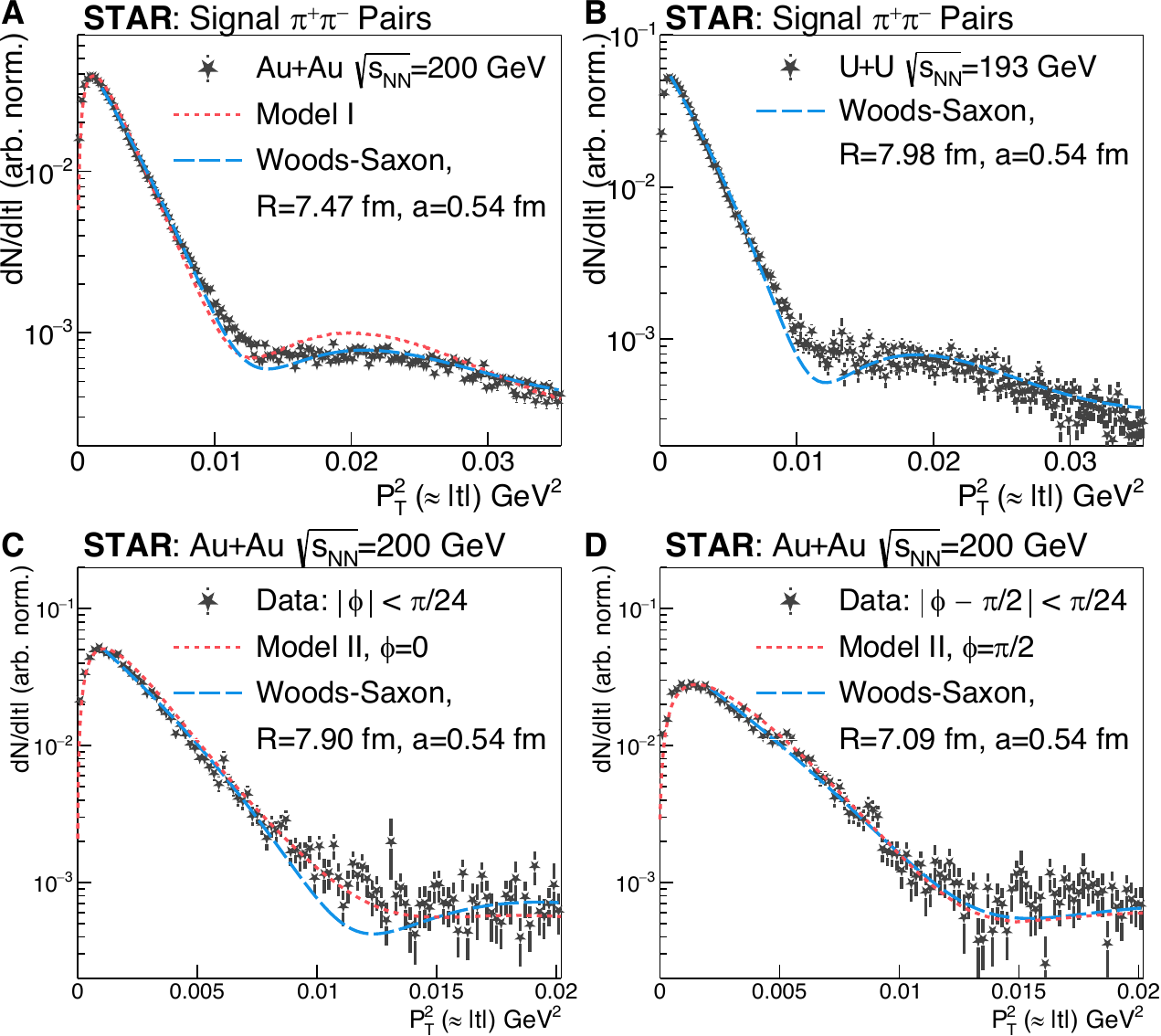}
    \caption{The distributions \( \frac{dN}{d|t|} \) as a function of \( |t| \) (approximately \( P_T^2 \)) for Au+Au collisions (A) and U+U collisions (B). Additionally, the \( \frac{dN}{d|t|} \) as a function of \( |t| \) for Au+Au collisions is displayed for $\phi$ bins at 0° (C) and 90° (D). In panels (A-D), the long-dashed curves represent the best-fit Woods-Saxon distributions. For the Au+Au collision data (A, C, D), red short-dashed curves depict results from two different models~\cite{Xing:2020hwh,Brandenburg:2022jgr,Zha:2020cst}. Vertical error bars denote the statistical uncertainties. From Ref.~\cite{STAR:2022wfe}.
}
    \label{fig:rho:spin-interference:figure3}
\end{figure}

\begin{figure}[htbp]
    \centering
    \includegraphics[width=0.95\textwidth]{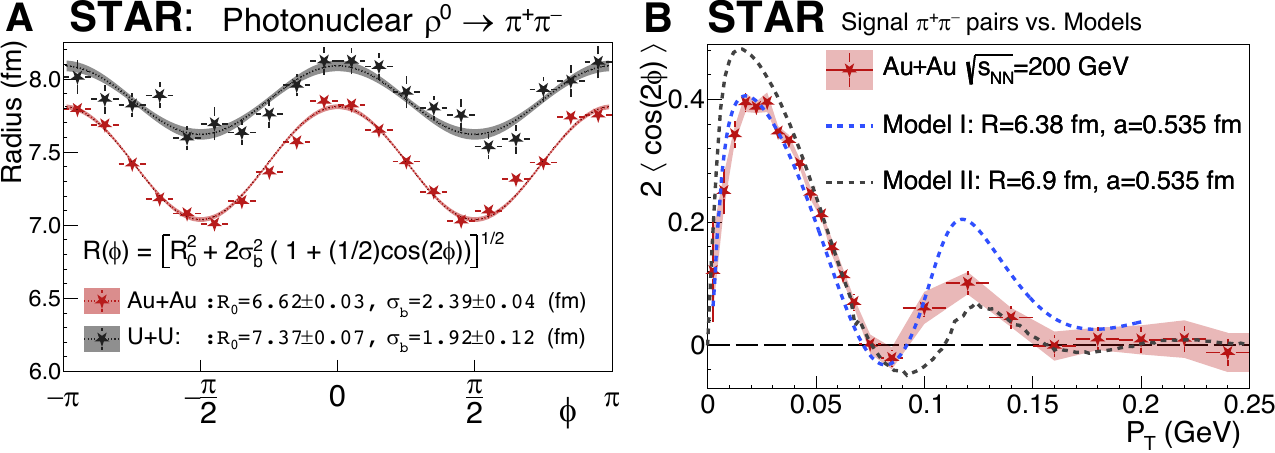}
    \caption{(Left panel) The radial parameter as a function of the $\phi$ angle for Au+Au and U+U collisions, along with a fit using an empirical second-order modulation. (Right panel) A comparison of the fully corrected Au+Au distribution with theoretical models~\cite{} that incorporate the effects of the photon's linear polarization and interference from two sources. From Ref.~\cite{STAR:2022wfe}.
}
    \label{fig:rho:spin-interference:figure4}
\end{figure}

An alternate approach to Fourier transforming $d\sigma/dt$ is to fit the measured $d\sigma/dt$.  The fit function can account for the experimental resolution and photon $p_T$.  The ALICE Collaboration measured the $|t|$ spectrum of coherent $J/\psi$ photoproduction, and found that the data was consistent with expectations from shadowing, but not with an unmodified lead nucleus \cite{ALICE:2021tyx}.  The collaboration has also applied this approach to incoherent $J/\psi$ photoproduction, finding a $d\sigma/dt$ spectrum that was better fit by models that included event-by-event fluctuations \cite{ALICE:2023gcs}, as is shown in Fig.  \ref{fig:incohJpsi_PbPb}. 

In the $\rho$ spin-interference measurements~\cite{STAR:2022wfe}, STAR also conducted a comprehensive analysis of the $t$ distributions. Figure~\ref{fig:rho:spin-interference:figure3} shows the measured $P_T^2 \approx |t|$ distribution for $\pi^+\pi^-$ pairs from $\rho$ mesons in Au+Au (A) and U+U (B) collisions, revealing two distinct classical diffractive peaks. A notable difference between the shapes of the two spectra, particularly for $|t| < 0.01 \ (\mathrm{GeV})^2$, is observed due to the different sizes and shapes of the two nuclei. Model calculations utilizing the dipole and VMD frameworks, shown as Model I~\cite{Zha:2020cst} and II~\cite{Xing:2020hwh} in the plots, accurately describe the entire spectrum, including the low $|t|$ dip. The measured $|t|$ spectra can also be modeled empirically, with a coherent contribution characterized by a Woods-Saxon form factor and an incoherent contribution characterized by a dipole form factor. Precise nuclear tomography requires corrections for the photon's transverse momentum. This correction can be achieved by deconvolution if the photon's transverse momentum distribution is decoupled from the production process, which is not the case in reality. This issue can be resolved by leveraging the linear polarization of the photoproduction process in heavy-ion collisions. Since the photons are 100$\%$ linearly polarized along their transverse momentum direction, the $\rho^0$ momentum perpendicular to the polarization has minimal contribution from the photon's transverse momentum in the $|t|$ distribution, making it most indicative of the nuclear matter density distribution. Figure~\ref{fig:rho:spin-interference:figure3} further shows the $|t|$ distribution for $|\phi| < \pi/24$ (C) and $|\phi - \pi/2| < \pi/24$ (D), illustrating the significant impact of the interference effect on the $|t|$ distribution. The extracted radius for the case with maximum interference ($\phi \approx 0$) is larger by $0.81 \pm 0.05(\text{stat}) \pm 0.03(\text{syst})$ fm compared to the case with minimum interference ($\phi \approx \pi/2$).

To quantitatively assess the impact of photon momentum, the $|t|$ distribution has been analyzed in different $\phi$ bins. Figure~\ref{fig:rho:spin-interference:figure4} presents the radius $R$ extracted from an empirical fit to data as a function of $\phi$. A second-order modulation in $R$ as a function of $\phi$ is observed. Thus, the equation
%\[
\begin{equation}
R = \sqrt{R_0^2 + \sigma_b^2} \times (1 + \epsilon_p \cos 2\phi)
\end{equation}
%\]
is used to extract the minimum radius $R_0$ at $\phi = 90^\circ$ with a second-order angular modulation parameter $\sigma_b$, where the parameter $\epsilon_{p} = \langle \cos 2\phi \rangle = 1/2$ accounts for the resolution of the measurement when using the daughter momentum as a proxy for the polarization direction. The free parameter $\sigma_b$ quantifies the observed strength of the interference effect, considering contributions from factors such as the range of impact parameters probed, the finite photon transverse momentum, and the resolution of the polarization measurement. Using this approach, which provides the true diffractive radius from the nuclear form factor, the obtained values are $R_0 = 6.62 \pm 0.03$ fm, $a = 0.5 \pm 0.1$ fm, and $\sigma_b = 2.38 \pm 0.04$ fm for Au+Au collisions, and $R_0 = 7.37 \pm 0.06$ fm, $a = 0.5 \pm 0.1$ fm, and $\sigma_b = 1.9 \pm 0.1$ fm for U+U collisions. Due to the finite size of the nuclei, the polarization direction of photons is not perfectly aligned with the reaction plane. The depolarization can be estimated from the average angle between the photon and the impact parameter as $1 - P = (R_0/\langle b \rangle)^2/4 \approx 0.03$. Contributions also arise from the finite size of the $\rho^0$ wavefunction, characterized by a transverse radius $R_T^\rho = 1.03$ fm. This value was extracted using HERA data on diffractive $\rho^0$ photoproduction within the dipole framework~\cite{Forshaw:2010py}, which may be model-dependent; however, no alternative results are currently available for comparison.
 The resulting true nuclear radii are $R_A^2 = R_0^2 - (R_T^\rho)^2 - (1/P - 1)\sigma_b^2$, yielding $R_{\text{Au}} = 6.53 \pm 0.03 (\text{stat.}) \pm 0.05 (\text{syst.})$ fm for $^{197}\text{Au}$ and $R_{\text{U}} = 7.29 \pm 0.06 (\text{stat.}) \pm 0.05 (\text{syst.})$ fm for $^{238}\text{U}$. These radii are systematically larger than the nuclear charge radii obtained from low-energy electron scattering~\cite{Shou:2014eya}. When the strong-interaction radius is taken as a weighted average of neutral and nuclear charge radii, the measurements yield: $S = R_n - R_p = 0.17 \pm 0.03(\text{stat.}) \pm 0.08(\text{syst.})$ fm for $^{197}\text{Au}$, consistent with observations at low energies~\cite{Centelles:2008vu}, while the measurement neutron skin value of $0.44 \pm 0.05(\text{stat.}) \pm 0.08(\text{syst.})$ fm for $^{238}\text{U}$ appears to be higher than the expectations.

\begin{figure}[htbp]
    \centering
    \includegraphics[width=0.45\textwidth]{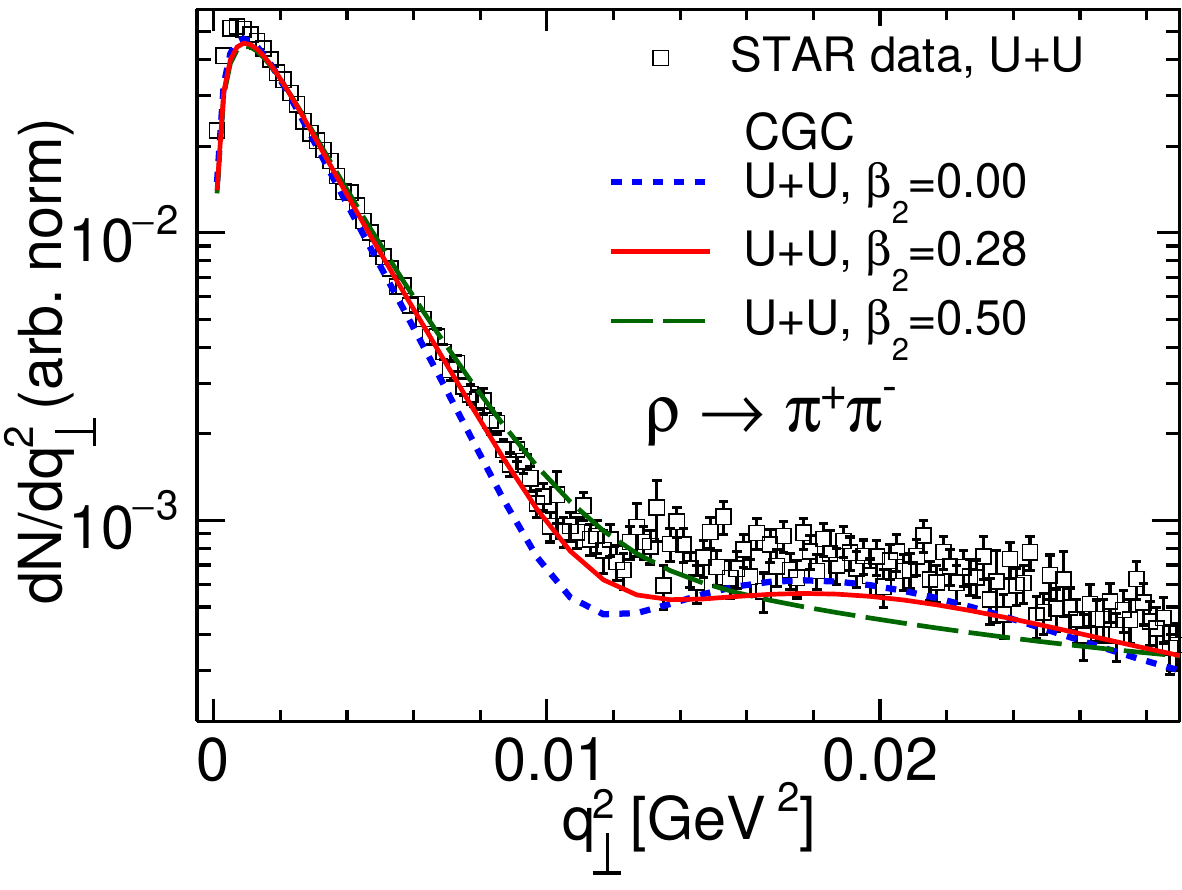}
    \includegraphics[width=0.45\textwidth]{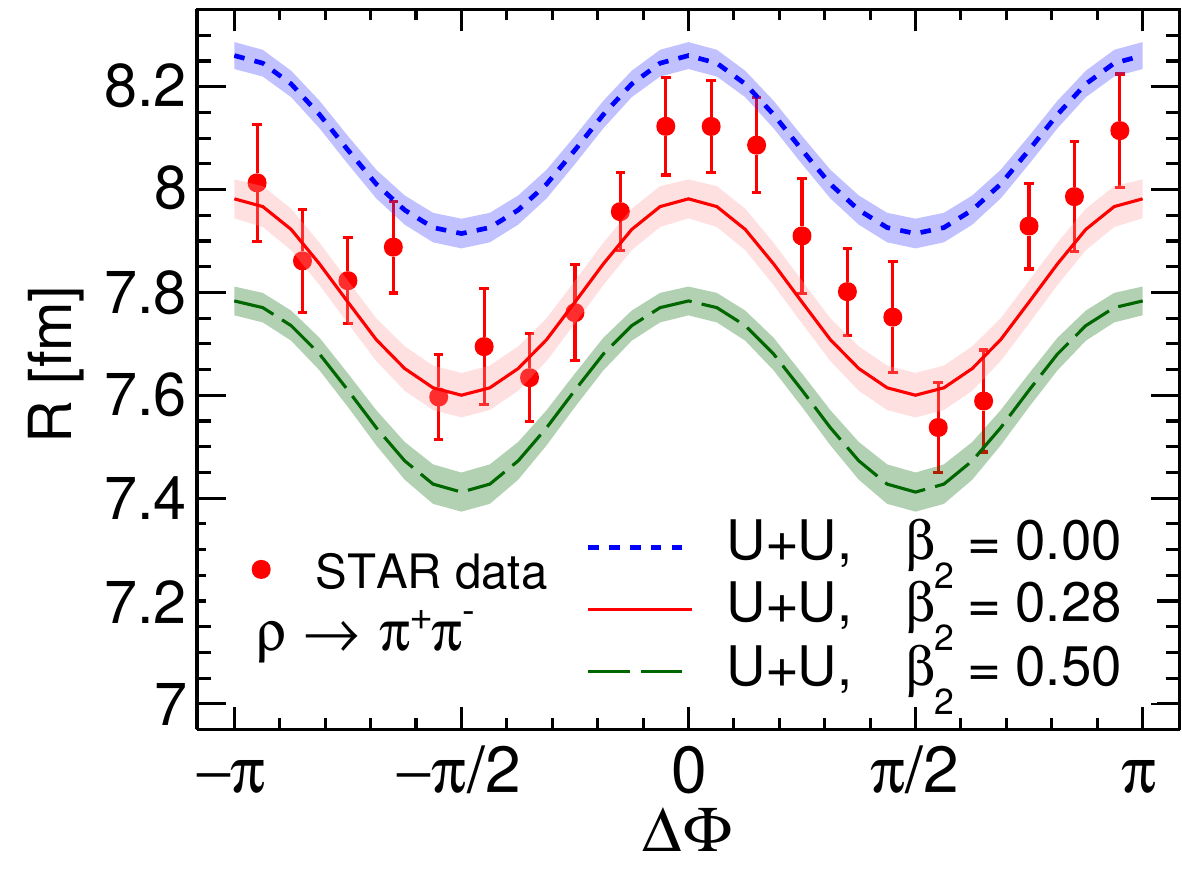}
    \caption{Left: The cross section as a function of $|t|$ ($q^2_{\perp}$) with theory curves of different value of nuclear deformation; right: Extracted apparent radius parameter as a function of $\Delta\Phi$ with heory curves of different value of nuclear deformation. Figures From Ref.~\cite{Mantysaari:2023prg}.
}
    \label{fig:deformed_UU}
\end{figure}

It is well known that $^{238}\text{U}$ is a highly deformed nucleus with a $\beta$ parameter as large as 0.3. To further investigate the effect of nuclear deformation on the measured $|t|$ distribution and interference, Ref.~\cite{Mantysaari:2023prg} performed a calculation with deformed nuclear Woods-Saxon distribution. Figure~\ref{fig:deformed_UU} shows the description of $|t|$ and $\Delta\Phi$ distributions with different degree of Uranium deformation and it appears that a deformation value of $\beta=0.28$ is preferred by the UPC measurement and is consistent with other measurements. Nevertheless, this exercise still requires a radius parameter which appears to be larger than other measurements~\cite{Mantysaari:2023prg}. The framework used in this result is also the dipole model, which introduces model dependence, including how the $\rho$ meson wave function is treated and how parton modifications in nuclei are handled, especially at low $x$. To ensure the reliability and consistency of the extracted nuclear structure parameters, future work should consider comparing results from different models and assessing the associated uncertainties. These newly developed experimental tools together with other new techniques recently developed from flow measurements in relativistic heavy-ion collisions provide a new approach to map the initial condition of heavy-ion collisions and nuclear structure across the nuclide chart~\cite{Achenbach:2023pba,Bally:2022vgo}.

        \label{sec:tomography}
        \subsection{The application to determine the reaction plane}

\hspace{2em}Understanding the properties of QGP requires precise information about the initial geometry of the collisions~\cite{Heinz:2013th}, including the reaction plane, which is defined by the impact parameter and the beam axis. Traditional methods for determining the reaction plane rely on the anisotropic flow of produced particles. However, these methods suffer from indirectness and potential biases due to non-flow correlations and event-by-event fluctuations. An alternative is to use photoproduction processes as a more direct probe of the reaction plane~\cite{Xiao:2020ddm,Wu:2022exl,Zhao:2022dac}, leveraging the linear polarization of photons produced in these collisions.

\begin{figure}[!htbp]
    \centering
    \includegraphics[width=0.5\textwidth]{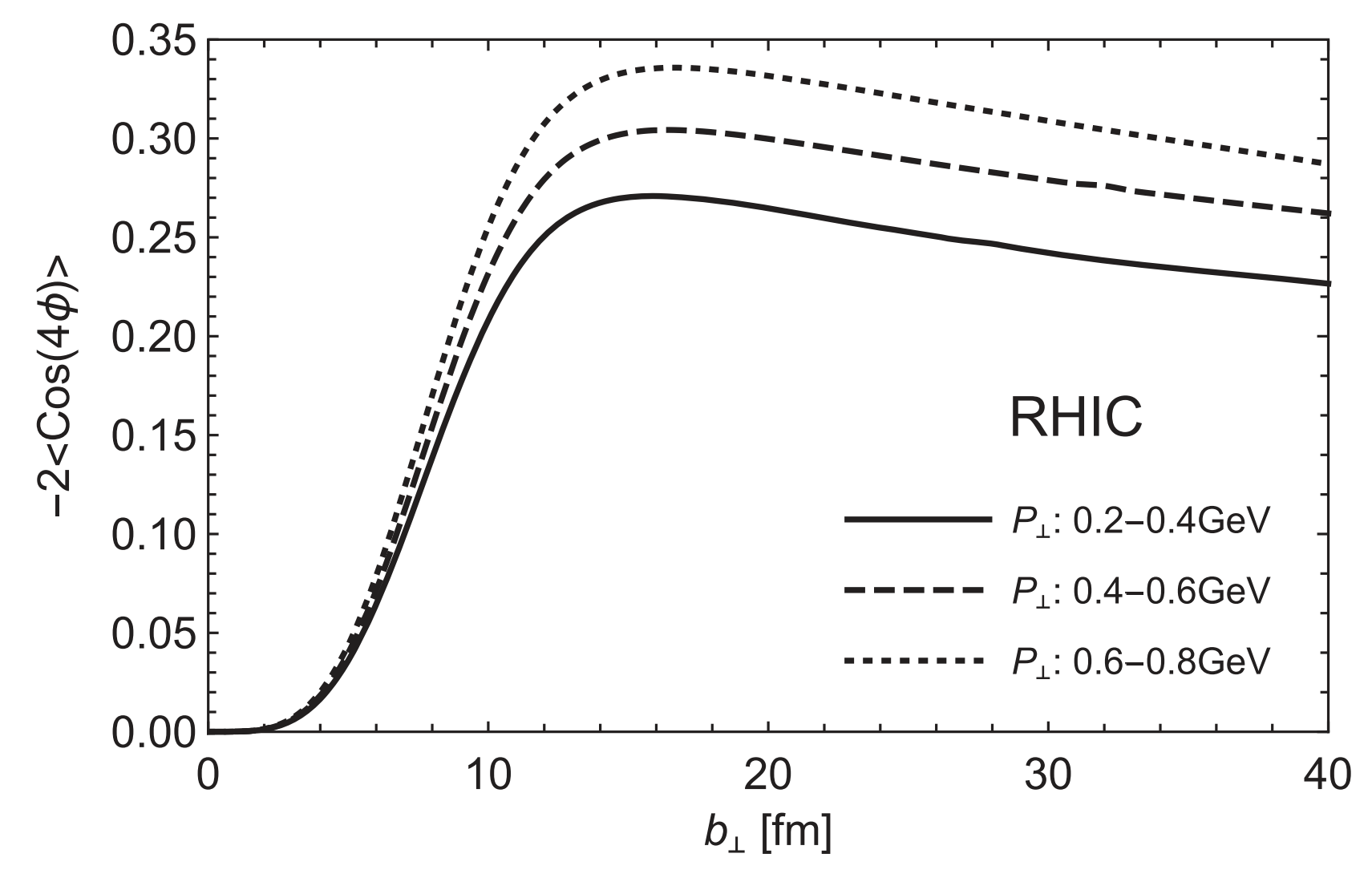} 
     \includegraphics[width=0.45\textwidth]{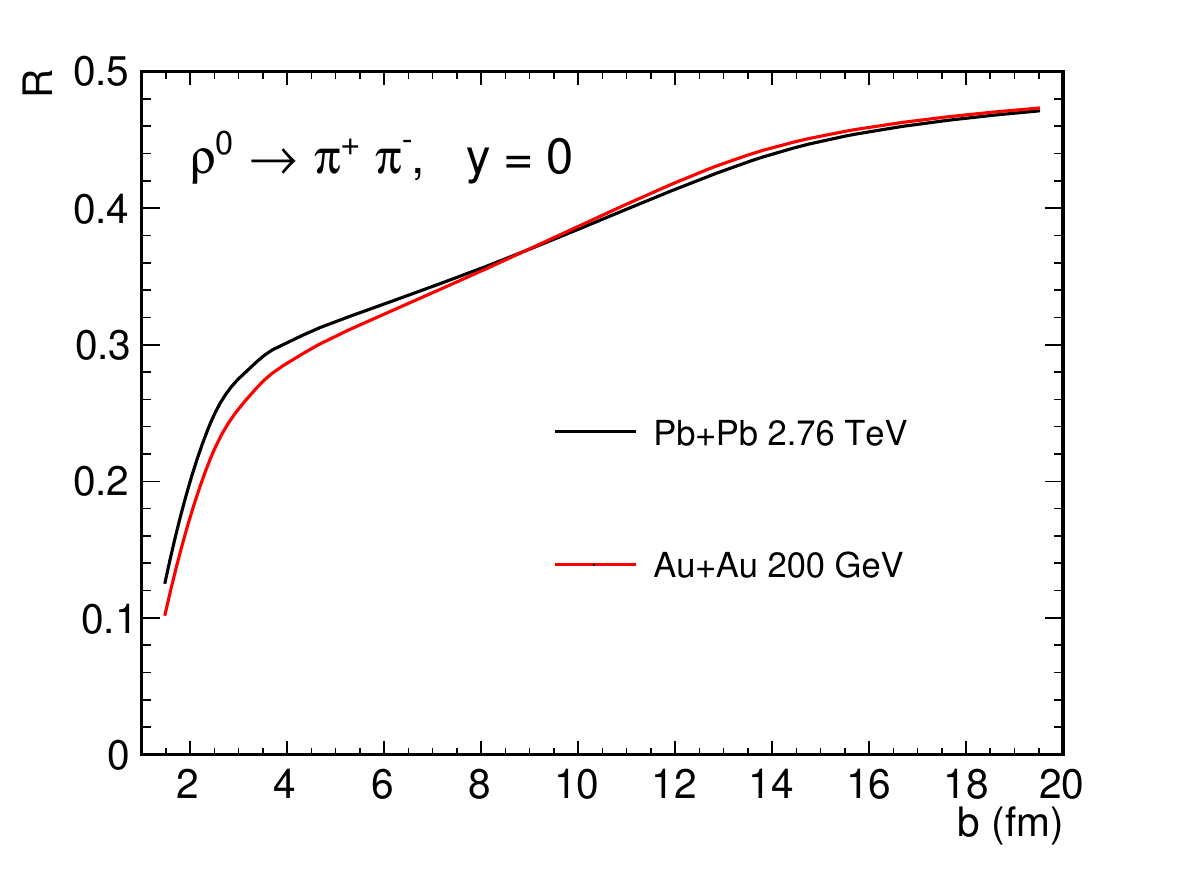} 
    \caption{(Left panel) Estimates of the $\cos4\phi$ asymmetry as the function of impact parameter ($b_{\perp}$) in Au+Au collisions at $\sqrt{s_{\rm{NN}}} =$ 200 GeV. The results are integrated for dilepton rapidities $[-1, 1]$, $P_{\perp}$ is defined as $P_{\perp} = (l_{1\perp}-l_{2\perp})/2$, where $l_{1\perp}$ and $l_{2\perp}$ are the transverse momenta for the final state two leptons. (Right panel) The estimated resolution of the reaction plane determination $R$ as a function of the impact parameter (b) for the coherent process of $\gamma + A \rightarrow \rho^0 + A \rightarrow \pi^{+} + \pi^{-} + A$ in Au+Au collisions at $\sqrt{s_{\rm{NN}}} =$ 200 GeV and Pb+Pb collisions at $\sqrt{s_{\rm{NN}}} =$ 2.76 TeV at midrapidity for $p_{T} < 0.1$ GeV/c. From Refs.~\cite{Xiao:2020ddm,Wu:2022exl}
    }
\label{fig:reaction-plane}
\end{figure}

In Ref.~\cite{Xiao:2020ddm}, Xiao et al. focus on the pure electromagnetic process of $\gamma + \gamma \rightarrow l^+ + l^-$ in heavy-ion collisions.They predict that the photon flux has a significant linear polarization along the impact parameter direction.  This will generate $\langle cos4\phi \rangle$ azimuthal asymmetries, where $\phi$ is the azimuthal angle between the lepton’s momentum and
the impact parameter $\vec{b}$. As shown in the left panel of  Fig.\ \ref{fig:reaction-plane}, the asymmetry increases with impact parameter until it reaches a maximal value when $|\vec{b}|$ is slightly larger than $R_A$. The strength of the asymmetry depends on lepton transverse momenta regions, impact parameter, collision energy, and collision species but remains sizable for peripheral collisions. The anisotropy of leptons can be measured through azimuthal angular correlations between the leptons and hadrons, similar to what has been done for hadron flow. Due to fluctuations, the fourth-order event plane may not be well aligned with the impact parameter, introducing considerable uncertainties in measuring $v_4$ for leptons in central collisions, while this issue becomes less severe for peripheral events.

Wu et al. ~\cite{Wu:2022exl} discuss how the coherent photoproduction of vector mesons, such as the $\rho^0$ meson, can be utilized to determine the reaction plane in heavy-ion collisions. The linear polarization of the photons leads to a preferential emission direction for the decay products of the vector mesons. Specifically, the pions resulting from the decay of the $\rho^0$ meson exhibit an angular distribution that is aligned with the reaction plane \cite{Baur:2003ar}. They derive that the resolution of reaction plane determination from this approach, defined as $R = \langle \cos(2\Delta \phi) \rangle$, where $\Delta \phi$ is the angle between the reaction plane and the decay plane of the vector meson, can reach an ideal limit of 0.5 in peripheral collisions. This value is a significant improvement over traditional methods, especially in peripheral and ultraperipheral collisions where the impact parameter is large, and the photon polarization is well-defined, as shown in the right panel of  Fig.\ \ref{fig:reaction-plane}.

The primary advantage of using photoproduction processes for reaction plane determination is the direct link to the initial collision geometry. Unlike traditional methods that rely on the anisotropy of final state particles, photoproduction processes are less affected by medium-induced effects and non-flow correlations. This direct probing reduces biases and increases the accuracy of the reaction plane determination. However, there are limitations to this approach. The significant background from hadronic interactions can reduce the effectiveness of photoproduction methods in central collisions. Additionally, the requirement for exclusive reactions may lead to lower statistics and potential biases in the impact parameter distribution. Despite these challenges, the unique advantages of photoproduction methods make them a valuable tool for studying the initial conditions of heavy-ion collisions.

Future research can focus on improving the statistical accuracy of photoproduction measurements and extending these methods to other photon induced products, such as dimuon and $J/\psi$. By analyzing the coherent photoproduction of different products, researchers can obtain detailed information about of the initial geometry and the properties of the QGP. Moreover, the comparison of anisotropy measurements between coherent photon products and hadrons can provide deeper insights into the different stages of heavy-ion collisions. The anisotropy measurements of coherent photon products offer a cleaner probe of the initial conditions of collisions, while hadrons reflect the collective behavior of the QGP. Studying both observables can help disentangle the effects of initial geometry and medium evolution.

        \label{sec:collectivity}
        \subsection{The Manipulation of "slits" in HICs}\label{sec6-3}

         \hspace{2em}In high-energy heavy-ion collisions, various conditions can be created to study the quantum phenomena at femtometer scale by manipulating the experimental parameters of collisions. This subsection focuses on the manipulation of "slits" in HICs. This includes controlling slit distance by selecting specific impact parameters, altering slit size and profile by colliding different nuclear species, and utilizing asymmetric slits in asymmetric heavy-ion collisions. By precisely adjusting these parameters, the complex quantum effects can be studied researchers in these intricate environments.
\begin{figure}[!htbp]
    \centering
    \includegraphics[width=0.5\textwidth]{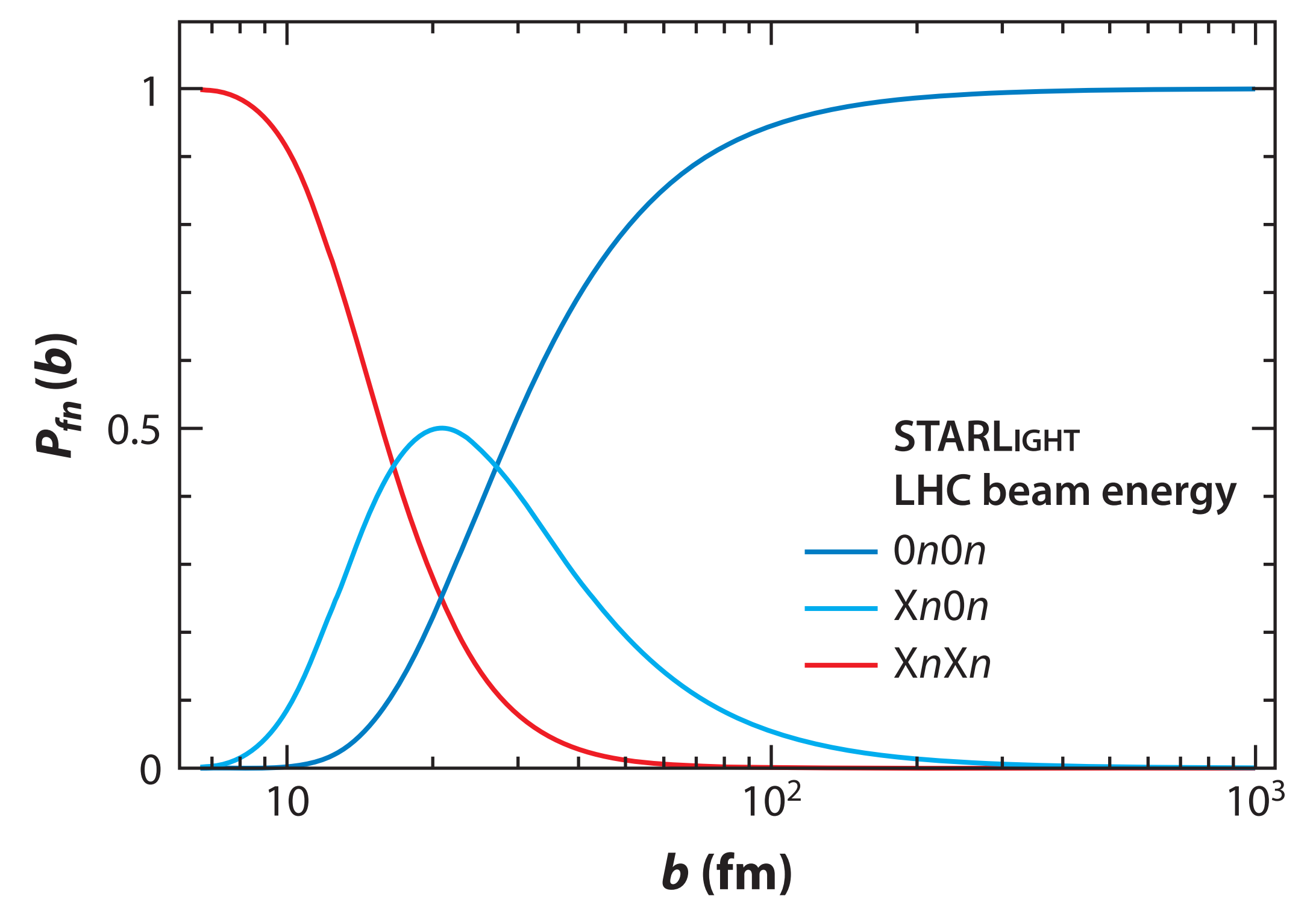} 
    \caption{Impact parameter dependence of the probabilities for the three primary forward neutron topologies, as calculated by STARLight for the lead-lead collisions at the LHC: 0$n$0$n$, indicating no neutron emission in either direction; X$n$0$n$, indicating neutron emission in only one direction; and X$n$X$n$, indicating neutron emission in both directions. From Ref.~\cite{Klein:2020fmr}.
    }
\label{fig:MCD:probability}
\end{figure}

In hadronic heavy-ion collisions, the separation of slits can be easily controlled by selecting the centrality. Centrality is determined by analyzing particle multiplicity, energy deposition, and spectator neutron counts, and calibrating these measurements with models like the Glauber model. However, the photoproduction rate in hadronic collisions is low, hindering precise measurements, and there is also a significant background contribution from the nuclear overlap region. Fortunately, centrality determination has been extended to UPCs via the nuclear Coulomb dissociation pattern. Figure~\ref{fig:MCD:probability} shows the impact parameter dependence of the probabilities, as calculated by STARLight for the LHC beam, for the three primary forward neutron topologies. These topologies are 0$n$0$n$, indicating no neutron emission in either direction, which dominates for impact parameters \(b > 40 \, \text{fm}\); X$n$0$n$, with neutron emission in only one direction, which is largest around \(b \approx 20 \, \text{fm}\); and X$n$X$n$, with neutron emission in both directions, which is important when \(b < 15 \, \text{fm}\). This behavior allows for the selection of UPC events over different impact parameter ranges based on neutrons detected by Zero Degree Calorimeters.

\begin{figure}[!htbp]
    \centering
    \includegraphics[width=0.75\textwidth]{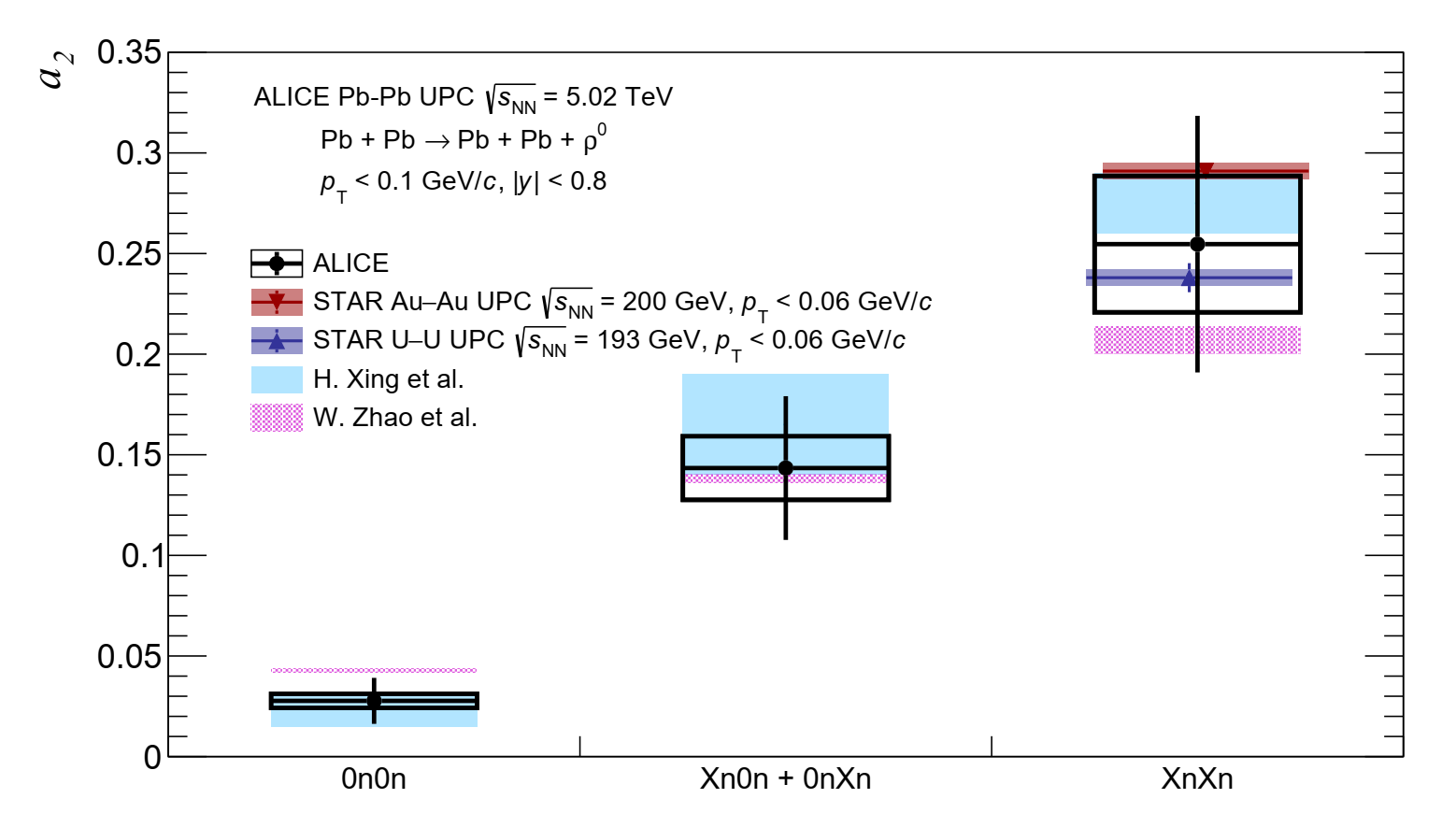} 
    \caption{The amplitudes of the \( \cos(2\phi) \) modulation of the \( \rho^0 \) yield in Pb+Pb collisions at \( \sqrt{s_{\mathrm{NN}}} = 5.02 \) TeV for all neutron emission classes. These results are compared with the model predictions by Xing et al.~\cite{Xing:2020hwh} and W. Zhao et al.~\cite{Mantysaari:2023prg}, and, for the XnXn class, with the STAR results~\cite{STAR:2022wfe} from Au-Au and U-U collisions at RHIC. Statistical uncertainties are indicated with bars, and systematic uncertainties with boxes for all experimental data points. From Ref.~\cite{ALICE:2024ife}.
    }
\label{fig:ALICE:rho:MCD}
\end{figure}

The neutron tagging technique was first employed by the STAR Collaboration to study the interference effect of $\rho^0$ photoproduction~\cite{STAR:2008llz}. This method was further refined by the CMS Collaboration to investigate the impact parameter dependence of two-photon processes in UPCs~\cite{CMS:2020skx}.
Recently, the ALICE Collaboration also utilized this method to investigate the spin interference in \(\rho\) photoproduction~\cite{ALICE:2024ife}. Figure~\ref{fig:ALICE:rho:MCD} presents the \( \cos(2\phi) \) modulation of the \(\rho^0\) yield in Pb+Pb collisions at \( \sqrt{s_{\mathrm{NN}}} = 5.02 \) TeV across all neutron emission classes. Spin interference was observed in all classes, with the measured anisotropy showing a clear anticorrelation with the impact parameter, increasing significantly by an order of magnitude from the 0n0n to the XnXn class. This phenomenon can be explained by a straightforward physical model: as the "slit" separation increases, the oscillations from spin interference also increase, leading to a decrease in the modulation strength of \( \cos(2\phi) \). Models~\cite{Xing:2020hwh,Mantysaari:2023prg} incorporating spin interference successfully reproduce this trend and reasonably describe the modulation magnitude.

 Altering the slit size and profile can also be done by changing nuclear species. At RHIC, nuclei such as gold (Au), uranium (U), oxygen (O), zirconium (Zr), and ruthenium (Ru) are used in collisions, while at the LHC, lead (Pb), xenon (Xe), and oxygen (O) are utilized. The differences in nuclear size and shape allow researchers to systematically study the impact of slit size and profile on the resulting photoproduction processes and interference patterns. Additionally, utilizing asymmetric slits in asymmetric heavy-ion collisions provides further control over the experimental conditions. At RHIC, asymmetric collisions, such as copper (Cu) colliding with gold (Au), could be utilized to explore how asymmetry affects the interference and diffraction behavior of the slits. These interference and diffraction behaviors can significantly alter the momentum distribution of photoproduced particles and the spin-interference pattern, providing valuable insights into the underlying quantum phenomena.

        \subsection{Tests of quantum entanglement utilizing photon-induced reactions}

        \hspace{2em}As previously mentioned, the $\rho^{0}$ meson spin-interference observed in the photoproduction process is closely related to the Einstein-Podolsky-Rosen (EPR) paradox. When a $\rho^{0}$ meson decays into a pair of pions ($\pi^{+} \pi^{-}$), their quantum states become entangled. The interference pattern in the angular distribution of the decay products arises from the overlap of the pions' wave functions over a considerable separation distance. This indicates that the momenta of the pion pairs are correlated such that they exhibit quantum interference despite being separated by significant distances, demonstrating nonlocality. This phenomenon exemplifies the EPR paradox, where measurements of one particle instantaneously collapse the state of another distant particle, regardless of the distance between them.

This continuous variable momentum-space quantum correlation is challenging to describe using entanglement language or to test using a Bell experiment. Entanglement is often observed through measurements of discrete properties such as spin or polarization. For instance, consider the entanglement of two spin-1/2 particles where the quantum state of each particle is described by a quantum state vector. The state of a particle can be represented as a linear combination of its basis states, typically denoted as $|0\rangle$ and $|1\rangle$ (corresponding to spin-up and spin-down states). An entangled state for two particles is a superposition of their joint states that cannot be decomposed into individual states. A common example is the Bell state: \(
|\psi \rangle = \frac{1}{\sqrt{2}} (|0, 1\rangle - |1, 0\rangle).
\)
In an entangled state, measuring the state of one particle instantaneously determines the state of the other, regardless of the physical separation between them.

Recently, the CMS Collaboration conducted a study to measure quantum entanglement~\cite{CMS:2024pts} in top quark-antiquark (\(t\bar{t}\)) events produced in proton-proton collisions at a center-of-mass energy of 13 TeV. They analyzed data recorded in 2016, corresponding to an integrated luminosity of 35.9 fb\(^{-1}\). The events were selected based on the presence of two oppositely charged high transverse momentum leptons. Top quark pairs originating from gluon-gluon fusion near threshold correspond to the Bell state, i.e., they are maximally entangled spin-singlets. An entanglement-sensitive observable \(D\) was derived from the spin-dependent parts of the \(t\bar{t}\) production density matrix, defined as:

\[
D = -\frac{\text{tr}[C]}{3} = -\frac{C_{11} + C_{22} + C_{33}}{3},
\]
where \(C\) is the spin correlation matrix. Values of \(D\) smaller than \(-1/3\) provide evidence of entanglement. The study found $D=-0.478^{+0.025}_{-0.027}$, with a significance of $5.1~\sigma$, demonstrating quantum mechanical entanglement in \(t\bar{t}\) pairs.

Theoretically, similar tests can be applied in UPCs for the process \(\gamma + \gamma \rightarrow t\bar{t}\). In this scenario, top quark pairs would correspond to the Bell state, satisfying the angular momentum conservation of the two linearly polarized photons. However, due to the large mass of top quarks, their production in UPCs at current collision energies is exceedingly rare. Consequently, researchers must turn to lower mass products, such as \(\tau\) pairs~\cite{Shao:2023bga}. Recently, ATLAS~\cite{ATLAS:2022ryk} and CMS~\cite{CMS:2024qjo,CMS:2022arf} Collaborations  reported the observation of the \(\gamma\gamma \rightarrow \tau\bar{\tau}\) process in ultraperipheral collisions. Events were selected based on the presence of one muon from a \(\tau\)-lepton decay and an electron or charged-particle tracks from the other \(\tau\)-lepton decay, with minimal additional central-detector activity and no forward neutrons. The \(\gamma\gamma \rightarrow \tau\bar{\tau}\) process was observed with a significance exceeding 5 standard deviations. In principle, an entanglement-sensitive observable \(D\) can also be derived from the spin-dependent parts of the \(\tau\bar{\tau}\) production density matrix. This provides a feasible method to probe entanglement proxies in photoproduction processes, and even to conduct a Bell test. Furthermore, other more abundant two-photon products could also be explored for this purpose.

        \subsection{Future opportunities}
        \subsubsection{Future measurements of vector meson photoproduction}
    
        \hspace{2em}The ongoing data taken from Run23 to Run25 at RHIC will provide the additional statistical precision necessary to conduct multi-differential analyses, offering stronger theoretical constraints. Specifically, multi-differential analysis of the $\cos{2\phi}$ modulation with respect to pair rapidity and pair mass is required. This analysis will help determine if pion pairs from $\rho^0$ production and from the continuum Drell-S\"oding production exhibit different interference effects.
Enhanced statistical precision is essential for measuring higher harmonics and rare photoproduction processes, which may be more sensitive to the gluonic structure of nuclei. Measurements of $J/\Psi \rightarrow l^+l^-$ at higher masses than that of $\rho^0$ mass offer better comparisons because the reference calculations are based on pQCD. Ultraperipheral A+A collisions, where photons generated by the Lorentz-boosted electromagnetic field of one nucleus interact with gluons inside the other nucleus, can provide 3D gluonic tomography measurements of heavy ions even before the operation of the future Electron-Ion Collider (EIC)~\cite{STAR:2022wfe}. The STAR collaboration has conducted experimental measurements of $J/\psi$ photoproduction at low transverse momentum in non-UPC heavy-ion collisions~\cite{STAR:2019yox}. Detailed studies with $p_{T}$ distributions have shown that the $|t|$ distribution in peripheral collisions is more consistent with a coherent diffractive process than an incoherent process. Although models~\cite{Zha:2017jch,Klusek-Gawenda:2015hja} incorporating partial coherent photon and nuclear interactions could explain the yields, it remains unclear how the coherent process occurs and whether final-state effects play any role~\cite{Zha:2018jin}. Resolving this puzzle with high-statistical data and detailed $|t|$ distributions at different centralities at RHIC, may be crucial for understanding the nature of coherent photoproduction, the formation of vector mesons, and the exclusivity of similar processes in future EIC experiments with forward neutron veto/tagging.

All four LHC detectors were substantially upgraded for the on-going LHC Run 3 (2023-2026).  These upgrades will allow them to take data at higher rates, and with increased precision.  For ALICE, one of the big improvements is a streaming data acquisition system which will save all of the data from lead-lead collisions \cite{Rohr:2022xcs}, eliminating the trigger that has been a major bottleneck for UPC physics.  This should allow for much higher UPC data collection rates, and allow studies of reactions that could not previously be studied.

Data projections for LHC Runs 3 and 4 predict that billions of $\rho^0$, $\rho'$ and $\phi$ should be produced, along with more than 10 million $J/\psi$ \cite{Citron:2018lsq}.  Although the LHC detectors do not have the full acceptance to observe all of these decays, samples of millions of $\rho$ and $\rho'$ should be collected, and the luminosity should permit the observation of a considerably wider range of mesons than have been studied to date.  Thousands of $\Upsilon\rightarrow l^+l^-$ should also be produced, although there may be significant irreducible broadband backgrounds from $\gamma\gamma\rightarrow l^+l^-$ which will need to be subtracted, increasing the statistical errors.  These data samples will allow for high-precision studies of interference observables.  As is discussed in the next section, it will also give access to events containing multiple vector mesons from a single ion-pair interaction. A brief oxygen-oxygen run is also planned, giving the LHC an opportunity to study two-source interference with smaller sources.  

Measuring $\phi$ photoproduction in UPCs has been challenging due to its soft decay into two kaons. The daughter kaons have an average transverse momentum of about $100~\mathrm{MeV/c}$, making tracking difficult at such low $p_{\rm T}$. Additionally, triggering UPC $\phi$ meson decay requires fast detectors, such as calorimeters or Time-of-Flight systems, located far from the vertex. Consequently, the soft kaons do not have enough momentum to reach these fast detectors, a challenge faced by both LHC and RHIC experiments. A potential solution has been proposed: utilizing the large cross section of $\phi$ and the high luminosity of the accelerator to reconstruct $\phi$ from zero-bias or ZDC-based triggers only, without the need for fast detectors in the barrel. This development may offer a unique opportunity to study $\phi$ photoproduction. Although the EIC is the ultimate machine for exclusive production, it will also struggle to reconstruct $\phi$ at low $p_T$ ({\it i. e.} at low $Q^2$) ~\cite{Arrington:2021yeb}.  Studies of $\phi$ production at larger $|t|$ (incoherent photoproduction) and/or $Q^2$ are likely to be easier, along with higher statistics studies of continuum $K^+K^-$ pairs, above the $\phi$ \cite{ALICE:2023kgv}.

For Run 4 at the LHC (2029--2032), the ALICE experiment will deploy the FoCal detector~\cite{ALICE:2020mso}, a high-granularity, compact silicon-tungsten (Si+W) sampling electromagnetic calorimeter with longitudinal segmentation, backed by a conventional high-granularity metal and scintillating hadronic calorimeter. While FoCal is designed for measuring direct photons at forward rapidity in pp and pPb collisions, studying UPCs is also an integral part of its physics program~\cite{Bylinkin:2022temp}. Future measurements of the energy dependence of $\sigma(\gamma \mathrm{p})$ and $\sigma(\gamma \mathrm{Pb})$ for vector meson photoproduction using the FoCal detector should provide clear signatures of gluon saturation, probe nuclear gluon shadowing, and test various theoretical approaches. Results from HERA indicated that the photoproduction cross section, $\sigma(\gamma~p~\rightarrow~J/\psi~p)$, increases with center-of-mass energy following a power law. ALICE~\cite{ALICE:2014eof,ALICE:2018oyo} extended this observation to $x \sim 10^{-5}$ without significant deviation from this behavior, though with large statistical error. The FoCal detector will access an unexplored kinematic regime at small-$x$ where deviations from the power-law trend might be observed as the saturation effect is widely believed to set in such small $x$ region. The quantitative predications have been made using various models,  such as the Hot Spot model~\cite{Cepila:2016uku}, the NLO BFKL~\cite{Bautista:2016xnp}, and CGC-based calculations~\cite{Armesto:2014sma}, predict saturation effects. Projections based on the NLO BFKL calculation  suggest that FoCal UPC measurements could provide the first observation of deviations from the power-law trend at high energies. Additionally, dissociative processes have garnered interest~\cite{Mantysaari:2016ykx,Mantysaari:2016jaz,Cepila:2016uku}. Moreover, it is feasible to probe the fluctuations of the proton target configurations for the  saturated gluonic matter with FoCal. Observing a significant reduction in the measured cross section with increasing energy would be a strong indication of gluon saturation. The ratio of dissociative-to-exclusive $J/\psi$ photoproduction as a function of $W_{\gamma \mathrm{p}}$, including H1 and ALICE data, and comparisons to BM and Hot Spot model predictions, will further shed new light on this phenomenon.

Looking ahead to the mid/late 2030s, the proposed ALICE 3 detector~\cite{ALICE:2022wwr} will offer acceptance for both charged and neutral particles over a wide solid angle, covering pseudorapidity $|\eta|<4$. This increased acceptance will enhance the study of more complex UPC final states. A STARlight simulation~\cite{Klein:2016yzr} indicated that the acceptance for photoproduced $\pi^+\pi^-\pi^+\pi^-$ would be 19 times higher than in the current ALICE detector. This improvement will significantly increase background rejection and allow for detailed studies of the energy dependence of photoproduction cross-sections.ALICE 3 will enable new studies, including the complete complete reconstruction photoproduction of open charm, where both the $c$-containing meson and the $\overline{c}$-containing hadron can be observed. These advancements will provide deeper insights into the underlying mechanisms of vector meson photoproduction and contribute to the broader understanding of QCD at high energies.

        \subsubsection{Double vector meson photoproduction}
    
          \hspace{2em}In near-grazing ($b\approx > 2R_A$) relativistic heavy-ion collisions, the probability of producing a single vector meson is significant - about 3\% chance to produce a $\rho^0$ with lead beams at the LHC \cite{Klein:1999qj}.  
Accordingly, one might expect that the probability of producing two $\rho^0$ (or heavier mesons) is also significant.   The cross section for multiple vector meson production is calculated in an impact-parameter dependent integral:
\begin{equation}
\sigma = \int d^2b \prod_{i=1}^n P_{Vi}(b).
\label{eq:multisigma}
\end{equation}
where the $P_{Vi}(b)$ are the probabilities to produce the given vector meson at impact parameter $b$.  It is also possible to add additional probabilities to account for the possibility of additional photon exchange to excite one nucleus, or two additional photons to account for both nuclei.  Each $P_{Vi}(b)$ scales as $1/b^2$, so the more photons, the smaller the typical impact parameter \cite{Baur:2003ar}.  It is possible to integrate Eq. \ref{eq:multisigma} to find the impact parameter distribution.  For single-photon exchange, the mean impact parameter depends on the minimum and maximum impact parameters:
\begin{equation}
\langle b \rangle = \frac{b_{\rm max}-b_{\rm min}}
{\ln{(b_{\rm max}/b_{\rm min})}}.
\end{equation}
Here, $b_{\rm min}\approx 2R_A$, while $b_{\rm max}$ depends on the maximum photon energy.  When $n$ photons are exchanged (with $n>1$), though, $b_{\rm ma}$ drops out, and the mean impact parameter becomes 
\begin{equation}
\langle b \rangle \approx b_{\rm min} \frac{2n-2}{2n-3},
\end{equation}
a small numerical multiple of $b_{\rm min}$.

This approach can be used to determine the cross-sections for multi-vector meson production \cite{Klein:1999qj}.  The rates depend on the base production probability - about 3\% to produce a $\rho$ when $b=2R_A$ with lead beams at the LHC.  Then the probability to produce two $\rho$ is just $(3\%)^2/2$, which leads to a cross-section large enough to be observable. 

There is one critical assumption here: that the production processes are independent.   It has been shown that photon emission is independent, as long as the total photon energy is small compared to the emitter energy \cite{Gupta:1955zza}.  The second concerns scattering.  Although it has not been proven that multiple simultaneous elastic scattering will occur independently, the fact that the target nuclei are not significantly affected leads to some confidence that this is true  \cite{Baur:2003ar}.  

The same approach has also been applied to calculate the joint rapidity distributions for the production of two vector mesons
\cite{Klusek-Gawenda:2013dka,Azevedo:2022ozu}. 

A system composed of two (or more) vector mesons offers a rich quantum mechanical phenomenology, with multiple opportunities for interference \cite{Klein:2024wos}.    In analyzing the situation, one must distinguish between two different cases: nonidentical mesons ({\it e. g. $\rho J/\psi$}) or identical mesons, like $\rho^0\rho^0$.   This is similar to interferometry involving two entangled photons, which admit a much richer phenomenology than single-photon interferometers \cite{RevModPhys.71.S274}.

\begin{figure}[!tbp]
    \centering
    \includegraphics[width=0.85\textwidth]{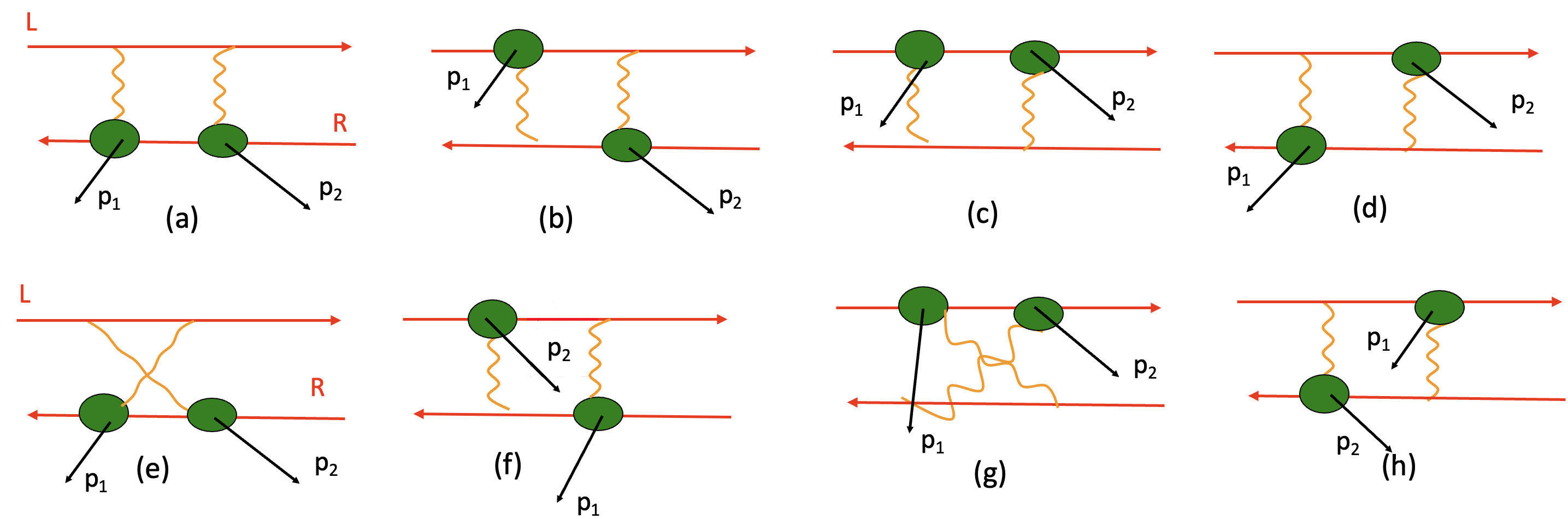} 
    \caption{(top) The four diagrams that contribute to the photoproduction of two non-identical mesons. (bottom) The four additional diagrams that contribute for the production of two identical mesons.  From Ref. \cite{Klein:2024wos}. }
\label{fig:diagrams}
\end{figure}

For non-identical mesons, the four diagrams shown on the top row of Fig. \ref{fig:diagrams} can contribute to the production of meson pairs.  In this case, the interference patterns factorize into the products of the two individual meson productions alone.  For example, for one meson at $y=0$, the cross section goes to zero as $p_T\rightarrow 0$, regardless of the kinematics of the other meson.   

The production of two identical particles is more interesting, since the four additional diagrams on the bottom row of Fig. \ref{fig:diagrams} can contribute.  The two mesons are indistinguishable, so each photon can lead to either meson.  

Here, we will consider one especially interesting case: forward production of two light mesons.  At forward rapidity, the two possible photon energies are very different.  Because the photon flux scales roughly as $1/k$ and the $\gamma A\rightarrow VA$ cross section varies slowly with energy, most of the production at forward rapidities corresponds to the lower photon energy solution.  So, the two right diagrams in Fig. \ref{fig:diagrams} dominate forward production.  These diagrams lead to a forward $P(b)$ that may be written

\newcommand*{\ppa}{\vec{p_1}}
\newcommand*{\ppb}{\vec{p_2}}

\begin{equation}
\begin{split}
P(b) = \bigg| \Sigma_i\Sigma_j
& A_L(\ppa,b)A_L(\ppb,b) \exp(i(\vec{k}_1\cdot\vec{b}_i+\vec{k}_2\cdot\vec{b}_j)) \\
& + A_L(\ppb,b)A_L(\ppb,a) \exp(i(\vec{k}_2\cdot\vec{b}_i+\vec{k}_1\cdot\vec{b}_j))
\bigg|^2.
\end{split}
\label{eq:forward}
\end{equation}
The sums over $i$ and $j$ run over the nucleons at positions $\vec{b}_i$ in the target to show the conditions for coherent emission from the nucleus.  $A_L$ is the amplitude to produce a meson with momentum $\ppa$ at a given $b$; this accounts for both the photon flux (at impact parameter $b$) and the photon-to-vector-meson conversion.

The exponential imposes a condition for coherent production which is essentially $|\vec{k}_1|,|\vec{k}_2| < \hbar/R_A$.  However, because the two terms in the exponential are added, the coherence requirement becomes tighter: $|\vec{k}_1|,|\vec{k}_2| < \hbar/2R_A$.  

When this condition is satisfied, the exponential term is one.  For meson production at similar rapidity ({\it i. e.} similar photon energy), the two amplitudes are equal, and $P(b)$ becomes
\begin{equation}
P(b) = |2N^2   A_L(\ppa,b)^2|^2 = 4 N^4 A_L(\ppa,b)^4.
\label{eq:twoforward}
\end{equation}
where $N$ is the number of nucleons in the target - double what it would be for non-identical mesons. The cross section scales as the fourth power of the amplitude and the number of nucleons because two photons are exchanged.

        \subsubsection{Vector meson photoproduction at very large photon energy}
    
        \hspace{2em}As photon energy rises, the minimum momentum transfer from the nucleus drops, and the coherence length, 
\begin{equation}
 l_c=\hbar/p_{min} = 2\hbar k /M_V^2 
 \label{eq:lc}
\end{equation} 
rises.   At high enough photon energies, a new coherence regime appears, where photoproduction can be coherent across multiple target nuclei.  For $k > \approx 3\times10^{14}$ eV, the momentum transfer is small enough that the coherence length becomes larger than the typical inter-atom separation in solids or liquids, so multiple nuclei may participate coherently in a single interaction \cite{Couderc:2009tq}.  

Multi-nuclear coherence requires that the transverse momentum transfer is also very small, so the photon interaction is not well localized transversely, and coherent scattering becomes increasingly peaked in the forward direction.

Several factors enhance this effect.  The cross-section continues to rises with increasing energy.  The rate of increase is not well known.  The Regge trajectory increase ($\sigma\propto W^{0.22}$) cannot continue indefinitely, because the target will become a black disk, and the rise in the cross section will be moderated. Still, a continued slow rise is expected, as the black disk continues to grow slowly as the energy rises. 

Also in this energy regime, the $e^+e^-$ pair production cross-section is suppressed by the Landau-Pomeranchuk-Migdal effect \cite{Klein:1998du} and photoproduction becomes the dominant photon interaction, with $\rho$ photoproduction the largest single component of photoproduction. 

At these very high energies, a photon can scatter from the bulk medium, effectively converting into a vector meson.  In the forward limit, the transverse coherence length goes to infinity as the transverse momentum transfer goes to zero.  The longitudinal coherence limit is the smaller of the coherence length from Eq. \ref{eq:lc} or the vector meson decay length $L=\gamma \beta c\tau=\hbar kc/M_V\Gamma$.  When the photon energy is large, this is a sizable distance, so this coherence should lead to a large,  increase in the forward cross section, and eventually photons will be dominated by this $\rho$ component.  The resulting $\gamma$-to-$\rho$ conversion probability has been calculated for the case where it is not too large - the beginning of dominance \cite{Couderc:2009tq}.  This occurs at energies around $10^{23}$ eV \cite{Couderc:2009tq}, which sadly seems beyond experimental reach.

A full calculation, suitable at even higher energies, would also include back-conversion, along with additional vector meson states.  For very high photon energies, dipoles with high masses and small transverse sizes become more important, so higher mass states may play a significant role \cite{Rogers:2009zza}.  This could lead to the need to include additional vector mesons in the calculation (in a manner similar to generalized vector meson dominance \cite{Frankfurt:1997zk}), although the $\rho$ should remain the most important \cite{Couderc:2009zz}

The two level ($\gamma$ and $\rho$) process is similar to kaon regeneration \cite{DOVER19821}, involving a transition between states with similar quantum numbers ($K_S$ vs. $K_L$ or photon vs. vector meson) which is induced by elastic scattering. For kaons, periodic targets have been proposed as a way to alter the regeneration probability \cite{Akhmedov:2001xt}.  It might be possible to use a periodic target to reduce the threshold energy for $\rho$ conversion, but probably not to the point where it could be studied in a terrestrial experiment.

	\section{Summary}\label{sec:sum}
         \hspace{2em}This review provides an in-depth overview of both experimental measurements and theoretical models of the quantum phenomena recently explored through photoproduction in relativistic heavy-ion collisions. We  introduce the historical key milestones and the paradigm shift initiated by quantum theory, such as double-slit interference, which are relevant to our subject. This review analyzes experimental measurements of photoproduction processes in electron/proton and heavy-ion collisions, detailing the mechanisms of high-energy photon production, the experimental setups at RHIC and LHC, and the ultra-peripheral collisions that utilize Weizsäcker-Williams photons. 
Significant emphasis is placed on the spin interference phenomena recently observed in the diffractive $|t|$ spectra of $\rho^0$ photoproduction. Theoretical frameworks such as the Vector Meson Dominance model and the color dipole approach are introduced to interpret these observations. This exploration links experimental data with underlying fundamental quantum mechanics principles and discusses new insight into how to quantitatively extract nuclear effect from the vector meson photoproduction. Furthermore, the review details the coherence requirements for interference, the quantum entanglement paradox and the corresponding observation effects in photoproduction in high-energy nuclear physics, and addresses future opportunities which could provide deeper insights into the gluon transverse spatial distributions within nuclei and the broader quantum mechanical framework.

In conclusion, this review synthesizes experimental results and theoretical models to offer a comprehensive snapshot of unique quantum phenomena in the photoproduction in relativistic heavy-ion collisions. It underscores the critical role of the polarization of the photon source, quantum interference and nuclear effect on the gluon distribution in the ongoing experimental advancements and theoretical developments, and paves a way for quantitatively probing the quantum nature of these high-energy nuclear collisions.
	
	\section{Notations and Conventions}\label{nota}

 \hspace{2em}In this review, we adopt natural units ($\hbar c$) with $\rm{GeV}$ as the default energy  scale. Our conventions for relativity follow all recent field theory texts. We use the metric tensor with Greek indices running over 0, 1, 2, 3 or $t$, $x$, $y$, $z$. Roman indices denote only the three spatial components.  Einstein's summation convention is employed to handle repeated indices. Four vectors are denoted by light italic type; three vectors are denoted by boldface type; unit three vectors are denoted by a light italic label with a hat over it.

 The selection of symbols is motivated by the need for precision and uniformity in conveying concepts throughout the review. Each symbol introduced serves a specific purpose within the context of our discussion. For clarity, we define the key symbols and provide illustrative examples to demonstrate their meaning. Readers are encouraged to refer to the symbol index at the end of this section for a comprehensive overview of the symbols used and their corresponding meanings.

  \begin{itemize}
        \item $\omega$: energy of a quasi-real photon
        \item $k$: four momentum of a quasi-real photon
        \item $\mathbf{k}$: three momentum of a quasi-real photon 
         \item $\mathbf{k_{\perp}}$: transverse momentum of a quasi-real photon  
        \item $\Delta$: the recoil four momentum from a target nucleus
        \item $\mathbf{\Delta}$: the recoil momentum from a target nucleus
        \item $\mathbf{\Delta_{\perp}}$: the recoil transverse momentum from a target nucleus
        \item $y$: the rapidity of the photoproduced vector meson or dilepton pair
        \item $\mathbf{p_{\perp}}$: the transverse momentum of the photoproduced vector meson or dilepton pair
        \item $\mathbf{\beta}$: the velocity of the relativistic nucleus, in the lab frame.
        \item $\gamma$: the Lorentz contraction factor of the relativistic nucleus in the lab frame.
        \item $\alpha$: the fine-structure constant
        \item $Z$: the charge carried by the nucleus
        \item $A$: the atomic number of the nucleus
        %added by SRK:
        \item $W_{\gamma p}$: photon-nucleon center of mass energy
        \item $d\sigma_{\rm coherent}/dt$: the differential cross-section for coherent production
        \item $R_A$: nuclear radius
        \item $m_p$: proton mass (or this could be changed to $m_N$, the nucleon mass)
     
    \end{itemize}
 
\section{Acknowledgements}
 \begin{comment}
     \begin{itemize}
        \item coherent vs. incoherent photoproduction - addition of amplitudes
        \item 0n0n xn0n and xnxn and selection of impact parameter 
        \item cross-sections and $p_T$ spectra
        \item Angular modulations in ee, rho, etc.
        \item angular modulations in $\gamam\gamma$ to dileptons
        \item Good Walker picture
        \item Experimental results that challenge Good-Walker
        \item Incoherent measurements and access to fluctuations (connection to GW)
        \item interference in asymmetric collision systems as a double-slit (with different width slits) - tests in RHIC CuAu, LHCb SMOG
        \item discussion of loss of coherence in non-UPC events (both in terms of the interference effect and in terms of the GW picture of the target)
        \item O+O and alpha cluster - can we use $\gammaA$ to study nuclear structure
        \item polarized targets with non-spherical target nuclei
        \item photon flux - coherent and incoherent (the latter is only potentially significant for peripheral collisions)
        
    \end{itemize}
    \end{comment}
\hspace{2em}The authors would like to thank Prof. Shi Pu, Prof. Qun Wang, Prof. Bowen Xiao, Prof. Hongxi Xin, Prof. Yu Jia and many members of the PHENIX, STAR, ATLAS, ALICE, CMS, LHCb and ePIC Collaborations. 

The work of WZ is supported in part by National Key R\&D Program of China with Grant No. 2022YFA1604900, CAS Project for Young Scientists in Basic Research with Grant No. YSBR-088, National Natural Science Foundation of China with Grant No. 12305145, Anhui Provincial Natural Science Foundation No. 2208085J23 and Youth Innovation Promotion Association of Chinese Academy of Sciences. The work of SY is supported in part by National Natural Science Foundation of China with Grant No. 12275091, GuangDong Basic and Applied Basic Research Foundation with Grant No. 2023A1515010416. The work of SK is supported in part by the U.S. Department of Energy, Office of Science, Office of Nuclear Physics, under contract numbers DE-AC02-05CH11231. The work of XZB is supported in part by the U.S. DOE Office of Science under contract Nos. DE-FG02-89ER40531, DE-SC0012704, DE-FG02-10ER41666, and
DE-AC02-98CH1088. The work of JDB is supported in part by the U.S. Department of Energy, Office of Science, Office of Nuclear Physics, under contract number DE-SC0024189 as part of the Early Career Research Program. The work of JZ has been supported by the National Science Foundations of China under Grant No. 12175118
and 12321005.

	\renewcommand*{\thesection}{\Alph{section}}

	\bibliography{mybibfile.bib}
	
\end{document}